\documentclass[pra,twocolumn,showpacs,superscriptaddress,amssymb]{revtex4}

\usepackage[colorlinks=true,linkcolor=magenta,citecolor=magenta,urlcolor=magenta,linktocpage=true]{hyperref}
\usepackage[utf8]{inputenc}
\usepackage[english]{babel}
\usepackage[T1]{fontenc}
\usepackage{amsmath}
\usepackage{amsthm}
\usepackage{cleveref}
\usepackage{svg}
\usepackage{tabularx}
\usepackage{array, makecell}
\usepackage{multirow}

\usepackage{float}

\usepackage{listings}
\usepackage{xcolor}

\lstdefinestyle{mystyle}{
    backgroundcolor=\color{backcolour},   
    commentstyle=\color{codegreen},
    keywordstyle=\color{magenta},
    numberstyle=\tiny\color{codegray},
    stringstyle=\color{codepurple},
    basicstyle=\ttfamily\footnotesize,
    breakatwhitespace=false,         
    breaklines=true,                 
    captionpos=b,                    
    keepspaces=true,                 
    numbers=none,                    
    numbersep=5pt,                  
    showspaces=false,                
    showstringspaces=false,
    showtabs=false,                  
    tabsize=2
}
\usepackage{psfrag,graphicx}
\usepackage{dcolumn}
\usepackage{bm}
\usepackage{amsfonts,amssymb,amsmath}        
\usepackage{slashed}
\usepackage[utf8]{inputenc}
\usepackage{graphicx}
\usepackage[caption=false]{subfig}
\usepackage{cancel}
\usepackage{xcolor}
\usepackage{amsmath}
\usepackage{cleveref}
\crefformat{footnote}{#2\footnotemark[#1]#3}
\definecolor{codegreen}{rgb}{0,0.6,0}
\setcitestyle{numbers,square}

\makeatletter

\renewenvironment{widetext@grid}{%
  \par\ignorespaces
  \setbox\widetext@top\vbox{%
   \vskip15\p@
   \hb@xt@\hsize{%
    \leaders\hrule\hfil
    \vrule\@height6\p@
   }%
   \vskip6\p@
  }%
  \setbox\widetext@bot\hb@xt@\hsize{%
    \vrule\@depth6\p@
    \leaders\hrule\hfil
  }%
  \onecolumngrid

  \let\set@footnotewidth\set@footnotewidth@ii
}{%
  \par

  \twocolumngrid\global\@ignoretrue
  \@endpetrue
}%

\makeatother

\begin{document}
\title{The Closed and Open Unbalanced Dicke Trimer Model: Critical Properties and Nonlinear Semiclassical Dynamics}

\author{Cheng Zhang}
\affiliation{Key Laboratory for Microstructural
Material Physics of Hebei Province, School of Science, Yanshan
University, Qinhuangdao 066004, China}

\author{Pengfei Liang}
\email{pfliang@gscaep.ac.cn}
\affiliation{Graduate School of China Academy of Engineering Physics, Haidian District, Beijing, 100193, China}

\author{Neill Lambert}
\email{nwlambert@gmail.com}
\affiliation{Theoretical Quantum Physics Laboratory, Cluster for Pioneering Research, RIKEN, Wakoshi, Saitama 351-0198, Japan}

\author{Mauro Cirio}
\email{cirio.mauro@gmail.com}
\affiliation{Graduate School of China Academy of Engineering Physics, Haidian District, Beijing, 100193, China}

\date{\today}

\begin{abstract}
We study a generalization of a recently introduced Dicke trimer model [Phys.~Rev.~Lett. 128, 163601, Phys.~Rev.~Research 5, L042016], which allows for cavity losses and unbalanced light-matter interactions (in which rotating and counter-rotating terms can be tuned independently). In the original description of a Dicke trimer, three Dicke models are coupled in a ring topology via a complex photon hopping whose complex phase describes a synthetic magnetic field threading the loop. This original model features several intriguing equilibrium phases and critical phenomena such as frustrated superradiance, two-critical scalings in the frustrated superradiant phase, and finite critical fluctuations in the anomalous normal phase. Here, we find that in the extreme unbalanced limit, where  only rotating terms are present,  the $U(1)$ symmetry of the Tavis-Cummings model is restored, qualitatively altering the critical phenomena  in the superradiant phase due to the presence of a zero-energy mode. To analyze this general regime, we develop a semiclassical theory based on a re-quantization technique. This theory also provides further physical insight on a recently reported anomalous finite critical fluctuations in the time-reversal broken regime. Moving to the open-Dicke case, by introducing local dissipation to the cavities, 
we observe the emergence of a rich range of nonequilibrium phases characterized by trivial and non-trivial dynamical signatures. In the former case, 
we identify,  when time-reversal symmetry is present, a new stationary phase that features superradiant states in two of the three cavities and a normal state in the other cavity.
In the latter case, we observe the emergence of dynamical phases in which the system exhibits superradiant oscillations, characterized by periodic or chaotic phase space patterns. 
The landscape of transitions associated with these dynamical phases features a wide range of qualitatively different behaviours such as Hopf bifurcations (followed by period-doubling cascades or quasiperiodic oscillations), anomalous Hopf bifurcations (with burst-oscillation-like post-bifurcation dynamics), collisions between basins of attraction (associated with different symmetry-broken equilibria), and exterior crises (featuring transient chaotic dynamics). We 
highlight how the two-critical-scalings feature of the closed model is robust under dissipation (with doubled critical exponents)
while the phenomenon of anomalous finite critical fluctuations becomes a mean-field scaling (as a consequence of Hopf bifurcations of the equilibria featuring the normal state) in the open model. 
\end{abstract}

\maketitle

\section{Introduction}\label{sec:intro}
The theory of critical phenomena lies at the heart of our understanding of quantum phase transitions (QPTs) \cite{RevModPhys.69.315,qptbook}. Continuous QPTs occur at zero temperature and exhibit a number of unique characteristics, including the presence of degenerate ground states with spontaneously broken symmetries and the closing of the spectral gap. 
A QPT is normally associated with some diverging length and time scales. As a consequence, both the statistics and the dynamics near the transition are characterized by universal scaling laws which are independent from the microscopic details of the model. In turn, this allows to classify different QPTs according to critical exponents which can be used to describe the scaling of the divergent properties \cite{scalingbook}. 

Due to considerable progress in the experimental control and manipulation of  quantum degrees of freedom, quantum systems made of bosonic modes, spins, and atomic ensembles have emerged as promising platforms for exploring QPTs and the associated critical phenomena. A paradigmatic example of such systems is the Dicke model~\cite{https://doi.org/10.1002/qute.201800043}, where a single bosonic mode is homogeneously coupled to a large ensemble of two-level atoms via  a dipole interaction. In this system, the atoms can coherently and constructively interact with light, leading to enhanced levels of radiation in the ground state~\cite{PhysRev.93.99} or the steady state~\cite{PhysRevA.7.831,PhysRevA.8.1440,PhysRevA.8.2517,PhysRevA.98.063815} of dissipative-driven systems. This prototypical Dicke model undergoes a superradiant phase transition (SPT) characterized by mean-field critical exponents when the light-matter coupling is comparable to the frequencies of the bosonic mode and the atomic ensemble. Experimental realizations of the Dicke model and SPT have been achieved in cavity QED systems~\cite{Baumann2010,PhysRevLett.107.140402,doi:10.1073/pnas.1417132112,Zhiqiang:17,PhysRevLett.121.163601,doi:10.1126/science.abd4385}, trapped ions~\cite{PhysRevLett.121.040503, doi:10.1126/science.abi5226} and ultracold atoms in a cavity~\cite{Zhiqiang2018DickemodelSV,PhysRevLett.91.203001}.

In this context, a recent paper by Zhao and Hwang~\cite{PhysRevLett.128.163601} proposes the realization of a "frustrated superradiance" phase in the so-called Dicke-lattice model, wherein an odd number of Dicke models are placed in a ring geometry and allowed to directly interact by photon hopping between cavities. In particular, the case of three Dicke models is called the Dicke trimer model. In the superradiant phase, the ground state energy function of each Dicke model is a double-well potential and the resulting macroscopic classical cavity field can, intuitively, be interpreted as an Ising spin with a variable amplitude. From this point of view, the photon hopping between cavities effectively acts as a magnetic exchange-coupling which introduces frustration in the antiferromagnetic case. Ultimately, this leads to the existence of a frustrated superradiant phase (FSP) characterized by the breaking of translational symmetry~\cite{PhysRevLett.128.163601}. Interestingly, the phase transition to the normal phase (NP) exhibits a novel two-scaling feature, in which critical exponents of mean-field and unconventional type coexist. 

In a follow-up work~\cite{zhao2022anomalous} (similar results were also reported in~\cite{PhysRevLett.129.183602}), the same authors further considered the case where the Dicke ring is threaded by a synthetic magnetic field, making the photon hopping amplitudes complex, thereby lifting the time-reversal symmetry. Interestingly, this model features an anomalous normal phase exhibiting finite critical fluctuations and multi-critical points hosting multiple critical modes with distinct critical exponents. These findings make the hopping-coupled Dicke ring model an intriguing platform for exploring new phases of matter and QPTs.
 
It is relevant to further analyze how these features describing closed quantum systems carry over to an open setting. This is particularly crucial in light-matter systems where  the interaction to the external electromagnetic environment is
often inevitable. In fact, dissipation can significantly affect the properties of quantum many-body systems, leading to exotic nonequilibrium phases~\cite{PhysRevLett.120.146402,PhysRevB.98.094308,PhysRevLett.124.086801,PhysRevLett.124.056802,PhysRevResearch.5.043004} and critical phenomena~\cite{PhysRevLett.110.195301,PhysRevLett.122.040604,PhysRevResearch.2.033018}. 
Interesting examples of these physics-rich models include lattices of bosonic gasses subject to engineered dissipation channels (characterized by a nonequilibrium phase transition into a steady state without any long-range order \cite{PhysRevA.83.013611}) and an atomic ensemble coupled to a lossy cavity (characterized by  anomalous multicritical phenomena and coexistence of phases~\cite{PhysRevLett.120.183603}). Furthermore, nonstationary dynamical phases can arise in dissipative many-body systems characterized by long-time states which display  unusual space-time order~\cite{PhysRevLett.121.035301,PhysRevB.103.184308} or complex nonlinear dynamics towards chaos~\cite{PhysRevResearch.2.033131}. 

\begin{figure}
\includegraphics[clip, width = 0.98\columnwidth]{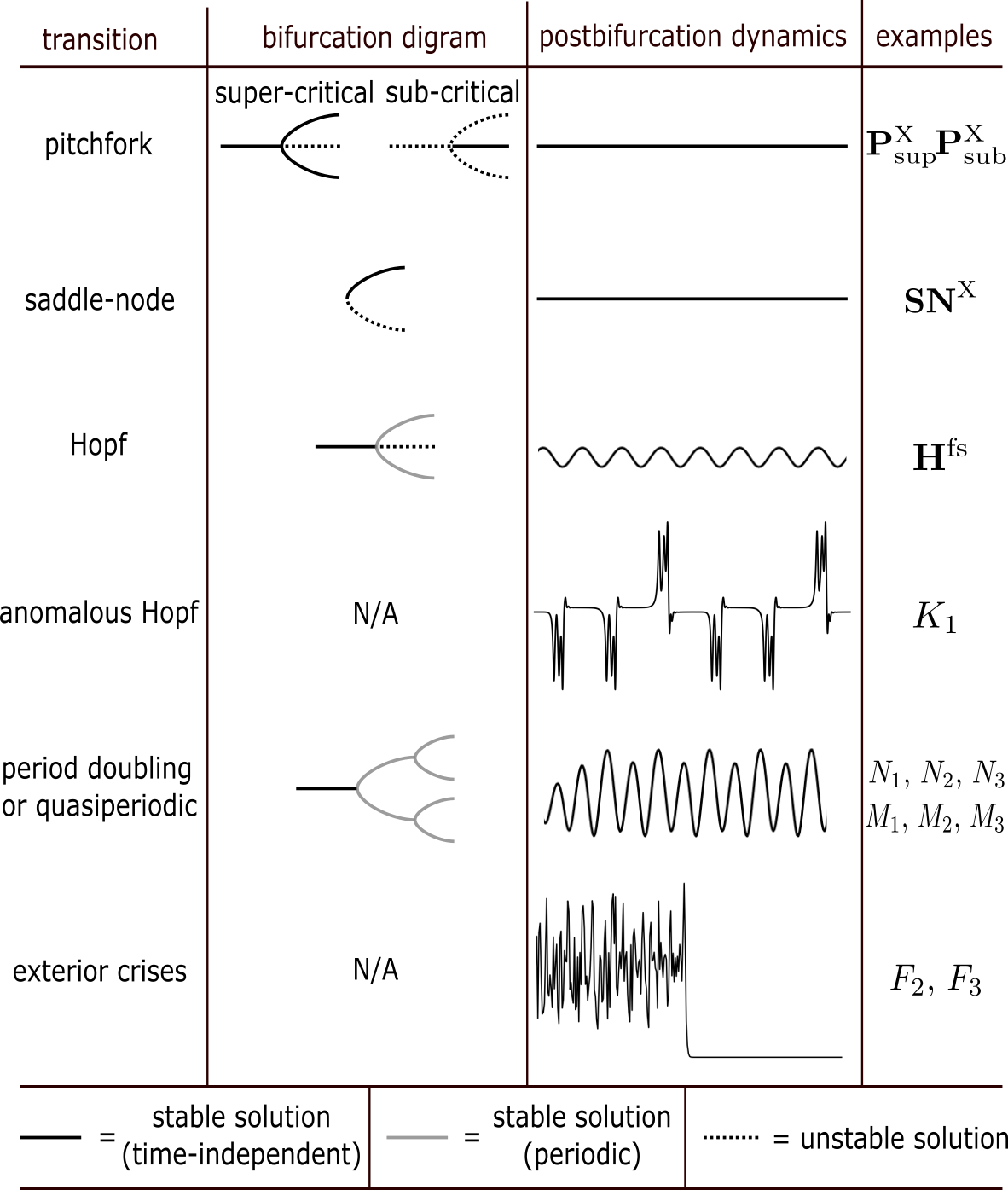}
\caption{Transitions classified by their bifurcation classes and/or dynamical features in the open Dicke trimer model. In the last column we give typical examples as they appear in Fig.~\ref{fig:openpd} for each type of transition. The superscript $\text{X}$ takes value in the set $\{\text{nfs},\,\text{fs},\,\text{mix}\}$. The label ``N/A'' means bifurcation analysis is not applicable. 
}\label{fig:tableofbifurcations}
\end{figure}

In this work, we consider the unbalanced version~\cite{PhysRevResearch.2.033131} of the Dicke trimer model with and without broken time-reversal symmetry and in both the closed and dissipative configurations (via cavity losses). 
We do this by allowing the rotating and counter-rotating terms in the light-matter interaction to be tuned independently through an anisotropy parameter (tuned to one to recover the standard Dicke model) similarly to its introduction for the Rabi model in \cite{PhysRevLett.119.220601}. 

The interplay between these terms is found to produce significant consequences, such as the appearance of interesting multi-critical points and unconventional critical scaling laws~\cite{PhysRevLett.119.220601,PhysRevLett.120.183603}. In the absence of cavity dissipation, i.~e.~, for the closed Dicke trimer model, we find the emergence of a zero-energy mode as a result of the $U(1)$ symmetry in the Tavis-Cummings limit when only rotating terms are present. Away from this limit, SPTs are controlled by the same fixed points as in the balanced, closed Dicke trimer model. Complementary to the method used in Ref.~\cite{zhao2022anomalous}, here we derive an effective semiclassical model to analyze the anomalous finite critical fluctuations in the unbalanced, closed model in all regimes. This is done by averaging over the matter degrees of freedom and by a canonical re-quantization of light.

We then discuss the open quantum dynamics of the Dicke trimer model in the presence of cavity losses. In the semiclassical (or thermodynamic) limit, the dynamics is generated by a set of nonlinear differential equations. 
These equations have both stable stationary solutions and dynamical ones which are non-stationary in the long time limit. We will use bifurcation diagrams to represent these solutions and analyze their transitions. 

In these diagrams, new equilibrium states are present which are not allowed in the equilibrium case. In particular, near the Tavis-Cummings line, we observe the emergence of a new type of equilibrium solution that features superradiant states in two cavities and a normal state in the remaining cavity. Most interestingly, we identify a variety of transitions, which are characterized by different dynamical signatures and are responsible for the arising of these nonequilibrium states.
 
To help navigate through the phase diagram, we list in Fig.~\ref{fig:tableofbifurcations} the transitions identified through a bifurcation analysis, together with their bifurcation classes, typical dynamical features and representative examples in the present model. Apart from those (super- and sub-critical pitchfork, saddle-node bifurcations related to equilibria, and Hopf bifurcations related to periodic solutions) reported in the open Dicke model in Ref.~\cite{PhysRevResearch.2.033131}, we also find anomalous Hopf bifurcations after which the dynamics shows unusual burst oscillations, and exterior crises that display transient chaos in the dynamics. The former occurs when the Jacobian matrix at the bifurcation point possesses a pair of zero eigenvalues, instead of a pair of conjugate purely imaginary ones seen in conventional Hopf bifurcations; while the latter arises from the collision of a chaotic attractor and the basin of attraction of an equilibrium point. Lastly, we numerically examine steady-state fluctuations, in particular, in the cases where the equilibrium setting exhibits the two-scaling feature and finite critical fluctuations. 

In this context, we have two main results. First,  we find that the presence of two critical scalings is robust under dissipation, which causes the critical exponents to double with respect to the equilibrium setting. Second, we find that the phenomenon of finite critical fluctuations no longer exists due to the Hopf bifurcations of the branch of the normal state equilibria.

The paper is organized as follows. In Sec.~\ref{sec:model} we outline the unbalanced Dicke trimer model and the methods we employed to solve it, including expansion of its Hamiltonian in terms of the number of atoms, its semiclassical approximation as a set of nonlinear differential equations, and direct evaluation of steady-state fluctuations. In Sec.~\ref{sec:lossless} we analyze the phase diagram, excitation spectra and critical exponents of the closed Dicke trimer model. The semiclassical theory for the interpretation of the anomalous finite critical fluctuations is then presented in Sec.\ref{sec:semitheory}. In Sec.~\ref{sec:lossy} results of the open Dicke trimer model, like the bifurcation diagrams, bifurcation analysis and scaling analysis of steady-state fluctuations, are elaborated. Conclusions and an outlook are given in Sec.~\ref{sec:conclusion}.

\section{Models and Methods}\label{sec:model}

\subsection{The Closed and Open Dicke Trimer Models}

We start by introducing the Hamiltonian for the generalized Dicke trimer model as
\begin{equation}
\label{eq:H1}
H_N = \sum_{n=1}^N H_n^{\mathrm{Dicke}} + J(e^{i\varphi}a_n^\dagger a_{n+1}+e^{-i\varphi}a_{n+1}^\dagger a_n),
\end{equation}
where $N=3$, and the Hamiltonian for the Dicke model as
\begin{equation}\label{eq:HDicke}
\begin{split}
H_n^{\mathrm{Dicke}} &= \omega_0 a_n^\dagger a_n + \omega_a J_n^z \\
&\quad+ \frac{2\lambda\eta_+}{\sqrt{N_a}}J_n^x(a_n+a_n^\dagger) + \frac{2\lambda\eta_-}{\sqrt{N_a}}iJ_n^y(a_n-a_n^\dagger). 
\end{split}
\end{equation}
with $\eta_\pm = (1\pm\eta)/2$. Here, $\eta$ and $\lambda$ tune the anisotropy and strength of the light-matter interaction, respectively. We take periodic boundary condition $a_{N+1} = a_1$,  where $a_n$ is the annihilation operator of the  bosonic resonant mode in the $n$th cavity. This mode has frequency $\omega_0$ and its position and momentum quadratures are defined by $q_n=(a_n+a_n^\dagger)/\sqrt2,~p_n=-i(a_n-a_n^\dagger)/\sqrt2$ such that they satisfy the commutation relations $[q_n, p_n]=i$. 

The atomic ensemble in each cavity is made of $N_a$ two-level atoms with frequency $\Omega$, and it is described by the collective spin operators $J_n^{x,y,z}=\sum_{m=1}^{N_a} s_m^{x,y,z}$ with $s_m^{x,y,z}$ representing a single spin-half. Without loss of generality, we assume $J>0$ and the phase $\varphi\in[0,\pi]$  to interpolate between the antiferromagnetic ($\varphi=0$) and ferromagnetic ($\varphi=\pi$) limits which were already thoroughly investigated in Ref.~\cite{PhysRevLett.128.163601}. 

The phase $\varphi$ effectively tunes the flux of the synthetic magnetic field threading the Dicke ring while the 
anisotropy parameter $\eta$ is introduced to go beyond the balanced case ($\eta=1$) and further interpolate between the isotropic ($\eta=0$) and anisotropic ($\eta\neq0$) cases. In the asotropic case $H_n^{\mathrm{Dicke}}$ owns a $U(1)$ symmetry defined by $G_\zeta=\prod_n \exp{[i\zeta(a_n^\dagger a_n + J_n^z + N_a/2)]}$ with $\zeta\in\mathbb{R}$,  while in the anisotropic case only the $Z_2$ parity symmetry, described by $G_\pi$, is left. We will see later this causes important effects in the excitation spectra. 

Apart from this symmetry, $H_N$ possesses the $Z_N$ translational symmetry $T$ which acts as $Ta_n(J_n^{x,y,z})T=a_{n+1}(J_{n+1}^{x,y,z})$ and satisfies $T^N=1$ due to periodic boundary condition. At $\varphi=0,\,\pi$, $H_N$ also has the time-reversal symmetry $K$ with $K$ representing the complex conjugate operation and the reflection symmetry $R$ defined by exchanging a pair of indices. We note that the balanced model in Ref.~\cite{zhao2022anomalous} is a special case of the Hamiltonian in Eq.~(\ref{eq:H1}) in which $\eta=1$, i.~e., only the standard Dicke terms in Eq.~(\ref{eq:HDicke}) are present. As shown in \cite{zhao2022anomalous}, in the time-reversal broken case ($\varphi\neq 0,\pi$) this closed Dicke trimer model  exhibits a number of appealing critical phenomena, such as the emergence of finite critical fluctuations in the anomalous NP and the re-appearance of two critical scalings at a multicritical point $\varphi_\text{tr}$. 

The Hilbert space of our model is $\mathcal{H}=\bigotimes_{n=1}^N\mathcal{H}_n$, where $\mathcal{H}_n$ is the Hilbert space of the $n$th Dicke model and is spanned by the basis $\{|c\rangle_n\otimes|j,m\rangle_n\}$. Here $|c\rangle_n$ ($c\in\mathbb{N}$) are Fock states  
satisfying $a_n^\dagger a_n|c\rangle_n=c|c\rangle_n$, 
and $|j,m\rangle_n$ ($j\in\{N_a/2,N_a/2-1,\cdots\}$, $m\in\{-j,\cdots,j\}$) are eigenstates of the collective spin operators satisfying $J_n^z|j,m\rangle_n=m|j,m\rangle_n$ and $[(J_n^x)^2+(J_n^y)^2+(J_n^z)^2]|j,m\rangle_n=j(j+1)|j,m\rangle_n$. In particular, the states $|N_a/2,m\rangle_n$ are called the Dicke states~\cite{PhysRevResearch.2.033131,PhysRevE.67.066203}. 

Competition between rotating and counter-rotating terms has profound consequences on the states of matter and SPTs in the open Dicke trimer model, i.e., when cavity losses are introduced. To illustrate this, we consider the open quantum dynamics described by the following Lindblad master equation
\begin{eqnarray}\label{eq:ME}
\frac{d\rho}{dt} = -i[H_N,\rho] + \kappa\sum_n \left(2a_n\rho a_n^\dagger - \{a_n^\dagger a_n,\rho\}\right), 
\end{eqnarray}
in terms of the anticommutator $\{a_n^\dagger a_n,\rho\} = a_n^\dagger a_n\rho+\rho a_n^\dagger a_n$ and the dissipation rate $\kappa$ assumed identical for all cavities. Before studying this open setting, we analyze the closed model described by the Hamiltonian in Eq.~(\ref{eq:H1}).

\subsection{Ground State Energy and Hamiltonian of Quantum Fluctuations}

Here we consider the closed Dicke trimer model, i.e., in the absence of cavity losses. To solve the model, we follow a common strategy~\cite{PhysRevLett.128.163601,zhao2022anomalous,PhysRevLett.122.193201} which consists in first displacing the cavity fields $D_n^\dagger(\alpha_n) a_n D_n(\alpha_n) = a_n+\sqrt{N_a}\alpha_n$, where $\alpha\in\mathbb{C}$ and in terms of the displacement operator $D_n(\alpha_n)=\exp(\sqrt{N_a}\alpha_na_n^\dagger-\sqrt{N_a}\alpha_n^*a_n)$, followed by a collective spin rotation $U_n(\theta_n,\phi_n)=\exp(-i\theta_nJ_n^y)\exp(-i\phi_nJ_n^z)$, and then the application of the Holstein-Pirmakoff (HP) transformation $J_n^z = b_n^\dagger b_n-N_a/2,\; J_n^+ = b_n^\dagger\sqrt{N_a-b_n^\dagger b_n}$ in terms of the bosonic modes $b_n$ in the new frame. Note that in applying the HP transformation, we restrict our discussion to the subspace spanned by the basis $\{\bigotimes_{n=1}^N(|c\rangle_n\otimes|N_a/2,m\rangle_n)\}$. 

In the thermodynamic limit, the resulting displaced, rotated Hamiltonian $\tilde{H}_N=\mathcal{U}^\dagger H_N\mathcal{U}$ with $\mathcal{U}=DU$, $D=\prod_nD_n,\,U=\prod_nU_n$ can be written, neglecting order $1/N_a$ terms, in terms of a $O(N_a)$ classical contribution and $O(1)$ quantum corrections. Specifically, we can write 
\begin{equation}
\label{eq:HEGSHq}
    \tilde{H}_N= E_{\text{GS}}+H_\text{q}+O(1/N_a)\;,
\end{equation} in terms of the the ground state energy 
\begin{eqnarray}\label{eq:EGSsimplifed}
\bar{E}_{\text{GS}}&=&\sum_n\bigg[\lvert\bar\alpha_n\rvert^2-\frac12 \sqrt{1+4g^2\bar A_n^2} \nonumber\\
&&\quad\quad+\bar J\left(e^{i\varphi}\bar\alpha_n^*\bar\alpha_{n+1}+\text{H. c.}\right) \bigg],\;
\end{eqnarray}
where, following Ref.~\cite{PhysRevLett.128.163601}, we defined the rescaled parameters $\bar\alpha_n = \sqrt{\omega_0/\omega_a}\alpha_n$, $g=2\lambda/\sqrt{\omega_0\omega_a}$, $\bar J=J/\omega_0$, $\bar A_n=\sqrt{\eta_+^2\Re^2\bar\alpha_n+\eta_-^2\Im^2\bar\alpha_n}$ and the rescaled ground state energy $\bar E_\text{GS}=E_\text{GS}/N_a\omega_a$, see Appendix.~\ref{sec:clq} for a detailed derivation.

The quantum fluctuations over the ground state are described by the quantum Hamiltonian
\begin{eqnarray}\label{eq:Hq}
H_\text{q} &=& \sum_n \bigg[ \omega_0a_n^\dagger a_n +\frac{\omega_a}{\cos\theta_n} b_n^\dagger b_n\nonumber\\
&&\quad+\lambda\eta_+\cos\theta_n\cos\phi_n(a_n+a_n^\dagger)(b_n+b_n^\dagger) \nonumber\\
&&\quad-i\lambda\eta_+\sin\phi_n(a_n+a_n^\dagger)(b_n-b_n^\dagger) \nonumber\\
&&\quad+i\lambda\eta_-\cos\theta_n\sin\phi_n(a_n-a_n^\dagger)(b_n+b_n^\dagger) \nonumber\\
&&\quad-\lambda\eta_-\cos\phi_n(a_n-a_n^\dagger)(b_n-b_n^\dagger) \nonumber\\
&&\quad+J(e^{i\varphi}a_n^\dagger a_{n+1}+e^{-i\varphi}a_{n+1}^\dagger a_n)\bigg],\;
\end{eqnarray}
where the rotation angles $\theta_n,\,\phi_n$ are specified by the equations 
\begin{equation}
\cos\theta_n=\frac{\omega_a}{\Omega_n},\;\sin\phi_n = -\frac{\eta_-\Im\alpha_n}{A_n},\;\cos\phi_n = \frac{\eta_+\Re\alpha_n}{A_n},
\end{equation}
with $A_n=\sqrt{\eta_+^2\Re^2\alpha_n+\eta_-^2\Im^2\alpha_n}$ and $\Omega_n=\sqrt{\omega_a^2+16\lambda^2A_n^2}$. 
Note that in the expansion of the Hamiltonian in Eq.~(\ref{eq:HEGSHq}), terms of order $\mathcal O(\sqrt{N_a})$ should vanish for consistency if the parameters $\alpha_n,\,\theta_n,\,\phi_n$ associated with the ground state are used in the displacement and rotation transformations. The expressions in Eq.~(\ref{eq:EGSsimplifed}) and Eq.~(\ref{eq:Hq}) constitute our starting point to study the properties,
such as equilibrium phases and critical scalings, 
of the generalized Dicke trimer model without dissipation. 
In the next section we build the corresponding semiclassical model in the presence of dissipation.

\subsection{Nonlinear Semiclassical Dynamics and Steady-state Fluctuations}

When cavity losses are considered, the dynamics
of the expectation of the observables $a_n, J_n^x, J_n^y, J_n^z$ can be written in closed form using Eq.~(\ref{eq:ME}) and {by assuming a mean-field factorization in the thermodynamic (or semiclassical) limit for the correlation between light and matter operators, e.~g., $\langle J_n^x(a_n+a_n^\dagger)\rangle = \langle J_n^x\rangle\langle a_n+a_n^\dagger\rangle$. This results in the following set of nonlinear differential equations
\begin{equation}\label{eq:EOMs}
\begin{split}
\frac{d\alpha_n}{dt} &=\displaystyle -(\kappa+i\omega_0)\alpha_n - 2i\lambda\eta_+X_n-2\lambda\eta_-Y_n \\
&\quad\displaystyle - i J e^{i\varphi}\alpha_{n+1} - i J e^{-i\varphi}\alpha_{n-1}, \\
\frac{dX_n}{dt} &=\displaystyle - \omega_a Y_n - 4\lambda\eta_- Z_n\Im\alpha_n, \\
\frac{dY_n}{dt} &=\displaystyle \omega_a X_n - 4\lambda\eta_+ Z_n\Re\alpha_n, \\
\frac{dZ_n}{dt} &=\displaystyle 4\lambda\eta_+ Y_n\Re\alpha_n + 4\lambda\eta_-X_n\Im\alpha_n,\;
\end{split}
\end{equation}
with $X_n=\langle J_n^x\rangle/N_a$, $Y_n=\langle J_n^y\rangle/N_a$, $Z_n=\langle J_n^z\rangle/N_a$. Our discussion of the open quantum dynamics of Eq.~(\ref{eq:ME}) will be restricted to the subspace involving the Dicke states $|N_a/2,m\rangle_n$. This ensures the spin conservation relations $X_n^2+Y_n^2+Z_n^2=1/4$. 
The $Z_2$ parity symmetry of the original Hamiltonian corresponds to the invariance of the differential equations above under the transformation 
\begin{equation}
(\alpha_n,\,X_n,\,Y_n,\,Z_n) \to (-\alpha_n,\,-X_n,\,-Y_n,\,Z_n). 
\end{equation}
The $Z_N$ translational symmetry $T$ of the Hamiltonian in Eq.~(\ref{eq:H1}) and identical cavity dissipation rates assumed in Eq.~(\ref{eq:ME}) guarantee the translational invariance of Eq.~(\ref{eq:EOMs}) as well. Furthermore, for $\varphi=0,\,\pi$, the Hamiltonian also enjoys a ``ring-reflection'' symmetry corresponding to the exchange of a pair of Dicke models leaving the third invariant. 

Many features of the long-time dynamics of the solutions of Eq.~(\ref{eq:EOMs}) can be characterized by classifying its attractors. Intuitively, an attractor $A$ is a subset of points in the phase space $\mathbb{P}=\bigotimes_3(\mathbb{C}\otimes \mathbb{S}^2)$ ($\mathbb C$ is the complex plane and $\mathbb{S}^2$ is the 2-sphere) of the model which is invariant under the dynamics and possesses a basin of neighborhood points evolving towards it. We refer to \cite{Nayfehbook} for a more formal definition. Importantly, the dynamics in $A$  can be (i) trivial or constituted by stationary solutions (or equivalently equilibria, equilibrium points) that are time-independent; (ii) oscillatory, featuring periodic or quasiperiodic persistent oscillation in time; or (iii) chaotic,  featuring irregular oscillations and exponential sensitivity to initial conditions.

For an equilibrium point denoted by $\{\alpha_n^\text{eq},X_n^\text{eq},Y_n^\text{eq},Z_n^\text{eq}\}$, to examine its local stability, we define the deviation from it as $\delta\alpha_n = \alpha_n-\alpha_n^\text{eq}$, $\delta X_n=X_n-X_n^\text{eq}$, $\delta Y_n=Y_n-Y_n^\text{eq}$, $\delta Z_n=Z_n-Z_n^\text{eq}$ whose equations of motion can be obtained by linearizing Eq.~(\ref{eq:EOMs}) around the equilibrium point and using the spin conservation law. This leads to the set of differential equation
\begin{eqnarray}\label{eq:linearizedEOMs}
\begin{split}
\frac{d\delta\alpha_n}{dt} &= -(\kappa+i\omega_0)\delta\alpha_n - 2i\lambda\eta_+\delta X_n - 2\lambda\eta_-\delta Y_n \\
&\quad -iJe^{i\varphi}\delta\alpha_{n+1} - iJe^{-i\varphi}\delta\alpha_{n-1}, \\
\frac{d\delta X_n}{dt} &= -\omega_a\delta Y_n - 4\lambda\eta_-(\Im\alpha_n^\text{eq}\delta Z_n + Z_n^\text{eq}\delta\Im\alpha_n), \\
\frac{d\delta Y_n}{dt} &= \omega_a\delta X_n - 4\lambda\eta_+(\Re\alpha_n^\text{eq}\delta Z_n + Z_n^\text{eq}\delta\Re\alpha_n), \\ 
\end{split}
\end{eqnarray}
alongside the relation $Z_n^\text{eq}\delta Z_n = -(X_n^\text{eq}\delta X_n + Y_n^\text{eq}\delta Y_n)$. Then the local dynamical stability of an equilibrium point can be examined by the Jacobian matrix, 
\begin{equation}\label{eq:Jacobian}
\mathcal{J} = 
\begin{pmatrix}
\mathcal{A}_1 & \mathcal{B} & \mathcal{C} \\
\mathcal{C} & \mathcal{A}_2 & \mathcal{B}  \\
\mathcal{B} & \mathcal{C} & \mathcal{A}_3 
\end{pmatrix},
\end{equation}
where the sub-matrices $\mathcal{A}_n$, $\mathcal{B}$, and $\mathcal{C}$ are written in terms of the parameters $\delta\alpha_n$, $\delta X_n$, $\delta Y_n$, $\delta Z_n$ defining Eq.~(\ref{eq:linearizedEOMs}) and their explicit expression is given in Appendix.~\ref{sec:app_Jacmat}. We denote the (complex) eigenvalues of $\mathcal J$ by $s_i$ with $1\le i\le4N$ ($N=3$) and arranged with descending real part, i.e., $\Re[s_i]<\Re[s_j]$ for $i>j$. A stable equilibrium point is defined as one satisfying $\Re s_i<0$ for all $i$. For periodic or chaotic attractors, we search {for them using a dynamical approach, i.~e., by evolving Eq.~(\ref{eq:EOMs}) with randomly chosen initial conditions.

Finally, steady-state fluctuations of an equilibrium point can be modeled by a set of closed algebraic equations for quadratic operators, namely $a_na_m$, $a_n^\dagger a_m^\dagger$, $a_n^\dagger a_m$, $b_nb_m$, $b_n^\dagger b_m^\dagger$, $b_n^\dagger b_m$, $a_nb_m$, $a_n^\dagger b_m^\dagger$, $a_n^\dagger b_m$, $b_n^\dagger a_m$, see Appendix.~\ref{sec:ssfluc}. In total there are $76$ independent equations but they can be expressed in a matrix form as $\mathcal{M}_f\mathbf{f}_\text{ss} + \mathbf{v}_f = 0$ where $\mathbf{f}_\text{ss}$ is a vector whose entries are the expectation values of the quadratic operators listed above, while $\mathcal{M}_f$ and $\mathbf{v}_f$ are the coefficient matrix and vector of the system observables, see Appendix.~\ref{sec:ssfluc} for more details. Steady-state quantum fluctuations are obtained by solving this equation, which gives a unique solution $\mathbf{f}_{ss}=\mathcal{M}_f^{-1}\mathbf{v}_f$ provided $\det\mathcal{M}_f\neq0$ is satisfied.


\section{Lossless Cavities}\label{sec:lossless}
\subsection{Phase Diagram}

\begin{figure}
\includegraphics[clip, width = 0.98\columnwidth]{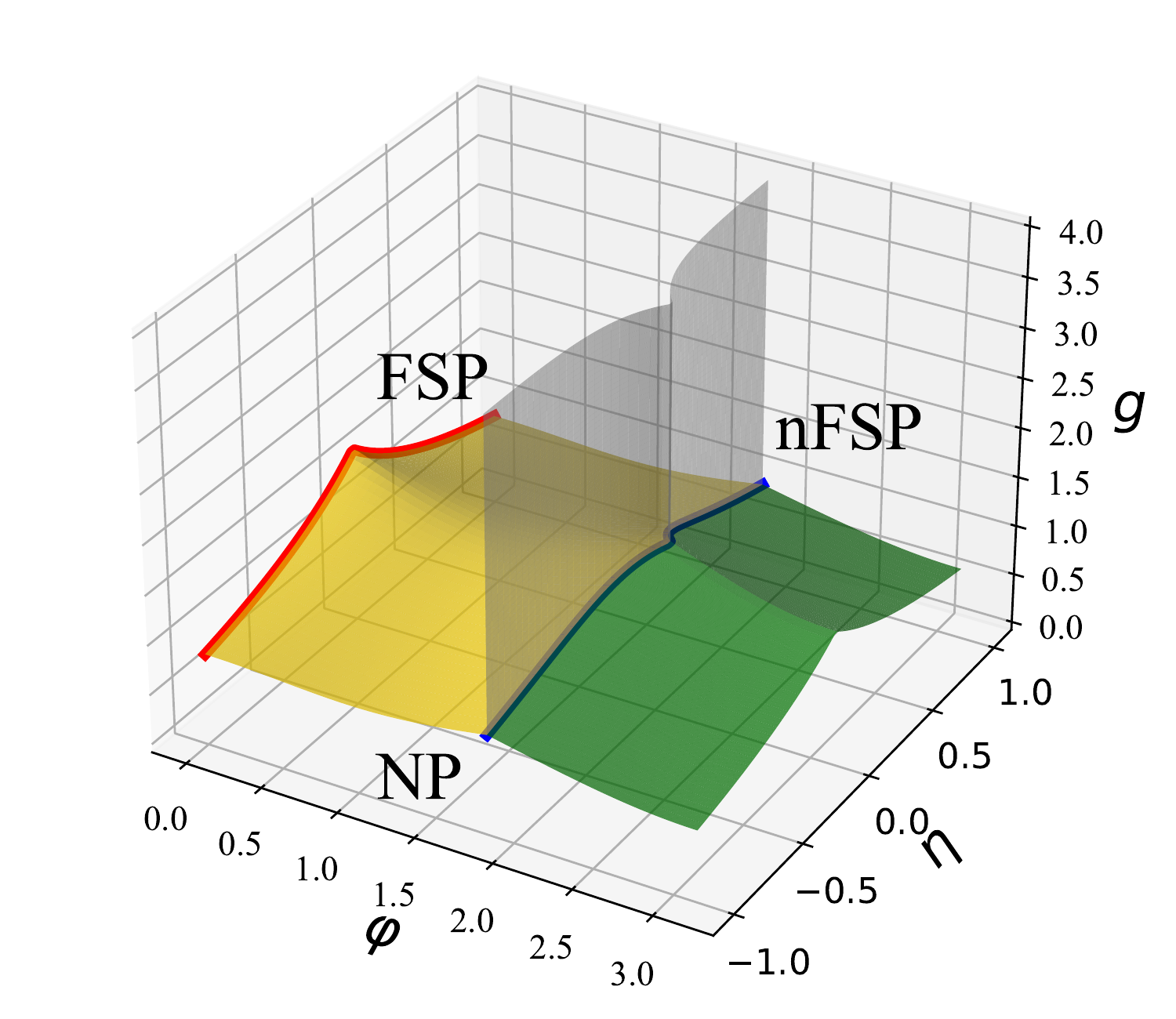}
\caption{3d plot of the equilibrium phase diagram. Colored surfaces and curves correspond to continuous phase transitions featuring different critical scalings; while gray surface corresponds to first-order phase transitions. Red curve is characterized by two critical scalings in the FSP and blue curve is made of tricritical points with two critical scalings on both sides of the transition. Here we use $\bar J=0.3$ to satisfy the constraints Eq.~(\ref{eq:Jphi}) such that for all values of $\varphi$ the NP is stable in numerical simulation.
}\label{fig:pd3d}
\end{figure}
We first calculate the phase diagram of the closed model. The ground state solution $\bar\alpha_n^\text{gs}$ is determined by minimizing $\bar E_{\text{GS}}$ with respect to $\bar\alpha_n$. 
Since the NP solution $\bar\alpha_n=0$ is always an extreme point of $\bar{E}_{\text{GS}}$, the phase diagram can be obtained by examining its stability. This is achieved by evaluating the eigenvalues of the $6\times6$ Hessian matrix $\partial^2\bar E_{\text{GS}}/\partial\bar\alpha_n\partial\bar\alpha_m$ at the origin $\bar{\alpha}_n=0$.  Interestingly, it is possible to find analytical expressions for the six eigenvalues $\xi_{1,2} = 2 + 4\bar J\cos\varphi - g^2(1\pm\eta)^2/2$, $\xi_{3,4} = 2-2\bar J\cos\varphi-g^2(1+\eta^2)/2\pm\sqrt{12\bar J^2\sin^2\varphi+g^4\eta^2}$ and $\xi_{5(6)}=\xi_{3(4)}$. 
Since a stable NP corresponds to having all eigenvalues positive, the phase boundary separating the NP  and the superradiant phase can be written as
$g_\text{c}=\min(g_\text{c}^\text{nf},g_\text{c}^\text{f})$ where $g_\text{c}^\text{nf}=\sqrt{1+2\bar J\cos\varphi}/\max(\lvert\eta_+\rvert,\,\lvert\eta_-\rvert)$ and 
\begin{eqnarray}\label{eq:gcf}
g_\text{c}^\text{f} = 
\begin{cases}
\sqrt{\frac{3(1-\bar J^2)}{1-\bar J\cos\varphi}-4\bar J\cos\varphi-2},~~\text{for}~\eta=\pm1,  \\
\min\bigg(\frac{\sqrt{M+\sqrt{N}}}{\lvert1-\eta^2\rvert},\,\frac{\sqrt{M-\sqrt{N}}}{\lvert1-\eta^2\rvert}\bigg),~~\text{for}~\eta\neq\pm1, 
\end{cases}
\end{eqnarray}
in terms of the variables $M=(1+\eta^2)(1-\bar J\cos\varphi)$ and $N=4\eta^2(1-\bar J\cos\varphi)^2+3\bar J^2\sin^2\varphi(1-\eta^2)^2$. Note that, since $g_\text{c}^\text{f},\,g_\text{c}^\text{nf}$ are real, the following constraints on $\bar J,\,\varphi$ should be imposed 
\begin{eqnarray}\label{eq:Jphi}
1+2\bar J\cos\varphi > 0,~~~~~ 1-2\bar J\cos(\varphi-\pi/3) > 0. 
\end{eqnarray} 
The corresponding equilibrium phase diagram is shown in Fig.~\ref{fig:pd3d}, where surfaces and curves with different colors are characterized by distinct critical scalings, as we will elaborate below. The phase boundaries of the FSP (yellow surface) and the nFSP (green surface) intersect at a tricritical line $\varphi_{tr}$ (blue curve), which vertically extends into a first-order phase boundary.
Note that we did not find any analytical expression for the first-order phase boundary, which we verified numerically. In the NP and nFSP, the values for $\bar\alpha_n^\text{gs}$ are identical in all cavities and they are twofold degenerate in the nFSP (see Appendix.~\ref{sec:analy_nFSP}). Furthermore, in Appendix.~\ref{sec:app_prop_FSP}, we show the following generic properties of the sixfold degenerate FSP solutions (which break translational symmetry): (i) $\Im\bar\alpha_n=0,\;\bar\alpha_{n+1}=\bar\alpha_{n-1}^*$ for $\eta>0$; (ii) $\Re\bar\alpha_n=0,\;\bar\alpha_{n+1}=-\bar\alpha_{n-1}^*$ for $\eta<0$; (iii) At $\eta=0$, there exist special solutions satisfying either (i) or (ii), and from which the remaining solutions can be obtained by multiplying a phase factor $e^{i\zeta}$. Property (iii) describes a phase redundancy of the ground state solutions at $\eta=0$ which can be understood by inserting $\eta=0$ and $\bar\alpha_n = \lvert\bar\alpha_n\rvert e^{i\zeta_n}$ into Eq.~(\ref{eq:EGSsimplifed}) to obtain
\begin{eqnarray}
\bar{E}_{\text{GS}}&=&\sum_n\bigg[\lvert\bar\alpha_n\rvert^2 - \frac12
\sqrt{1+g^2\lvert\bar\alpha_n\rvert^2} \nonumber\\
&&+ 2\bar J\lvert\bar\alpha_n\bar\alpha_{n+1}\rvert
\cos(\varphi+\zeta_{n+1}-\zeta_n)
\bigg]. 
\end{eqnarray}
From this expression, we can appreciate how $\bar E_{\text{GS}}$ only depends on the phase difference $\zeta_{n+1}-\zeta_n$ of neighbouring cavity fields, implying that once a minimum $\bar\alpha_n^\text{gs}$ of $\bar E_{\text{GS}}$ is found, all other minima can be constructed as $\bar\alpha_n^\text{gs}e^{i\zeta}$. In the following, we will analyze the  consequences of this phase redundancy on the excitation spectra.
We finish this section noting that the phase diagram is symmetric with respect to $\eta=0$ as a consequence of the symmetric coupling in the Dicke Hamiltonian.

\subsection{Excitation Spectra and Critical Exponents}

\begin{figure*}
\includegraphics[clip,width = 13cm]{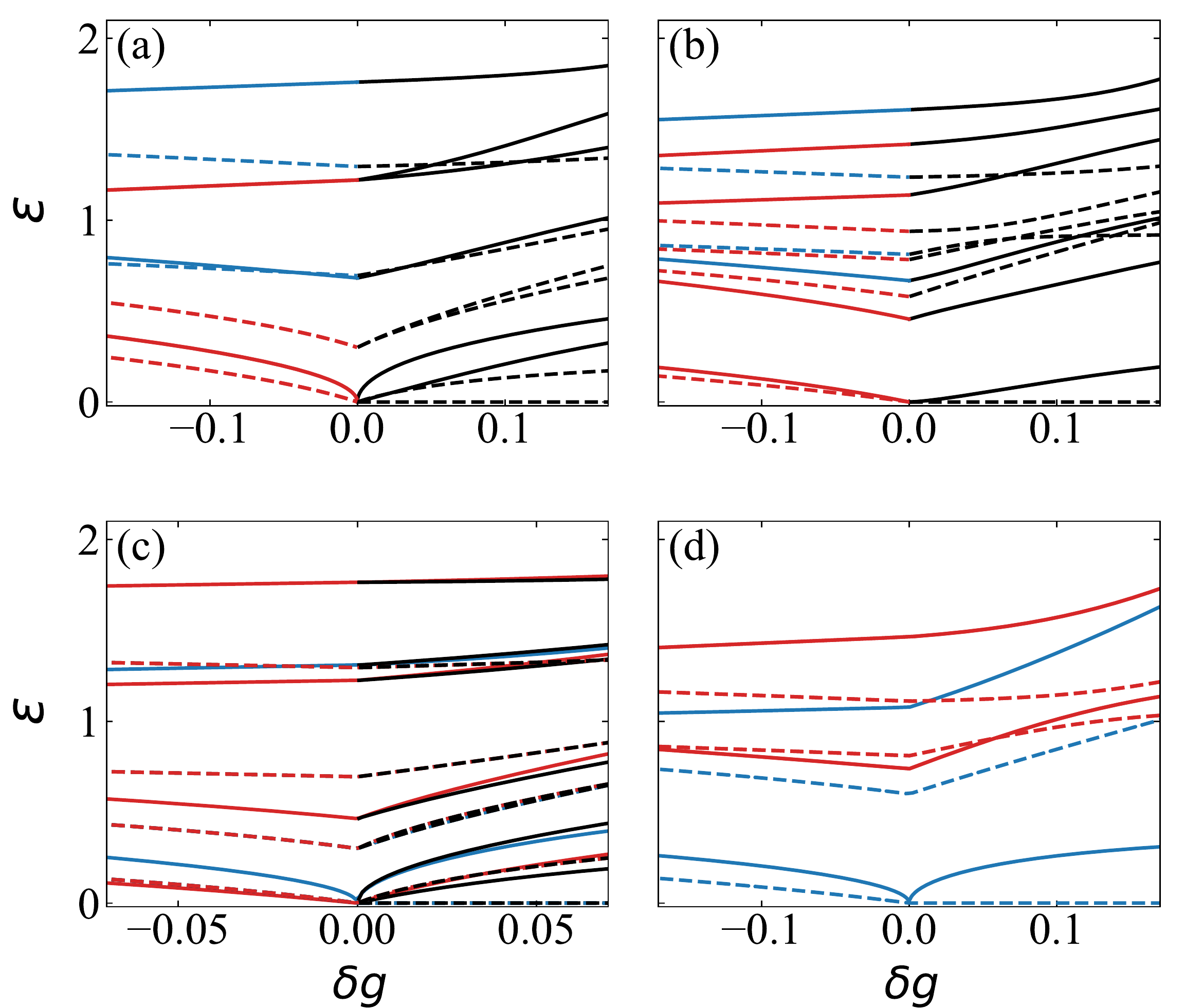}%
\caption{Excitation spectra for $\eta=1$ (solid lines) and $\eta=0$ (dashed lines) calculated at (a) $\varphi=0$, (b) $\pi/4$, (c) $\varphi_{tr}$ and (d) $\pi$ respectively. The two zero momentum modes (blue) are plotted using Eq.~(\ref{eq:dispersion0}) while the four finite momentum modes (red) and all modes in the FSP (black) are from numerics. In panel (c) the spectra in the FSP are plotted along the lines $\varphi_+=\varphi_\text{tr}+(g-g_{\text{c}})$ (red and blue lines) and $\varphi_-=\varphi_\text{tr}-(g-g_{\text{c}})$ (black lines). Other parameters are the same as in Fig.~\ref{fig:pd3d}. 
}
\label{fig:excspec}
\end{figure*}

In this section, we analyze the quantum properties of the model on top of the classical energy landscape described by $E_{\text{GS}}$. To do this, we rewrite $H_q$ in terms of the quadrature operators of the bosonic modes. In the superradiant phase this allows us to write
\begin{widetext}
\begin{eqnarray}\label{eq:HqsimplifiedSP}
H_\text{q}&=&\sum_n\bigg[ \frac{\omega_0}{2}(q_n^2+p_n^2) +  \frac{\omega_a}{2\cos\theta_n}(Q_n^2+P_n^2) + J\cos\varphi(q_nq_{n+1}+p_np_{n+1})-J\sin\varphi(q_np_{n+1}-p_nq_{n+1}) \nonumber\\
&&+\frac{g\sqrt{\omega_0\omega_a}}{\bar A_n}\Big(
\frac{\eta_+^2\Re\bar\alpha_n}{\sqrt{1+4g^2\bar A_n^2}}q_nQ_n 
- \eta_+\eta_-\Im\bar\alpha_n q_nP_n
+ \frac{\eta_-^2\Im\bar\alpha_n}{\sqrt{1+4g^2A_n^2}}p_nQ_n 
+ \eta_+\eta_-\Re\bar\alpha_n p_nP_n\Big)
\bigg],  
\end{eqnarray}
\end{widetext}
where $Q_n=(b_n+b_n^\dagger)/\sqrt2,\,P_n = -i(b_n-b_n^\dagger)/\sqrt2$ are the  position and momentum quadratures of collective spins. In the NP, $H_\text{q}$ is obtained by applying the replacements  $\eta_+\Re\bar\alpha_n/\bar A_n \to -1,\,\eta_-\Im\bar\alpha_n/\bar A_n \to 0$ and then setting $\bar A_n=0$ in the previous expression. In both cases, the quadratic Hamiltonian is a bilinear form so it can be written as 
$H_\text{q}/\omega_0 = \mathbf{r}^T \mathcal{H}_\text{q} \mathbf{r}/2$, 
where $\mathbf{r} = (q_1,p_1,Q_1,P_1,q_2,p_2,Q_2,P_2,q_3,p_3,Q_3,P_3)^T$, in terms of the following $12\times12$ real symmetric matrix 
\begin{eqnarray}
\mathcal H_\text{q} = 
\begin{pmatrix}
\mathcal H_1 & \mathcal H_J & \mathcal H_J^T \\
\mathcal H_J^T & \mathcal H_2 & \mathcal H_J \\
\mathcal H_J & \mathcal H_J^T & \mathcal H_3 \\
\end{pmatrix},
\end{eqnarray}
where the explicit expressions for the submatrices $\mathcal{H}_n,\;\mathcal{H}_J$ are given in Appendix.~\ref{sec:app_HnHJ}. Since $\mathcal{H}_\text{q}$ is positive definite, by Williamson’s theorem~\cite{Serafini}, we can find a symplectic matrix $S$ which simultaneously satisfies both $S^T\Omega_0S=\Omega_0$ and $S^T\mathcal{H}_\text{q}S = \Sigma$. Here, we defined the symplectic form $\Omega_0=\bigoplus_{n=1}^6 i\sigma_y$ (in terms of  the Pauli matrix   $\sigma_y$) and the matrix $\Sigma=\mathrm{diag}(\epsilon_1,\epsilon_1,\epsilon_2,\epsilon_2,\epsilon_3,\epsilon_3,\epsilon_4,\epsilon_4,\epsilon_5,\epsilon_5,\epsilon_6,\epsilon_6)$ whose parameters $\epsilon_i>0$ are the absolute value of the $12$ eigenvalues of the matrix $i\Omega_0\mathcal H_\text{q}$. Importantly, the commutation relations are invariant under this simplectic transformation. 

The translational invariance of the NP and nFSP solutions, i.~e., $\bar\alpha_n=\bar\alpha,\,\bar A_n=\bar A=\sqrt{\eta_+^2\Re^2\bar\alpha+\eta_-^2\Im^2\bar\alpha}$, guarantees the translational invariance of $H_\text{q}$. This allows to make some progress by defining the 
Fourier transformation $q_n = \sum_k e^{-ikn}q_k/\sqrt N$, $p_n = \sum_k e^{ikn}p_k/\sqrt N$
in terms of the quasimomentum $k=2\pi m/N$ with $m=0,\pm1$ which results in   
\begin{widetext}
\begin{equation}
\begin{array}{l}
H_\text{q} = \displaystyle\sum_k \Bigg[
\frac{\omega_0}{2}\Big(q_kq_{-k}+p_kp_{-k}\Big) + 
\frac{\omega_a}{2}\sqrt{1+4g^2A^2}\Big(Q_kQ_{-k}+P_kP_{-k}\Big) +  J\cos\varphi\Big(e^{ik}q_kq_{-k}+e^{-ik}p_kp_{-k}\Big) \nonumber\\
-\displaystyle 2iJ\sin\varphi\sin k\;q_kp_k + 
\frac{g\sqrt{\omega_0\omega_a}}{\bar A}\left(
\frac{\eta_+^2\Re\bar\alpha}{\sqrt{1+4g^2\bar A^2}}Q_kq_{-k} 
- \eta_+\eta_-\Im\bar\alpha P_kq_{-k} 
+ \frac{\eta_-^2\Im\bar\alpha}{\sqrt{1+4g^2\bar A^2}}Q_kp_{-k} 
+ \eta_+\eta_-\Re\bar\alpha P_kp_{-k}\right)
\Bigg].
\end{array}
\end{equation}
\end{widetext}
The $k=0$ subspace is now decoupled from the finite momentum sector and diagonalization of the corresponding Hamiltonian $H_\text{q}^{k=0}$ leads to two analytical solutions for the NP and nFSP
\begin{eqnarray}\label{eq:dispersion0}
\epsilon_{1,2}^{k=0} = 
\sqrt{F\pm\sqrt{F^2+G}}, 
\end{eqnarray}
where $F=(D_1^2+D_2^2-2R_1R_2-2I_1I_2)/2$, $G=D_1D_2(R_1^2+R_2^2+I_1^2+I_2^2)-D_1^2D_2^2-(R_1R_2+I_1I_2)^2$ 
with 
$D_1 = \omega_0+2J\cos\varphi$, 
$D_2 = \omega_a\sqrt{1+4g^2\bar A^2}$, 
$R_1 = g\sqrt{\omega_0\omega_a}\eta_+^2\Re\bar\alpha/\bar A\sqrt{1+4g^2\bar A^2}$, 
$R_2 = g\sqrt{\omega_0\omega_a}\eta_+\eta_-\Re\bar\alpha/\bar A$, 
$I_1 = g\sqrt{\omega_0\omega_a}\eta_-^2\Im\bar\alpha/\bar A\sqrt{1+4g^2\bar A^2}$, $I_2=g\sqrt{\omega_0\omega_a}\eta_+\eta_-\Im\bar\alpha/\bar A$. 
We do not find any closed-form solutions in the remaining momentum subspace. 

In Fig.~\ref{fig:excspec} we show the full spectra obtained by numerical diagonalization (black and red lines)
and the two branches (blue lines) using Eq.~(\ref{eq:dispersion0}) for the anisotropic case $\eta=1$ and the isotropic case $\eta=0$ at four representative values of $\varphi$, which are $\varphi=0,\;\pi/4,\;\varphi_{tr},\;\pi$. The main features of the spectra for $\eta=1$ was reported in Ref.~\cite{zhao2022anomalous} and we now summarize them for completeness: (i) Lifting of the time-reversal symmetry destroys the two critical scalings found at $\varphi=0$. At the same time, one soft mode with critical exponent $3/2$ in the FSP and one soft mode with exponent $1$ in the NP emerge; (ii) There exists a tricritical point $\varphi_\text{tr}$ characterized by two soft modes present on both sides of the transition. Note that here the critical point can be approached either from the nFSP or from the FSP and, in both cases, the exponents are the same, as shown in Fig.~\ref{fig:excspec}(c). These features persist in the whole parameter space except for the isotropic case $\eta=0$ where in the superradiant phase a zero energy mode emerges as a result of the phase redundancy of $\bar\alpha_n^\text{gs}$ mentioned above. In fact, this phase redundancy indicates that the system can be excited without any extra energy-cost, in turn implying the appearance of a zero-energy mode in the excitation spectra. 

\subsection{A Semiclassical Model for the Understanding of the Anomalous Finite Critical Fluctuations}\label{sec:semitheory}

In Ref.~\cite{zhao2022anomalous} the authors also reported the intriguing observation that the soft mode in the NP for $0<\varphi<\varphi_\text{tr}$ exhibits a finite critical fluctuation which can be explained using the fact that the Dicke model shares the same scaling limit as the Rabi model ($N_a=1$) in the infinite frequency limit $\Omega/\omega_0\to\infty$. Here, in order to gain more intuition about this phenomenon, we seek a different route using in the semi-classical picture of the Dicke model. In particular we consider a semi-classical model in which all matter degrees of freedom are evaluated at their energy minimum and all quantum effects are encoded by light. A similar analysis, involving both light and matter, has been developed to investigate the role of quantum chaos in the Dicke model, c.f.~Sec.~II in~Ref.~\cite{PhysRevA.44.1022} and Refs.~\cite{PhysRevA.50.2040,PhysRevE.54.1449,PhysRevLett.80.5524,PhysRevA.64.043801,PhysRevE.67.066203} for the discussion of the semiclassical limit of the Dicke model. 

Here, we show that a classical treatment of the matter degrees of freedom is sufficient for a self-consistent estimation of the critical exponents and for an easy understanding of the phenomenon of finite critical fluctuations. To build this model, we will operate a quantization procedure over the light degrees of freedom in the potential in Eq.~(\ref{eq:EGSsimplifed}) which can be interpreted as a classical Hamiltonian $\bar E_\text{GS}(\{\mathfrak q_i,\mathfrak p_i\})$,  where we performed the replacements $\Re\bar\alpha_i\to\mathfrak q_i,\;\Im\bar\alpha_i\to\mathfrak p_i$ to simplify the notation.
In this context, we can also define the Poisson bracket acting over two generic phase-space functions $\mathfrak Q(\{\mathfrak q_i,\mathfrak p_i\})$, $\mathfrak P(\{\mathfrak q_i,\mathfrak p_i\})$ as  
\begin{eqnarray}
\{\mathfrak Q,\mathfrak P
\}=\sum_l\bigg(
\frac{\partial\mathfrak Q}{\partial\mathfrak q_l}
\frac{\partial\mathfrak P}{\partial\mathfrak p_l} 
- 
\frac{\partial\mathfrak Q}{\partial\mathfrak p_l}
\frac{\partial\mathfrak P}{\partial\mathfrak q_l} 
\bigg), 
\end{eqnarray}
in terms of the conjugate pair $\mathfrak p_i, \mathfrak q_i$ which satisfy 
$\{\mathfrak q_i,\mathfrak p_j\}=\delta_{ij}$, $\{\mathfrak q_i,\mathfrak q_j\}=\{\mathfrak p_i,\mathfrak p_j\}=0$. Near the local energy minimum $\{\mathfrak q_i^m, \mathfrak p_i^m\}$, $\bar E_\text{GS}$ can be expanded as 
\begin{eqnarray}
\bar E_\text{GS}\approx\bar E_\text{GS}^\text{m} + \sum_{i,j=1}^3 \mathfrak D_{ij} \delta\mathfrak q_i\delta\mathfrak p_j,   
\end{eqnarray}
where the deviations $\delta\mathfrak q_i=\mathfrak q_i-\mathfrak q_i^\text{m}$, $\delta\mathfrak p_i=\mathfrak p_i-\mathfrak p_i^\text{m}$ are conjugate with respect to the Poisson bracket, where $\mathfrak D_{ij} = \partial^2\bar E_\text{GS}/\partial\delta\mathfrak q_i\partial\delta\mathfrak p_j$, and where $\bar E_\text{GS}^\text{m}=\bar E_\text{GS}(\{\mathfrak q_i^\text{m},\mathfrak p_i^\text{m}\})$. Since the matrix $\mathfrak D$ is real symmetric, invoking Williamson’s theorem again, one can find a symplectic matrix $S$ which brings it into a diagonal form while preserving the Poisson brackets. We can then write
\begin{eqnarray}\label{eq:Enormal}
\bar E_\text{GS}\approx \bar E_\text{GS}^\text{m} + \frac12\sum_{i=1}^3 (k_{\mathfrak q_i'}\delta \mathfrak q_i'^2 + k_{\mathfrak p_i'}\delta\mathfrak p_i'^2), 
\end{eqnarray}
where $\pm\sqrt{k_{\mathfrak q_i'}k_{\mathfrak p_i'}}$ are eigenvalues of the matrix $i\Omega_0\mathfrak D$ with $\Omega_0=\bigoplus_{n=1}^3 i\sigma_y$. 
We remark that, since we are interested in analyzing the presence of diverging behaviour at the critical point, we can always operate the substitutions $\delta\mathfrak q_i'\to\sqrt c\delta\mathfrak q_i',\;\delta\mathfrak p_i'\to\delta\mathfrak p_i'/\sqrt c,\;k_{\mathfrak q_i'}\to k_{\mathfrak q_i'}/c,\;k_{\mathfrak p_i'}\to ck_{\mathfrak p_i'}$ in Eq.~(\ref{eq:Enormal}), in terms of a $\delta g$-independent positive constant  $c$. To fix this freedom we assume, without loss of generality, that $\lVert\delta\mathfrak q_i'\rVert_2=1$ (in terms of the Euclidean 2-norm $\lVert\cdot\rVert_2$). 
The quantum ground state and low energy excited modes can be obtained by applying the quantization condition $\oint\delta\mathfrak p_i'd\delta\mathfrak q_i'=n_i\omega_0/\omega_a N_a$ for each conjugate pair $\delta\mathfrak q_i', \delta\mathfrak p_i'$, where $n_i\in\mathbb{N}$ and the effective Planck's constant $\omega_0/\omega_a N_a$ originates from the rescaling of $\bar{\alpha}_n$. Here, the presence of $N_a$ in the denominator is crucial to recover the mean-field nature of the Dicke model in the thermodynamic limit. 
Integrating the left side, one finds the relation $\epsilon_i\sim\sqrt{k_{\mathfrak q_i'}k_{\mathfrak p_i'}}/N_a$. 

To appreciate the validity of our theory, we employ this formalism to compute $\gamma$ in the NP and FSP for the special case $\eta=1$, and to further explain the finite critical fluctuations observed in the range $0<\varphi<\varphi_\text{tr}$. In the NP, the positive eigenvalues of $i\Omega_0\mathfrak D$ can be found analytically as $d_1=2g_\text{c}^\text{nf}\sqrt{((g_\text{c}^\text{nf})^2-g^2)}$ and $d_{2,3}=2\sqrt6\bar J\sin\varphi\sqrt{1-D\big(g^2-(g_\text{c}^f)^2\big)\pm\sqrt{1-2D\big(g^2-(g_\text{c}^\text{f})^2\big)}}$, where $D=(1-\bar J\cos\varphi)/6\bar J^2\sin^2\varphi$. For $\varphi>\varphi_\text{tr}$ ($0<\varphi<\varphi_\text{tr}$), only $d_1$ ($d_3$) vanishes at $g_\text{c}$ resulting in $\gamma=1/2$ ($1$). For $\varphi=0$, $d_{2,3}=2\sqrt{(1-\bar J)(1-\bar J-g^2)}$ leading to two soft modes, both characterized by $\gamma=1/2$. For $\varphi=\varphi_\text{tr}$, both $d_1$ and $d_3$ vanish at $g_\text{c}$ originating the two critical scalings at the tricritical point. In the FSP, despite the absence of analytical solutions, we found approximations valid near the critical point up to order $\mathcal O(\delta g^{3/2})$, see Appendix.~\ref{sec:app_fspsol}. Since only one critical mode exists for $0<\varphi<\varphi_\text{tr}$, it is possible  to  compute its correspondent determinant (in alternative toits diagonalization) which scales as $\det i\Omega_0\mathfrak D\sim\delta g^3$, implying that $\gamma=3/2$, in agreement with Ref.~\cite{zhao2022anomalous}.  

We recall that in the paradigmatic Landau's picture, a continuous phase transition is triggered when one, or several, of the curvatures $k_{\mathfrak q_i'}$, $k_{\mathfrak p_i'}$ in the Landau potential described in Eq.~(\ref{eq:Enormal})
(forth-order terms are neglected) vanish at $g_c$. As a consequence, it is possible to distinguish two different scenarios depending on the leading expansion of the curvatures, $k_{\mathfrak q_i'(\mathfrak p_i')}\sim\lvert\delta g\rvert^{d_{\mathfrak q_i'(\mathfrak p_i')}}$. When $d_{\mathfrak q_i'}\neq d_{\mathfrak p_i'}$ (for example $d_{\mathfrak q_i'} > d_{\mathfrak p_i'}$) the ground state wavefunction of the quantized Hamiltonian $H_i/k_{\mathfrak p_i'} \sim ({k_{\mathfrak q_i'}}/{k_{\mathfrak p_i'}})\delta\mathfrak q_i'^2+\delta\mathfrak p_i'^2$ is 
\begin{equation}
\psi(\delta\mathfrak q_i') = \left(\frac{\sqrt{k_{\mathfrak q_i'}/k_{\mathfrak p_i'}}}{\pi}\right)^{1/4} e^{-\sqrt{k_{\mathfrak q_i'}/k_{\mathfrak p_i'}}\delta\mathfrak q_i'^2/2},
\end{equation}
which becomes infinitely wide in the $\delta\mathfrak q_i'$ direction, leading to divergent variances in the original quadratures $\mathfrak q_i,\;\mathfrak p_i$ (note that the constraint $\lVert\delta\mathfrak q_i'\rVert_2=1$ is necessary) and thus divergent photon numbers. On the contrary, when $d_{\mathfrak q_i'}= d_{\mathfrak p_i'}$,  the quantized Hamiltonian can be written as $H_i/k_{\mathfrak p_i'} \sim c'\delta\mathfrak q_i'^2 + \delta\mathfrak p_i'^2$ (in terms of $c'$ which is constant in $\delta g$) and its ground state can be represented by the wavefunction 
\begin{equation}
\label{eq:semiclass_wf}
\psi(\delta\mathfrak q_i') = \Big(\frac{\sqrt{c'}}{\pi}\Big)^{1/4} e^{-\sqrt{c'} \delta\mathfrak q_i'^2}.
\end{equation}
As a consequence, neither of the variances of the original quadratures or the average photon numbers diverge at the critical point. In Fig.~\ref{fig:kqkp}(a), we show that a numerical evaluation of the variables $k_{\mathfrak q_i'},\;k_{\mathfrak p_i'}$ for $\varphi=0,\;\pi/4$ at $\eta=1$ is consistent with the above intuitive interpretation. In Fig.~\ref{fig:kqkp}(b), we show the variance $\langle q_n^2\rangle$ of the ground state is indeed finite for $\varphi=\pi/4$, in sharp contrast to the divergent behavior for $\varphi=0$. A comparison between the quantity $\langle q_n^2\rangle$ calculated via the  wavefunction in Eq.~(\ref{eq:semiclass_wf}) and via numerical diagonalization, gives further evidence of the consistency of our semiclassical model.

\begin{figure}
\includegraphics[clip,width=8.5cm]{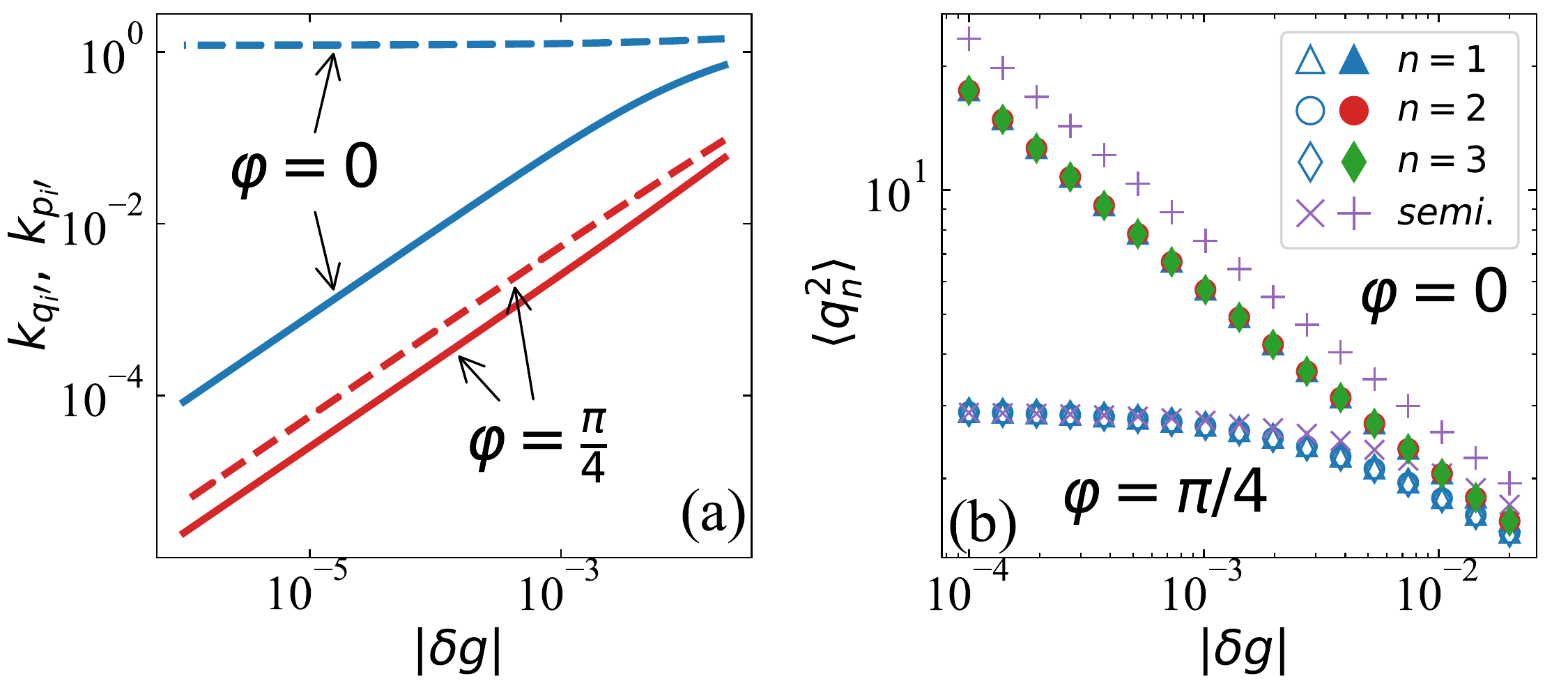}%
\caption{(a) Coefficients $k_{\mathfrak q_i'}$ (solid lines),  $k_{\mathfrak p_i'}$ (dashed lines) as a function of $\delta g$ with the constraint $\lVert\delta\mathfrak q_i'\rVert_2=1$. For $\varphi=0$, $k_{\mathfrak q_i'}$ vanishes while $k_{\mathfrak p_i'}$ remains finite at  the critical point, consistent with a divergent fluctuation of the ground state wavefunction in the $\delta\mathfrak q_i'$ direction. In contrast, for $\varphi=\pi/4$, both coefficients vanish for $\delta g\rightarrow 0$ at the same rate, as demonstrated by the identical slopes of the two red lines. This implies that the width of the Gaussian ground state wavefunction remains finite even at the critical point. See more detailed discussion in the main text. (b) Variances of the quadrature $q_n$ as functions of $\delta g$ in the NP for two values of $\varphi$ and $\eta=1$. Approximate results from the semiclassical method are marked by cross and plus symbols. In both panels we used $J=0.1\omega_0,\;\omega_a=\omega_0$. 
}\label{fig:kqkp}
\end{figure}

\section{Lossy Cavities}\label{sec:lossy}

\begin{figure*}[ht!]
\includegraphics[clip, width = 1\textwidth]{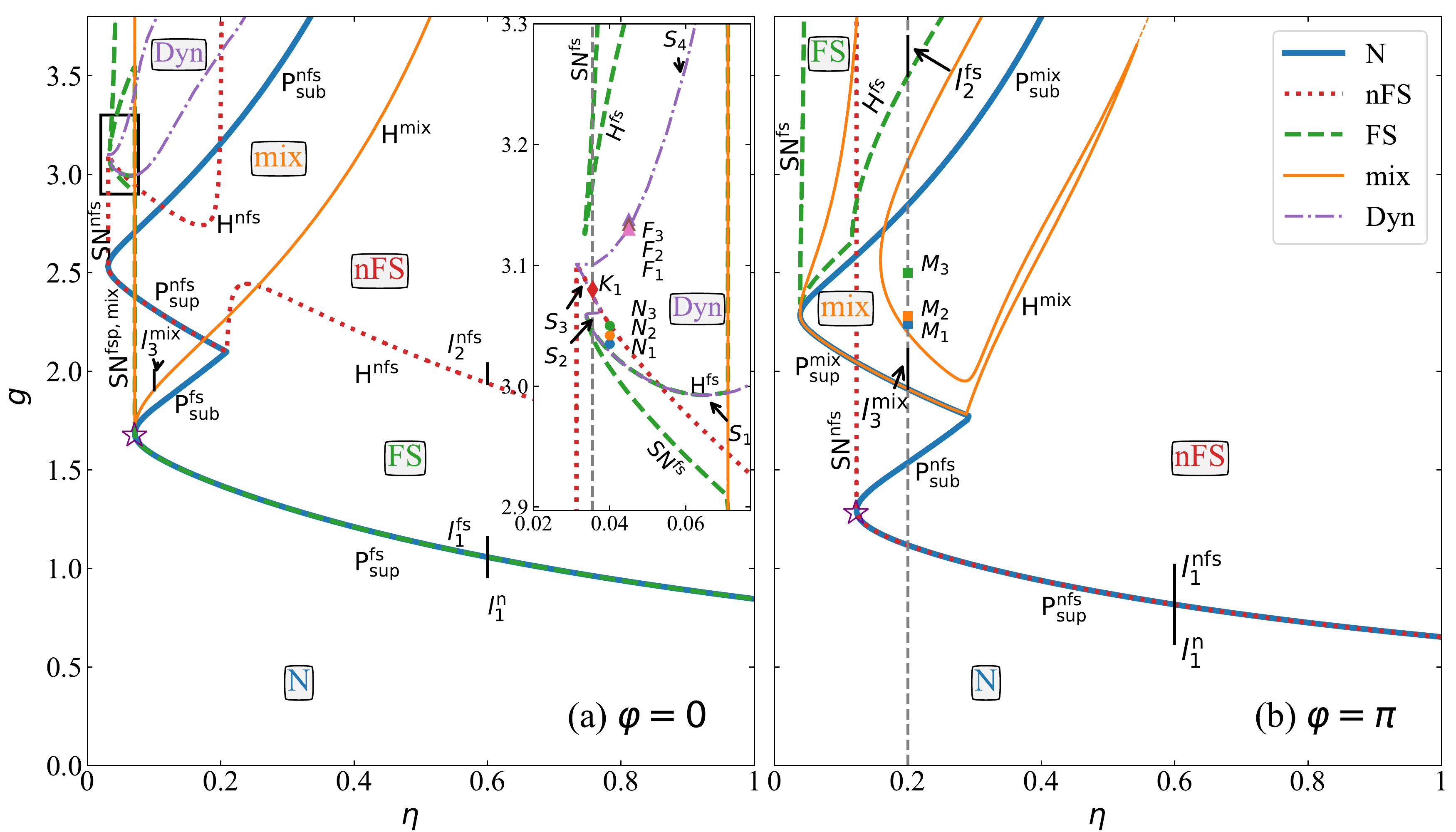}      
\caption{\label{fig:openpd}Bifurcation diagrams of the open Dicke trimer model for (a) $\varphi=0$ and (b) $\varphi=\pi$, respectively. Because of symmetry, these diagram can be extended to the range $\eta\in[-1,0]$ by reflection. The N, nFS, FS, mixed, and dynamical boundaries are identified by blue-solid, red-dotted, green-dashed, orange-solid, purple-dashdotted curves, respectively. The attractors are identified by the labels ``N'', ``nFS'', ``FS'', ``mix'' and ``Dyn'' inside their regions of existence. The inset in (a) magnifies the black-box region in the upper-left corner. In this region, the Dyn-boundary is partitioned by four sub-curves denoted by $\mathbf{S}_1$, $\mathbf{S}_2$, $\mathbf{S}_3$, $\mathbf{S}_4$, see Sec.~\ref{sec:phi=0andpi}. Different bifurcation classes of the boundaries are specified by the labels $\mathbf{P}_\text{sup(sub)}^X$ (for supercritical/subcritical pitchfork bifurcations), $\mathbf{SN}^\text{X}$ (for saddle-node bifurcations), $\mathbf{H}^\text{X}$ (when a branch of equilibria flips stability).   
The nonequilibrium tricritical points reported in Ref.~\cite{PhysRevLett.120.183603} are marked by empty stars. The line cuts used in Fig.~\ref{fig:openscaling} are, here, labeled by $l_i^\text{X}$ with $i\in\{1,2,3\}$, and where the superscript $\text{X}$ takes value in the set $\{\text{nfs},\,\text{fs},\,\text{mix}\}$. We have used $\omega_0=\omega_a=1$ throughout this section. 
}
\end{figure*}

\begin{figure}
\includegraphics[clip, width = 0.98\columnwidth]{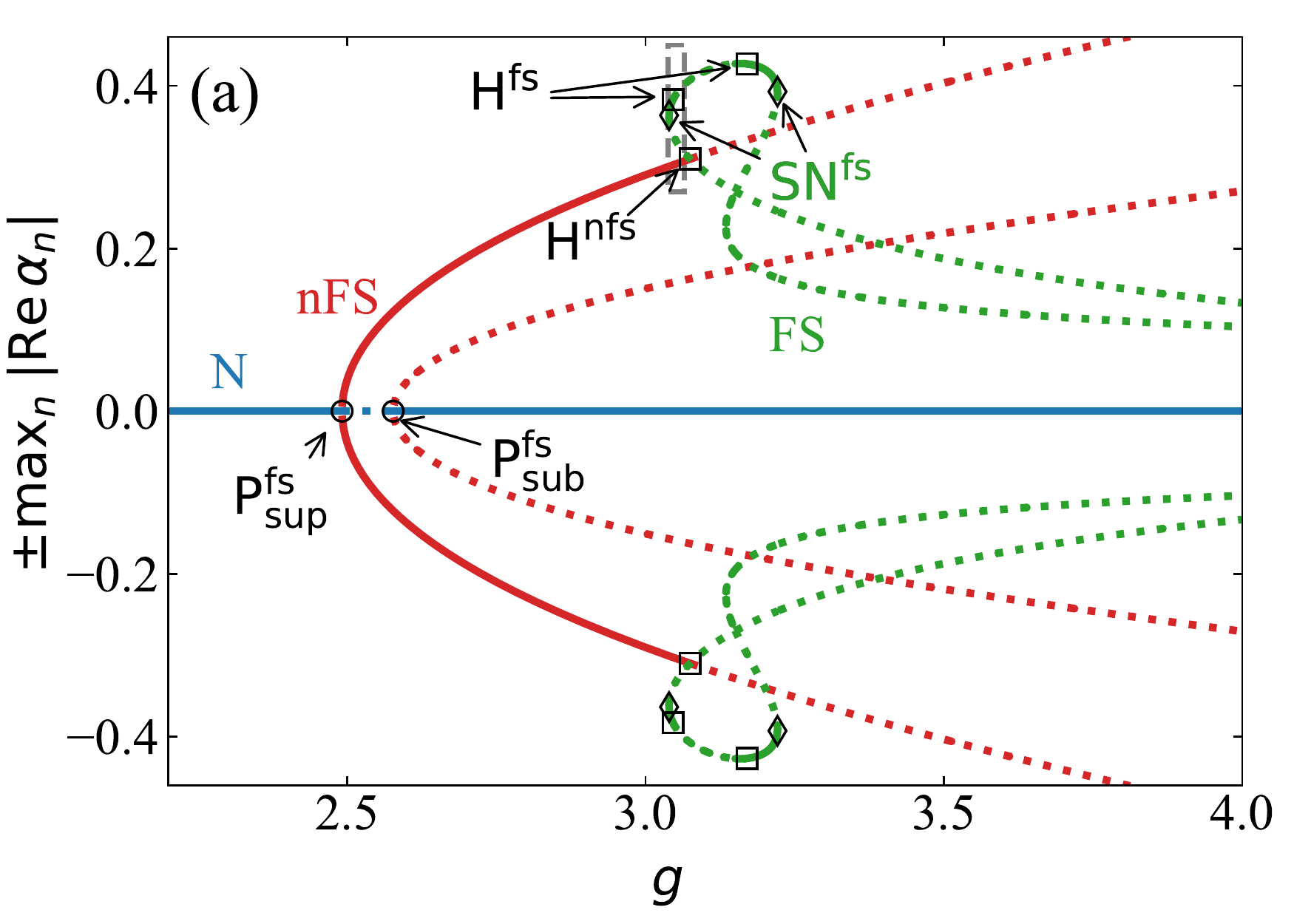}
\includegraphics[clip, width = 0.98\columnwidth]{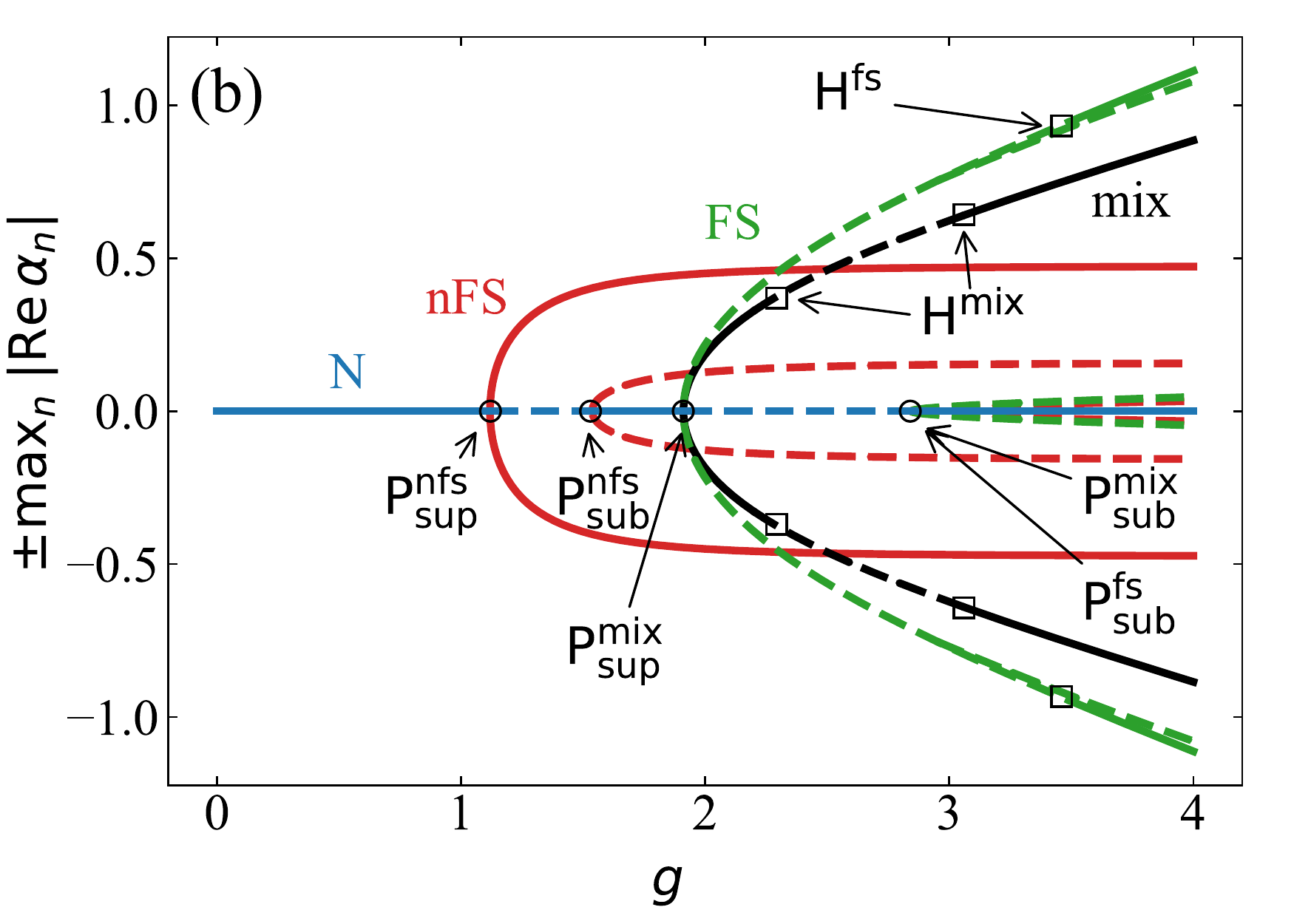}
\caption{Bifurcation diagrams corresponding to the equilibria along the vertical dashed gray lines in Fig.~\ref{fig:openpd}, i.e. for the parameters (a) $\eta=0.0355,\,\varphi=0$ and (b) $\eta=0.2,\,\varphi=\pi$. Solid (Broken) curves correspond to stable (unstable) equilibria. The pitchfork, saddle-node, and Hopf bifurcation points are marked by empty circles, diamonds and squares respectively. Specifically, at $\mathbf{H}^{\text{fs}}$ in (a) and  $\mathbf{H}^\text{mix}$ in (b) Hopf bifurcations occur,  followed by (a) a period-doubling cascade and (b) quasiperiodic oscillations respectively, see further discussion in Fig.~\ref{fig:period_doubling} and Fig.~\ref{fig:M1M2M3}. At $\mathbf{H}^\text{nfs}$ an anomalous Hopf bifurcation occurs where a pair of zero eigenvalues of the Jacobian matrix in Eq.~(\ref{eq:Jacobian}) emerges at the bifurcation point and the post-bifurcation dynamics displays burst oscillations, see Fig.~\ref{fig:burst_osc}.
}
\label{fig:bifurcation}
\end{figure}

In this section we study the semiclassical nonlinear dynamics, steady-state fluctuations and their scaling behavior of the open Dicke trimer model. Specifically, we are interested in the robustness of the equilibrium phases and the anomalous critical scalings under the inclusion of cavity losses. For example, the unbalanced, open Dicke model ($N=1$) shows a dynamical behavior  which is drastically different with respect to the closed case \cite{PhysRevResearch.2.033131}. In fact, Ref.~\cite{PhysRevResearch.2.033131} shows that, for $\lvert\eta\rvert\le1$, the semiclassical dynamics of the unbalanced, open Dicke model always evolves towards a stable equilibrium in the long-time limit $t\to\infty$ while, for $\lvert\eta\rvert>1$, persistent oscillations and even chaos are observed in certain parameter regimes. Throughout this section, we explore the nonlinear feature in the dynamics of the  open Dicke trimer model for $\eta\in[-1,1]$ and leave the complexity of more general regimes to a future work.

We have identified four classes of equilibrium solutions of Eq.~(\ref{eq:EOMs}) which are, 
\begin{itemize}
    \item ``N'' (normal). In this class, all cavity fields vanish ($\alpha_n=0$) while all the spins collectively point to either the north or the south pole ($\{X_n,Y_n,Z_n\}=\{0,0,\pm1/2\}$) of the Bloch sphere.
    
    \item ``nFS'' (non-frustated). This class is characterized by identical, nonvanishing cavity fields ($\alpha_n=\alpha$), and identical spin polarizations ($\{X_n,Y_n,Z_n\}=\{X,Y,Z\}$). 

    \item ``FS'' (frustrated). In this class, both the cavity fields and the spin polarizations are nonvanishing and not identical.

    \item ``Mixed''. This class is found for $\varphi=0,\,\pi$ and its equilibria  satisfy $\alpha_n=0,\, \alpha_{n+1}=-\alpha_{n+2}$ for $n=0,1,2$, where the subindexes are intended modulo $3$. In other words, one of the three cavity fields vanishes, while the other two are opposite to each other
\end{itemize}
Further details regarding the classification can be found in Appendix.~\ref{sec:app_classification}. 
In the next sections, we discuss the bifurcation diagrams of the model in the ferromagnetic ($\varphi=0$) and anti-ferromagnetic ($\varphi=\pi$) regimes. While an explicit numerical analysis for more general values is rather involved, we further present results for $\varphi$ close to these two limiting cases.

\subsection{Case I: $\varphi=0$ and $\varphi=\pi$}\label{sec:phi=0andpi}

We start by considering the two limiting cases for the phase of the photon hopping, i.e., $\varphi=0$ and $\varphi=\pi$ corresponding to a ferromagnetic and antiferromagnetic interaction. In Fig.~\ref{fig:openpd} we show the corresponding bifurcation diagrams in the $(g,\eta)$-plane. Despite their visual similarities, the regions defined by the five types of curves in these diagram do not carry the same meaning as they would in a standard equilibrium case, where boundaries between phases are associated with different classes of phase transitions. We emphasis that for each curve in Fig.~\ref{fig:openpd} the associated region we refer to here is the one bounded by the curve with its label inside. In fact, each of these regions in Fig.~\ref{fig:openpd} is labeled by a type of stable  (stationary or dynamical) solutions of Eq.~(\ref{eq:EOMs}) and allows the presence of all the solutions appearing in the regions below. 

These diagrams can be understood more easily from bottom to top, i.e., from weak to strong coupling $g$. For a fixed $\eta$, the bifurcation diagram of the equilibira can be constructed and the corresponding bifurcation points can be extracted. The collection of these bifurcation points constitutes the various curves in Fig.~\ref{fig:openpd}, which define regions with intricate shape where the system dynamics can evolve to different attractors, depending on the initial conditions. Specifically, in Fig.~\ref{fig:openpd}, we identify four distinct stable equilibrium solutions (labeled as ``N'', ``FS'', ``nFS'', ``mix'') and several regions (labeled as ``Dyn'') where the long-time semiclassical dynamics of Eq.~(\ref{eq:EOMs}) tends towards an oscillatory or chaotic behavior. As an illustration of the interpretation of Fig.~\ref{fig:openpd}(a), at $\eta=1$, one first find only the stable N solutions at $g=0$ while, by increasing $g$, stable FS and nFS solutions emerge successively. We note that for the stable N solutions identified here, the collective spins point to the south pole of the Bloch sphere~\cite{PhysRevResearch.2.033131}.

In general, we observe that, starting from the usual Dicke case ($\eta=1$), the N region is contiguous to either the FS region [Fig.~\ref{fig:openpd}(a)] or the nFS region [Fig.~\ref{fig:openpd}(b)], until a multicritical point \cite{PhysRevLett.120.183603} (marked by a star) is reached along their common boundary. Beyond this point, the behaviour becomes much richer with the presence of different coexisting stable equilibrium solutions. Cavity dissipation also allows the emergence of two extra stable symmetry-broken equilibria not present in the closed model. Specifically, these belong to the mixed/nFS class for $\varphi=0$, and to the mixed/FS  class for $\varphi=\pi$.

In the following, we are going to separately classify boundaries between regions in which only equilibrium solutions exist (Sec.~\ref{sec:equilibrium}) and boundaries involving dynamical solutions (Sec.~\ref{sec:dynamical}). We now present an overview of this classification.

We use $\mathbf{P}_\text{sup(sub)}^\text{X}$, $\mathbf{SN}^\text{X}$ and $\mathbf{H}^\text{X}$ with $\text{X}\in\{\text{nfs},\,\text{fs},\,\text{mix}\}$ to distinguish the bifurcation properties of the boundaries of stable equilibria. 
We numerically checked that the boundaries of stable equlilibrium solutions still belong to these three classes even for a more general parameter range $\varphi\in[0,\pi]$.

The model also supports the existence of regions involving stable dynamical solutions (which we characterize by the generic label ``Dyn'').
For $\varphi=0$, at large $g$, and close to the isotropic line $\eta=0$, we find the existence of a region (see the upper left corner of Fig.~\ref{fig:openpd}(a) and the corresponding inset box) whose boundary can be divided into four curves $\mathbf{S}_1$, $\mathbf{S}_2$, $\mathbf{S}_3$, $\mathbf{S}_4$, each associated with a distinct type of dynamical transition. For $\varphi=\pi$, near the $\mathbf{H}^\text{mix}$ line we also find one such region whose boundary is too intricate to be drawn at this level of detail. Nonetheless, we present an example of different dynamical transitions realized at a selection of  different points $M_1,\,M_2,\,M_3$ shown in Fig.~\ref{fig:bifurcation}(b). Qualitatively different dynamical solutions might emerge in more general parameter regimes. We numerically checked the existence of dynamical solutions
also for more general values of $\varphi$.

In the following we analyze the boundaries described above in more detail.

\subsubsection{Boundaries of the Stable Equilibria}
\label{sec:equilibrium}

We classified the boundaries of stable equilibria using the following bifurcation classes.
\begin{itemize}
\item$\mathbf{P}_{\text{sup}(\text{sub})}^\text{X}$: Supercritical (subcritical) pitchfork bifurcations. Across $\mathbf{P}_\text{sup(sub)}^\text{X}$, a branch of stable (unstable) equilibria flips its stability, while two other branches of stable (unstable) X equilibria emerge after the bifurcation. For example, in Fig.~\ref{fig:openpd} the N boundary is made of  two supercritical pitchfork bifurcation lines $\mathbf{P}_\text{sup}^\text{X}$ and two subcritical pitchfork bifurcation lines $\mathbf{P}_\text{sub}^\text{X}$, with $\text{X}=\text{fs},\,\text{nfs}$ [Fig.~\ref{fig:openpd}(a)] and $\text{X}=\text{nfs},\,\text{mix}$ [Fig.~\ref{fig:openpd}(b)]. 

\item$\mathbf{SN}^\text{X}$: Saddle-node bifurcations. Crossing $\mathbf{SN}^\text{X}$, one stable and one unstable branche of $\text{X}$ equilibria emerge whereas no $\text{X}$ equilibria exist before the bifurcation. This class can be found on all the three symmetry-broken boundaries $\mathbf{SN}^{\text{fs}}$, $\mathbf{SN}^{\text{nfs}}$, and $\mathbf{SN}^{\text{mix}}$.

\item$\mathbf{H}^\text{X}$: Hopf-like bifurcations. Here, we use the term ``Hopf-like'' because crossing these curves bears similarities with Hopf bifurcations without strictly being classified as such (since no periodic solutions are present). Specifically, while 
crossing $\mathbf{H}^\text{X}$, an X equilibrium branch flips its stability. In other words, the real part of one eigenvalue of the Jacobian matrix in Eq.~(\ref{eq:Jacobian}) changes sign, leading to a change of stability in the corresponding X equilibrium branch. It is important to mention that some parts of these $\mathbf{H}^\text{X}$ curves could be the boundaries of the Dyn-region, e.g., $\mathbf{S}_1,\,\mathbf{S}_2$ in Fig.~\ref{fig:openpd}(a) where an actual Hopf bifurcation could occur, thereby justifying our choice of the name ``Hopf-like''. We refer to the next section for a more detailed discussion of dynamical transitions. 
\end{itemize}

To clarify the nature of these transitions, in Fig.~\ref{fig:bifurcation} we present a bifurcation diagram for a specific range of parameters in Fig.~\ref{fig:openpd}. Specifically, Fig.~\ref{fig:bifurcation}(a) corresponds to the (dashed grey) vertical cut at $\varphi=0,\,\eta=0.0355$
in the inset of Fig.~\ref{fig:openpd}(a) while Fig.~\ref{fig:bifurcation}(b) corresponds to the (dashed grey) vertical cut at $\varphi=\pi,\,\eta=0.2$ in Fig.~\ref{fig:openpd}(b).

\subsubsection{Boundaries of the Stable Dynamical Solutions}
\label{sec:dynamical}
Here, we describe the boundaries involving stable dynamical solutions. We labeled different types of transitions as $\mathbf{S}_i$ with $i=1,2,3,4$. The line cut at $\varphi=0,\,\eta=0.0355$ in Fig.~\ref{fig:bifurcation}(a) is selected such that it intersects with all of $\mathbf{S}_i$. 
In Fig.~\ref{fig:collision_nfsp_fsp}, we present a more detailed analysis  of the FS branches for  the boxed region in Fig.~\ref{fig:bifurcation}(a) which includes both periodic and chaotic solutions. 
In Fig.~\ref{fig:collision_nfsp_fsp}(a) we see that after a saddle-node bifurcation (marked by a diamond), one stable and one unstable branches of FS equilibria appear. The stable branch then undergoes a Hopf bifurcation at a point (marked by a square) belonging to $\mathbf{S}_1$.

After the Hopf bifurcation a branch of periodic solutions emerges and further bifurcates into a period-doubling cascade, which eventually leads to the formation of chaotic attractors at around $g\approx3.057$. Therefore, we identify $\mathbf{S}_1$ as the onset of the oscillating behavior for the FS branches via Hopf bifurcations. At around $g\approx3.060$, the previously identified chaotic attractor is replaced by a new periodic solution, whose attractor collides with the basin of attraction of the nFS equilibrium point at $g\approx3.063$, resulting in the transition indicated by $\mathbf{S}_2$. In order to analyze the remaining $\mathbf{S}$ transitions, we switch back to Fig.~\ref{fig:bifurcation}(a). There, we can observe how the stable nFS branches emerge via a supercritical pitchfork bifurcation (marked by a circle), and they exist until they reach the $\mathbf{H}^\text{nfs}$ lines, where anomalous Hopf bifurcations with unusual busrt-oscillation-like post-bifurcation dynamics take place. This is the transition which we identify as $\mathbf{S}_3$. By further increasing $g$, the periodic attractor present beyond $\mathbf{S}_3$ turns into a chaotic attractor (not shown), which eventually hit the basin of attraction of the N equilibria, defining the transition labeled by $\mathbf{S}_4$. 

Now we are going to discuss each of these dynamical transitions in more detail by analysing their specific dynamical signatures.  
\begin{itemize}

\begin{figure}[H]
\includegraphics[clip, width = 0.98\columnwidth]{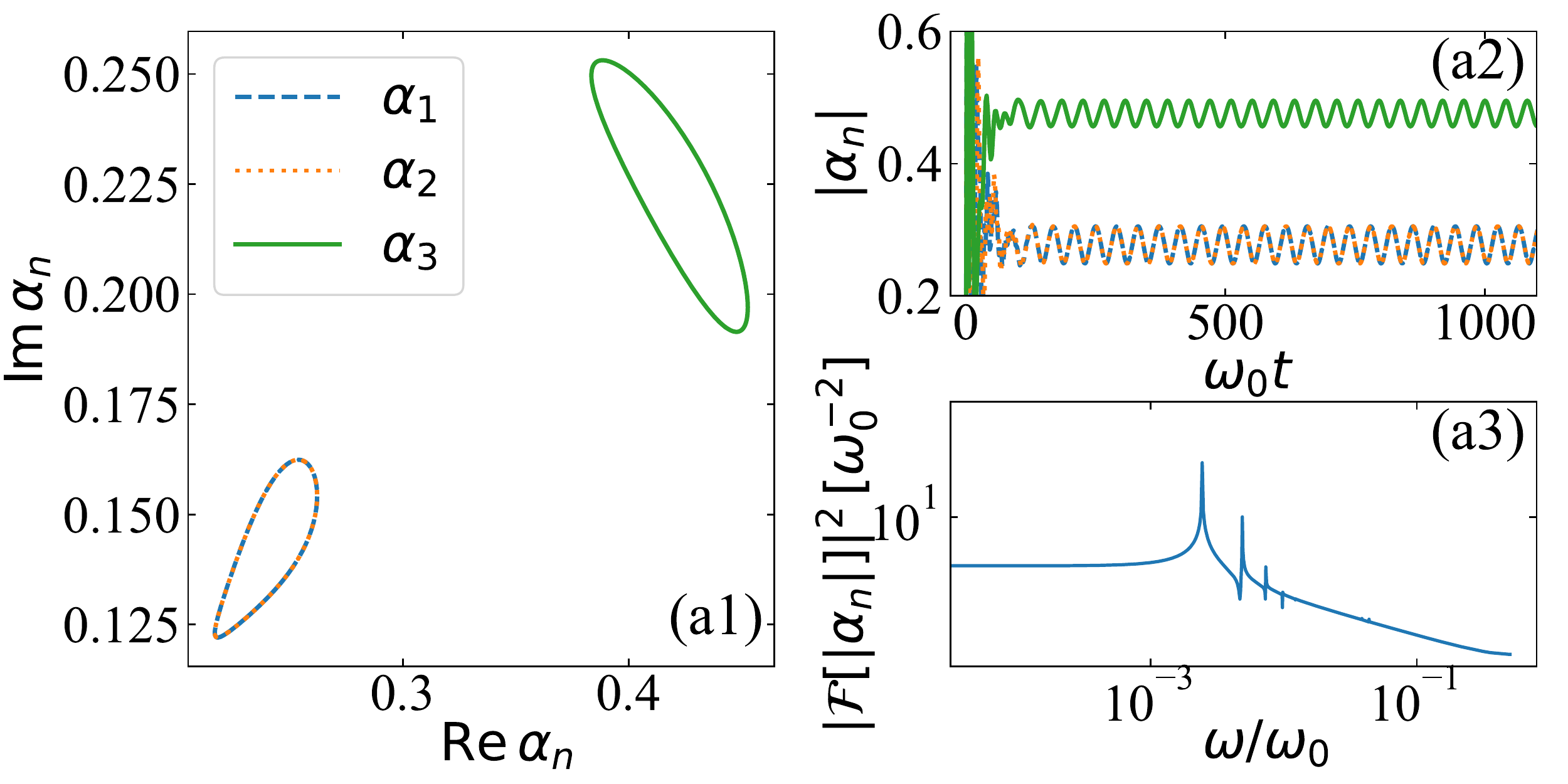}
\includegraphics[clip, width = 0.98\columnwidth]{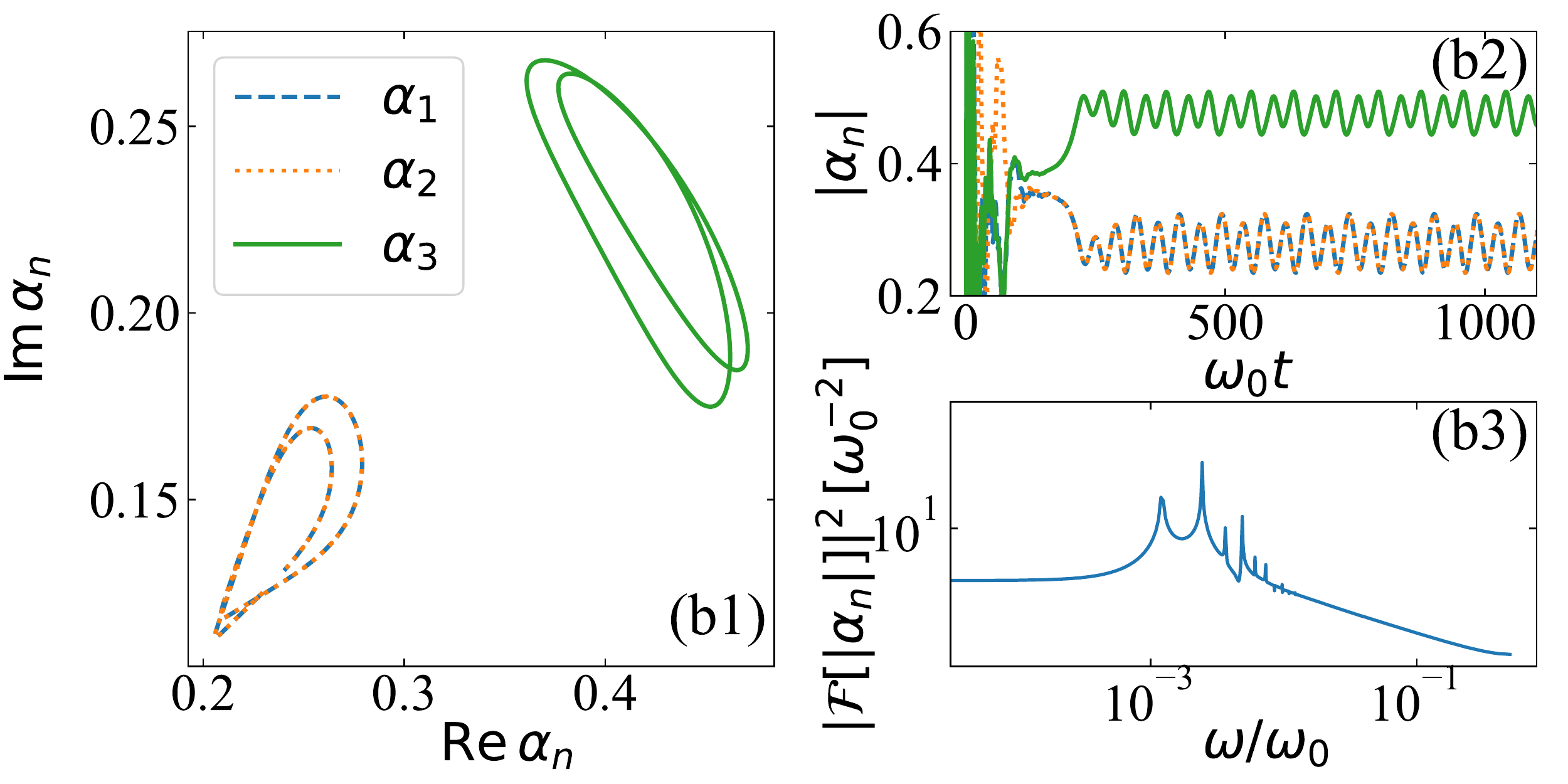}
\includegraphics[clip, width = 0.98\columnwidth]{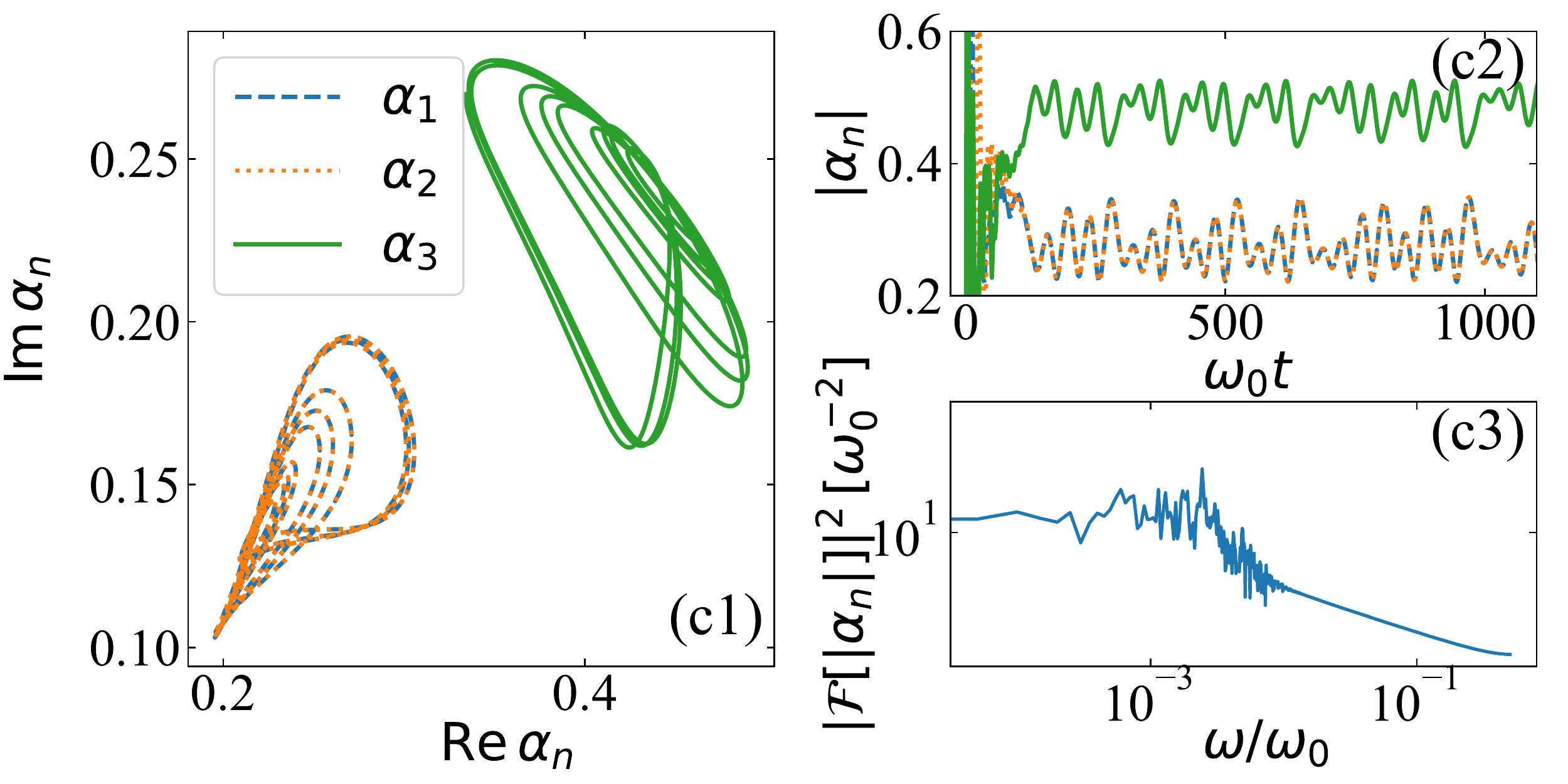}
\caption{[$\mathbf{S}_1$] Period doubling cascade after the Hopf bifurcation line $\mathbf{S}_1$ in the inset of Fig.~\ref{fig:openpd}(a). Top, middle, bottom panels correspond to the three points $N_1$, $N_2$, $N_3$ with $\eta=0.04$ and $g=3.035,\,3.042,\,3.050$ respectively. In each row, projections of the periodic attractor (left), temporal traces of its cavity field components (right top) and the corresponding (unnormalized) power spectra $\lvert\mathcal{F}[\lvert\alpha_n\rvert]\rvert^2$ (right bottom) are shown.
}\label{fig:period_doubling}
\end{figure}

\item\emph{$\mathbf{S}_1$.} This curve is a collection of Hopf bifurcation points for the stable FS branch. 
At the bifurcation points corresponding to the $\mathbf{S}_1$ line,  the Jacobian matrix in Eq.~(\ref{eq:Jacobian}) has a pair of  purely imaginary conjugate eigenvalues $\pm i\omega_\text{osc}$ with $\omega_\text{osc}>0$~\cite{Nayfehbook}, indicating the appearance of a periodic attractor.
This periodic attractor further experiences a sequence of period-doubling bifurcations (period-doubling cascade) and eventually the dynamics becomes chaotic. This can be seen in Fig.~\ref{fig:period_doubling} which shows cavity $\alpha$-dependent features of the periodic attractors and of the dynamics. It also shows the  (unnormalized) power spectra $\lvert\mathcal{F}[\lvert\alpha_n\rvert]\rvert^2$ with $\mathcal{F}$ representing the Fourier transform of $\alpha_n$ at three points $N_1$, $N_2$, $N_3$ (see the inset of Fig.~\ref{fig:openpd}) selected in proximity of $\mathbf{S}_1$ (for increasing coupling strength). At $N_1$ (closest to $\mathbf{S}_1$) the long-time dynamics features a single main harmonics, see Fig.~\ref{fig:period_doubling}(a2) and Fig.~\ref{fig:period_doubling}(a3). At $N_2$, the  power spectrum shows a second main harmonics with frequency half of the one at $N_1$, see Fig.~\ref{fig:period_doubling}(b3),  as a consequence of a period-doubling bifurcation. By further increasing $g$, an infinite sequence of period-doubling bifurcations takes place until chaos ensues, as shown in Fig.~\ref{fig:period_doubling}(c1-c3). We note that, in the oscillatory solution, two of the three cavity fields are synchronized, i.e. $\alpha_1(t)=\alpha_2(t)$ and not equivalent to the field in the remaining cavity, indicating that the Hopf bifurcations originate from a stable FS branch, see Fig.~\ref{fig:period_doubling}(c1-c3).

\begin{figure}[H]
\includegraphics[clip, width = 0.98\columnwidth]{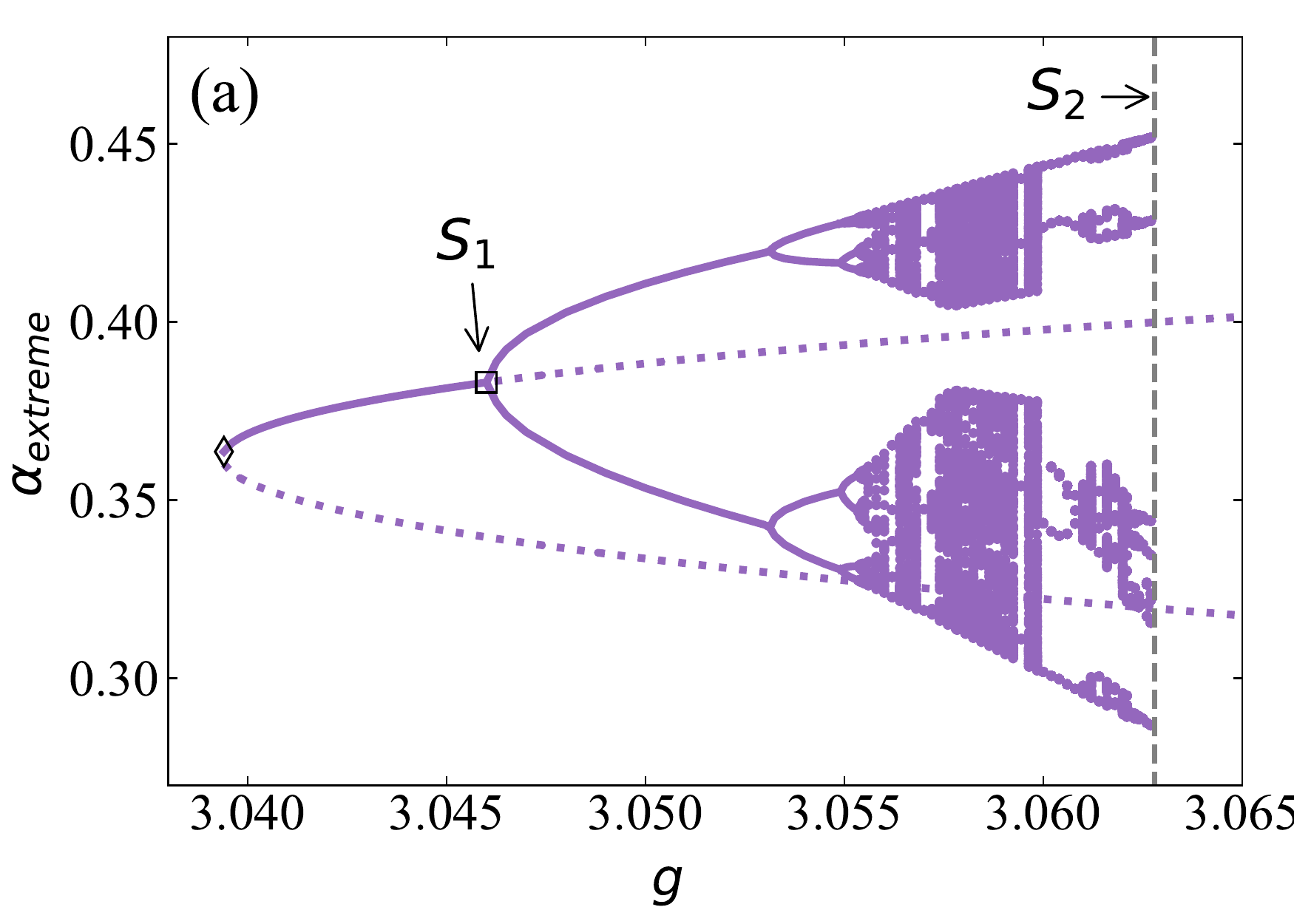}
\includegraphics[clip, width = 0.98\columnwidth]{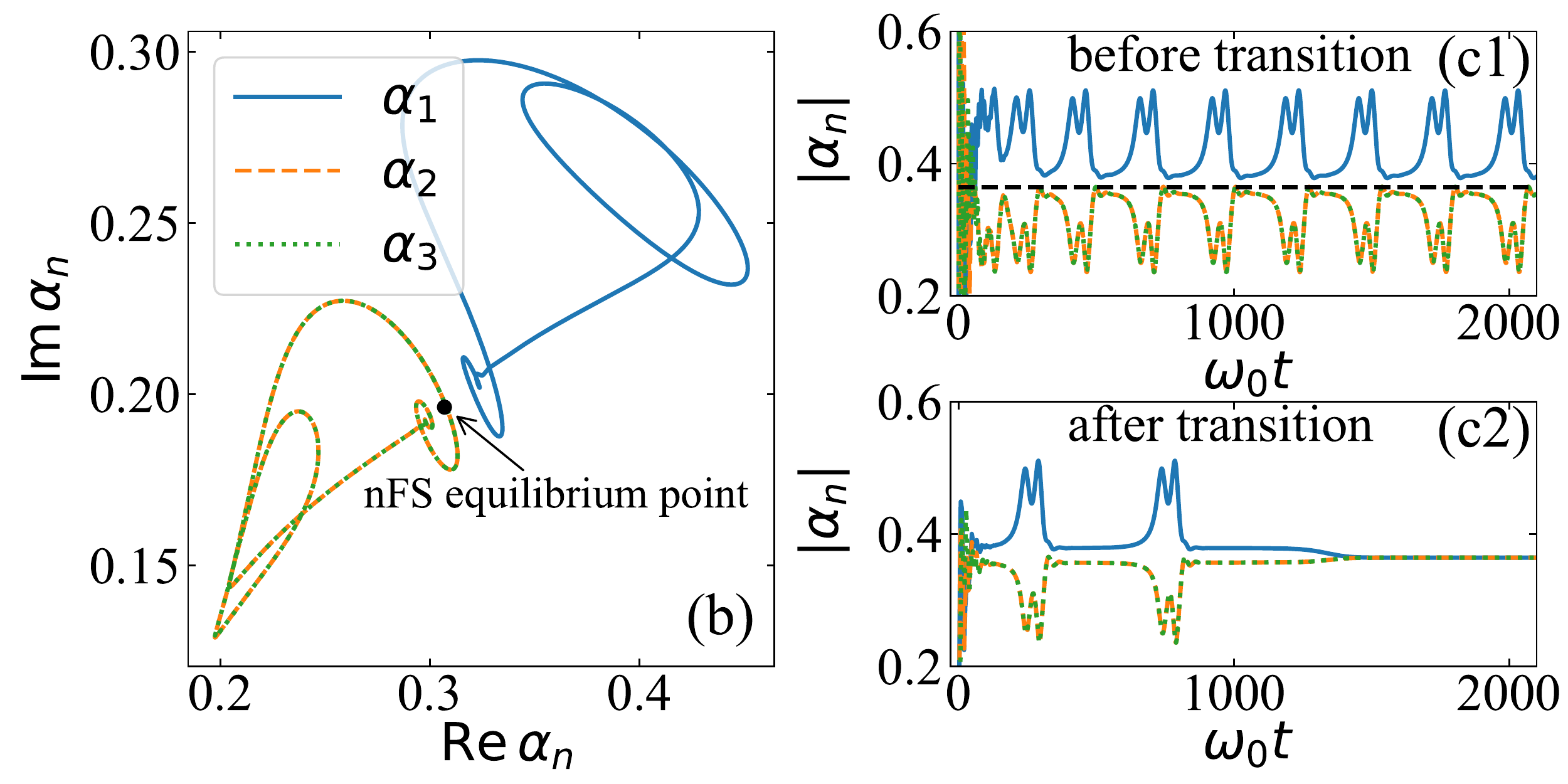}
\caption{[$\mathbf{S}_2$] (a) Full bifurcation diagram of the FS branch in the region marked by the dashed box in Fig.~\ref{fig:bifurcation}(a). Here $\alpha_\text{extreme}$ denotes the local extreme points of periodic and chaotic attractors. The diagram starts with a saddle-node bifurcation point (diamond) and ends with a periodic solution at the vertical dashed line where the transition to $\mathbf{S}_2$ occurs. (b) Projections of the periodic attractor onto the $\alpha$ plane slightly before the transition, for $\eta=0.355,\,g=3.0627$. Closer to the transition, the periodic attractor approaches the nFS equilibrium point as well, giving evidence to the collision of their basins of attraction on $\mathbf{S}_2$. (c1) Temporal traces of cavity fields at the same values for $\eta$ and $g$ as in (b). The value of the cavity fields at the nFS equilibrium point in (b) is marked by the horizontal line. (c2) Temporal traces of the  cavity fields after the transition, for the values $\eta=0.355,\, g=3.062780103$.
}\label{fig:collision_nfsp_fsp}
\end{figure}

\item\emph{$\mathbf{S}_2$.} This curve corresponds to the disappearance of the FS periodic attractors, induced by their collision with the basins of attraction of the nFS equilibria. To understand this mechanism, in Fig.~\ref{fig:collision_nfsp_fsp}(b) and (c1) we show (near the transition) projections of the periodic attractor  on the complex $\alpha$-plane and the associated temporal traces of the cavity fields respectively. We observe that,  in phase space, this periodic attractor moves closer to the stable nFS equilibrium point and its basin of attraction collides with that of the stable nFS equilibrium point at $\mathbf{S}_2$, thereby eliminating the periodic attractor after collision. Evidence for this collision lies in trajectories which stay located in the vicinity of the broken periodic attractor for a finite amount of time before transitioning towards the stable nFS equilibrium point, as exemplified in Fig.~\ref{fig:collision_nfsp_fsp}(c2).
We observed that this kind of collision only happens between solutions with different broken symmetries. 

\begin{figure}[H]
\includegraphics[clip, width = 0.98\columnwidth]{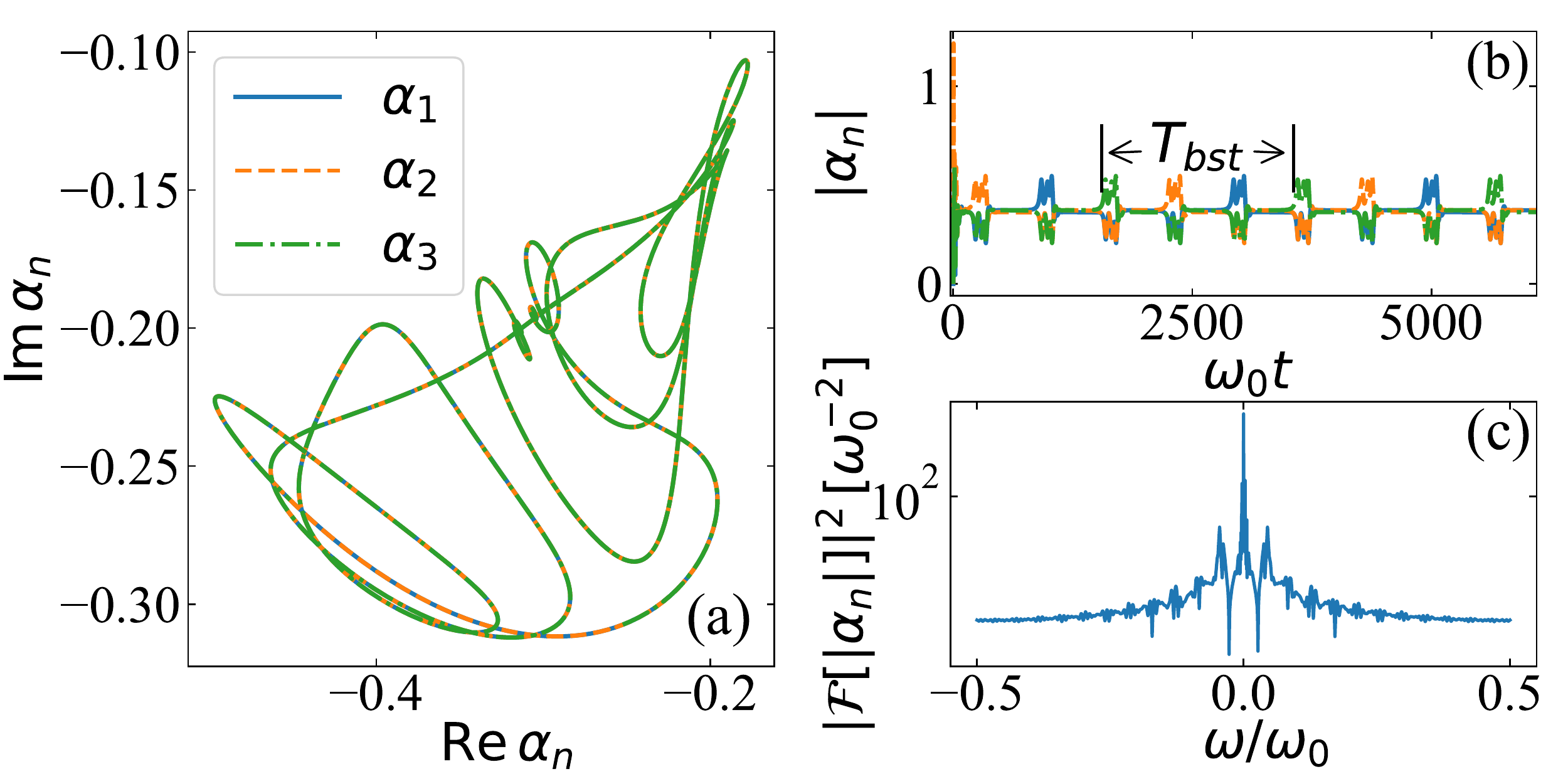}
\caption{[$\mathbf{S}_3$] Burst oscillations observed at the point $K_1$ with $\eta=0.0355,\,g=3.08$ near the segment $\mathbf{S}_3$. Shown in (a-c) are projections of the periodic attractor, temporal traces of the cavity fields and the associated (unnormalized) power spectra $\lvert\mathcal{F}[\lvert\alpha_n\rvert]\rvert^2$ respectively. 
}\label{fig:burst_osc}
\end{figure}

\item\emph{$\mathbf{S}_3$.} This curve is a collection of anomalous Hopf bifurcation points. Across $\mathbf{S}_3$, the system undergoes a Hopf bifurcation with two zero eigenvalues at the transition point, i.e., $\omega_\text{osc}=0$. As a consequence, the post-bifurcation dynamics is drastically different from that presented in Fig.~\ref{fig:period_doubling} and Fig.~\ref{fig:M1M2M3}, where regular Hopf bifurcations characterized by two conjugate purely imaginary eigenvalues at the transition point occur. To illustrate this, in Fig.~\ref{fig:burst_osc}(a), we show that projections of the periodic attractor at the point $K_1$ in Fig.~\ref{fig:openpd}(a) into each cavity field component are the same. 
This implies that they originate from a nFS branch as some pre-bifurcation structure carries over to the post-bifurcation solutions.
Interestingly, in the time domain this periodic attractor displays an unusual periodic burst-oscillation dynamics, which we show in Fig.~\ref{fig:burst_osc}(b). Within a period $T_\text{bst}$, the dynamics undergoes an approximately constant evolution, interrupted by some fast oscillations or bursts.
Furthermore, bursts in different cavities are synchronized with a characteristic time-lag $T_\text{bst}/3$, e. g., $\alpha_{3}(t)=\alpha_1(t+T_\text{bst}/3)$ and  $\alpha_{2}(t)=\alpha_1(t-T_\text{bst}/3)$ in Fig.~\ref{fig:burst_osc}(b). We also find that the period $T_\text{bst}$ diverges while the burst amplitude remains finite when approaching $\mathbf{S}_3$.
Burst oscillations are usually found in type III intermittency routes to chaos and are related to bifurcations of periodic solutions~\cite{Nayfehbook}. We emphasise that, here, this behavior is instead observed in periodic solutions that are generated through unusual Hopf bifurcations.

\begin{figure}[H]
\includegraphics[clip, width = 0.98\columnwidth]{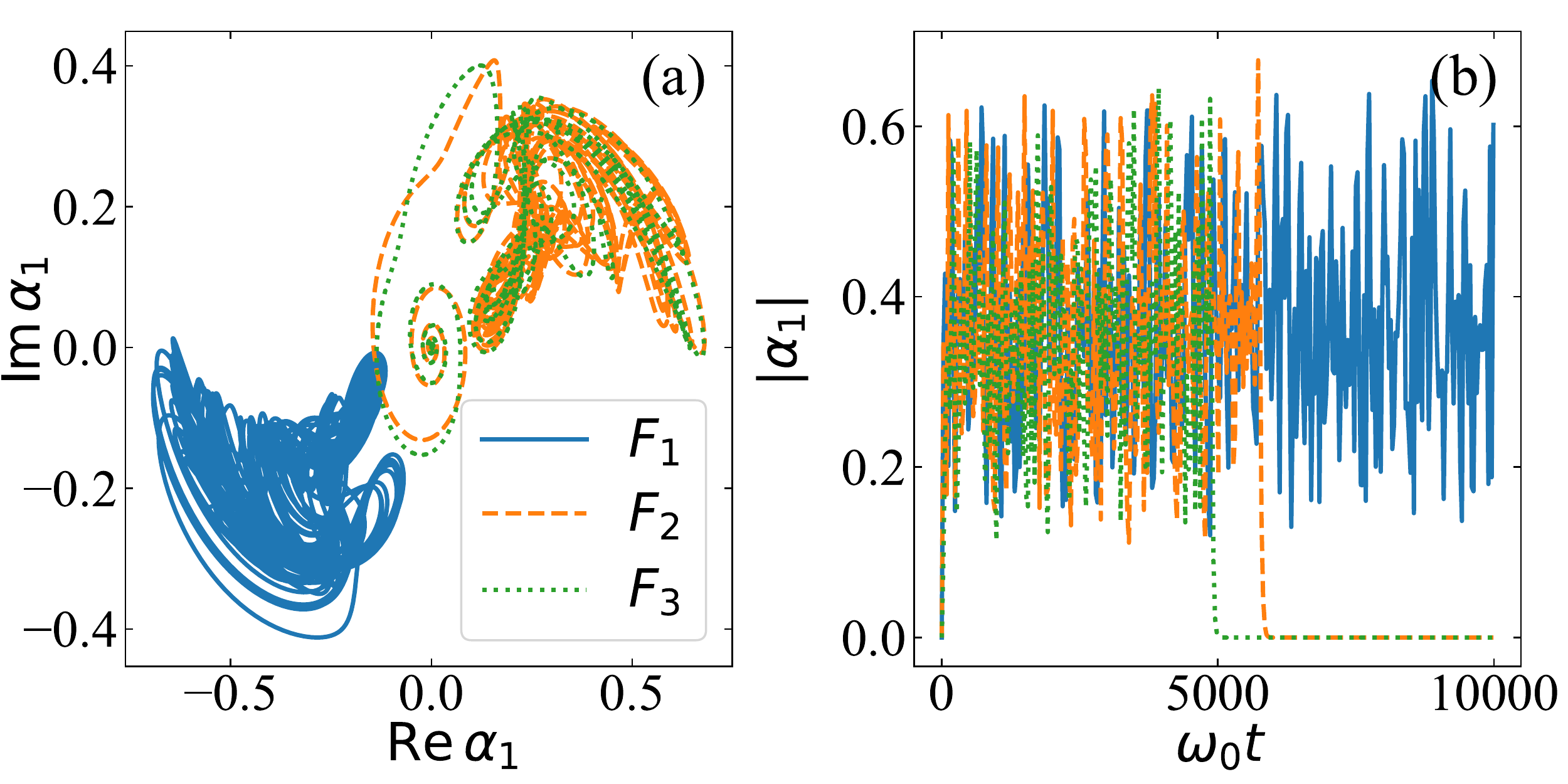}
\caption{[$\mathbf{S}_4$] Destruction of the chaotic attractor via an exterior crisis. Trajectories (a) and temporal traces (b) of the cavity field $\alpha_1$ for the three points $F_1,\,F_2,\,F_3$ in the inset of Fig.~\ref{fig:openpd}(a) with $\eta=0.045$ and $g=3.130, 3.134, 3.138$. At the point $F_1$, the chaotic attractor is alive, which however is destroyed suddenly at the points $F_2,\,F_3$. In (b) the dynamics at $F_2,\,F_3$ show typical signatures of transient chaos.  
}\label{fig:crisis}
\end{figure}

\item\emph{$\mathbf{S}_4$.} The transitions crossing $\mathbf{S}_4$ are associated with the disappearance of the chaotic attractors via exterior crises. To illustrate this, in Fig.~\ref{fig:crisis}, we show the dynamics at three points $F_1$, $F_2$, and $F_3$ near the $\mathbf{S}_4$ line. Specifically, we chose $F_1$ to be located inside the Dyn-region while the points $F_2$ and $F_3$ are outside. An intriguing feature in the dynamics of the cavity fields at $F_2$ and $F_3$ is the abrupt vanishing of the field to zero after a transient chaotic motion, see Fig.~\ref{fig:crisis}(b). This phenomenon was dubbed transient chaos~\cite{Nayfehbook}. This abrupt change of the cavity fields to zero happens at a random time which is a function of the initial conditions. Dynamically, exterior crises occur when a chaotic attractor collides with the basin of attraction of another (point, periodic or quasiperiodic) attractor and after the collision the chaotic attractor merges with the basin of the regular one. Transient chaos can then be observed in dynamics right after the collision since the broken chaotic attractor can still trap the evolution for some time. In our case, it is the collision between the chaotic attractors and the basins of attraction of the N equilibria that leads to the boundary curve $\mathbf{S}_4$.

\item\emph{Points $M_1,\,M_2,\,M_3$.} In Fig.~\ref{fig:M1M2M3} we show three typical (quasi)periodic attractors found in the case $\varphi=\pi$ near one of the $\mathbf{H}^\text{mix}$ curves. At $M_1$ we find a periodic attractor characterized by the oscillation of the cavity fields oscillate around different values for $\alpha$, see Fig.~\ref{fig:M1M2M3}(a). Therefore, near $M_1$ the system undergoes a Hopf bifurcation of the mixed equilibria \cite{Nayfehbook}. Interestingly, we can observe that one of the cavity fields oscillates around the origin of the $\alpha$ plane while trajectories of the other two fields are symmetric to each other with respect to the origin, akin to the mixed equilibria satisfying $\alpha_n=0,\, \alpha_{n+1}=-\alpha_{n+2}$ (with indexes intended modulo $3$). By further increasing the coupling strength $g$, quasiperiodic attractors emerge at the point $M_2$, as illustrated in Fig.~\ref{fig:M1M2M3}(b). At $M_3$, we find the appearance of an interesting type of periodic attractors, see Fig.~\ref{fig:M1M2M3}(c), whose phase space trajectories of all cavity field components coincide. This feature can be interpreted as the deformation and reconnection of the three isolated closed curves in Fig.~\ref{fig:M1M2M3}(a) as $g$ increases. In this case, the oscillations of different cavity fields in the time domain are synchronized only up to a certain time, similarly to the burst oscillations shown in Fig.~\ref{fig:burst_osc}.  
\end{itemize}

\begin{figure}
\includegraphics[clip, width = 0.98\columnwidth]{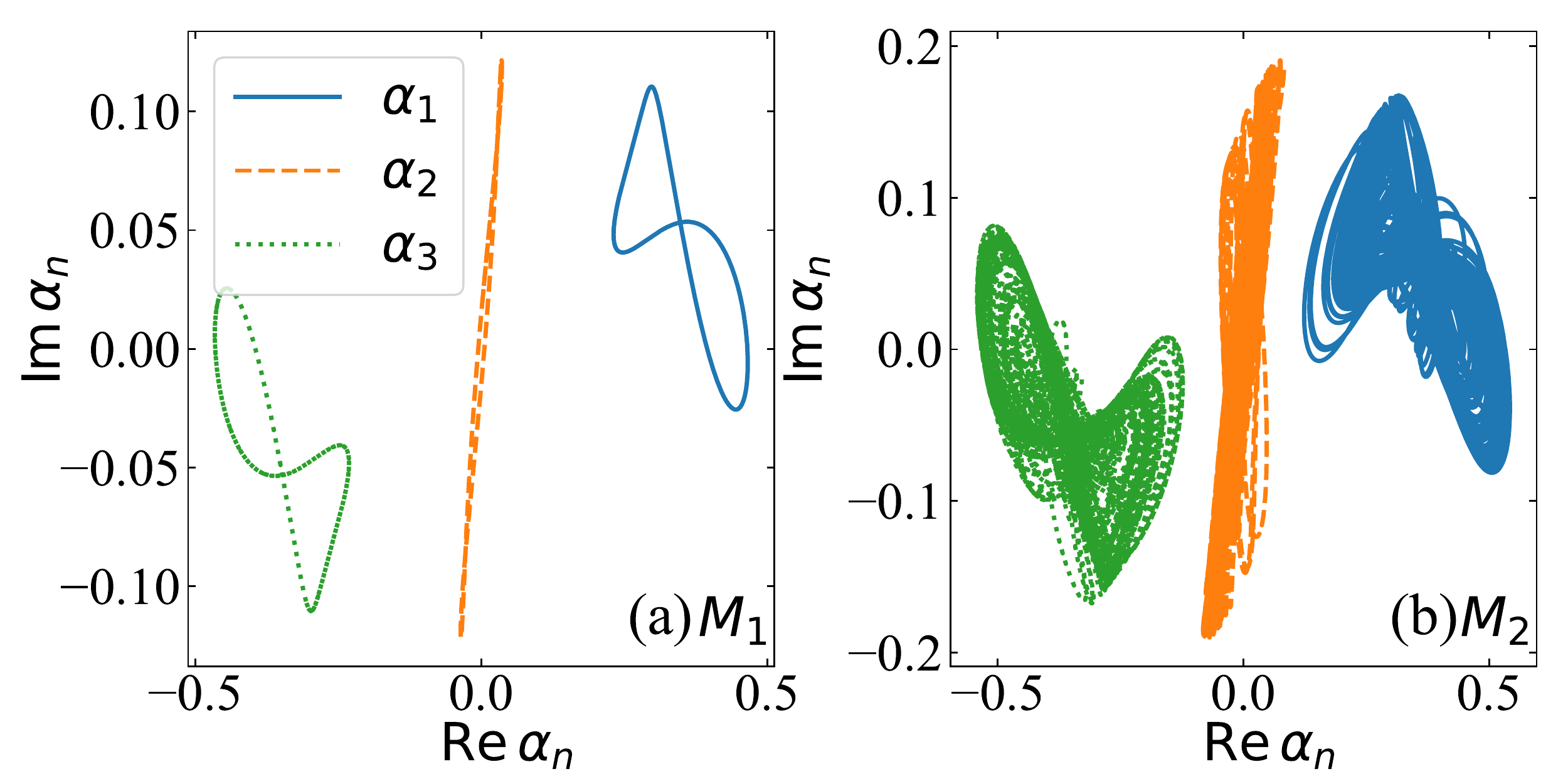}
\includegraphics[clip, width = 0.98\columnwidth]{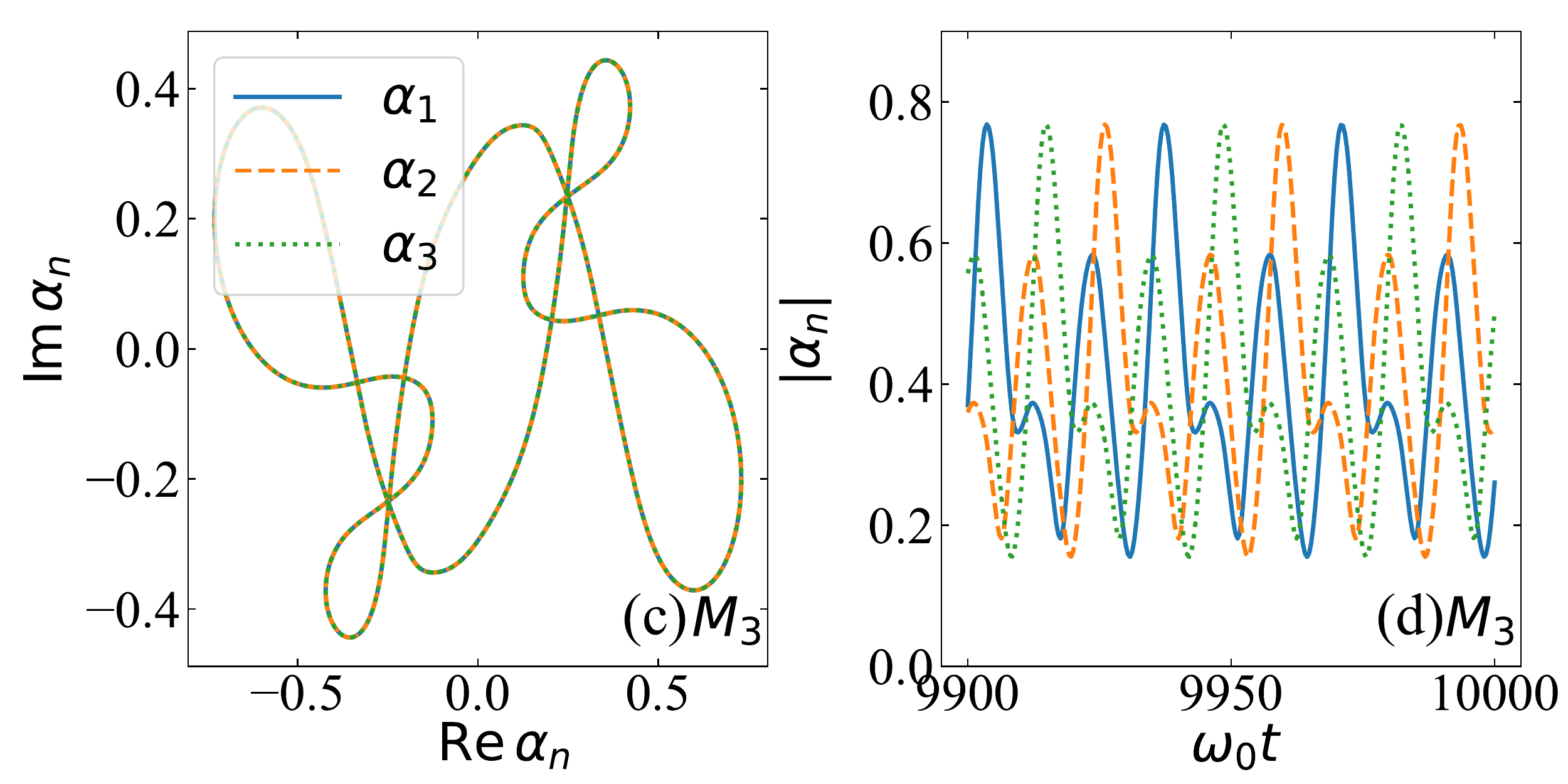}
\caption{[$M_1,\,M_2,\,M_3$] (a-c) Projections of the (quasi)periodic attractors for the points $M_1,\,M_2,\,M_3$ in Fig.~\ref{fig:openpd}(b), with the parameters $\eta=0.2$ and $g=2.24,\,2.28,\,2.5$ respectively. (d) Temporal traces of the cavity fields at $M_3$. 
}\label{fig:M1M2M3}
\end{figure}

\begin{figure}
\includegraphics[clip, width = 0.98\columnwidth]{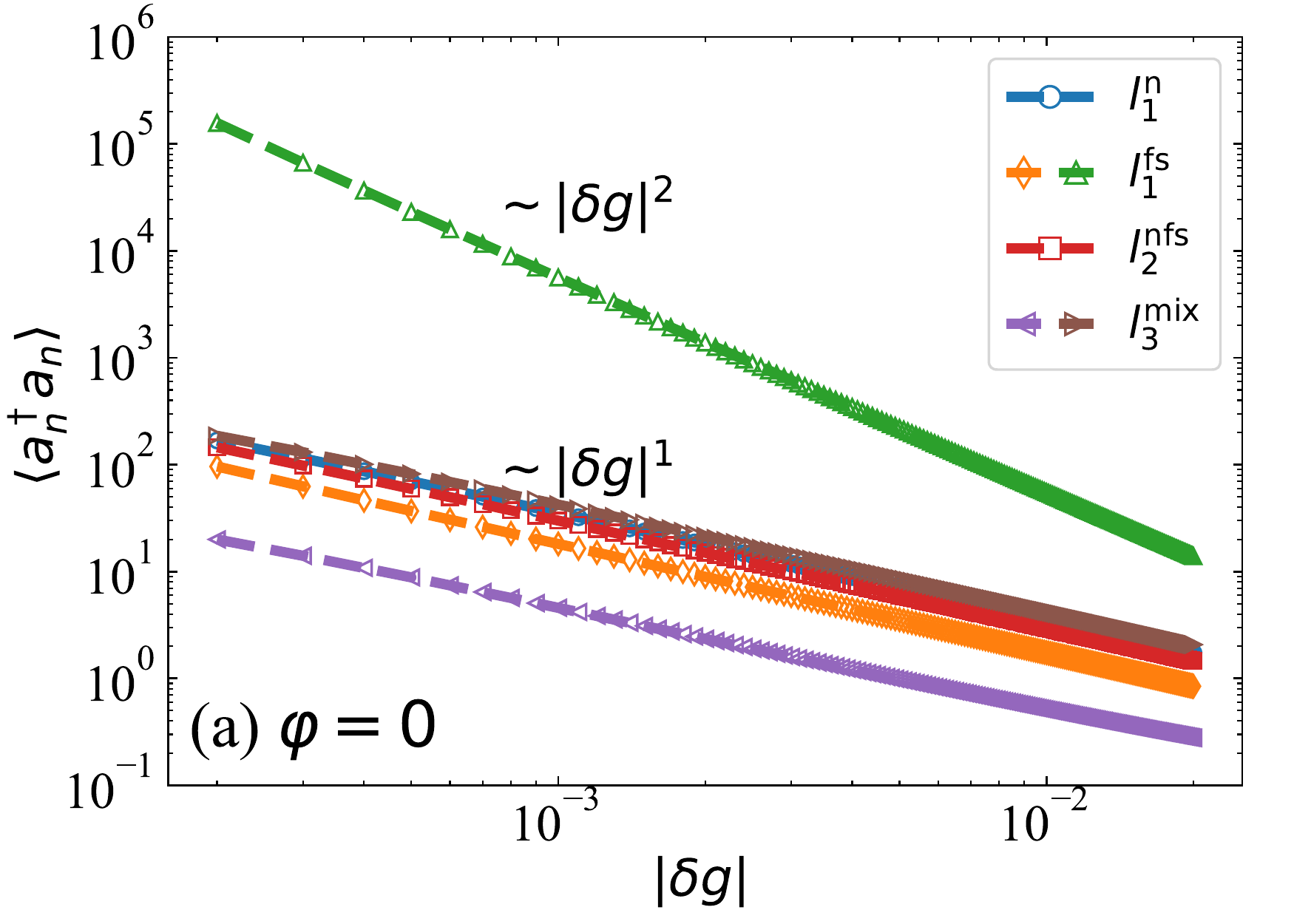}
\includegraphics[clip, width = 0.98\columnwidth]{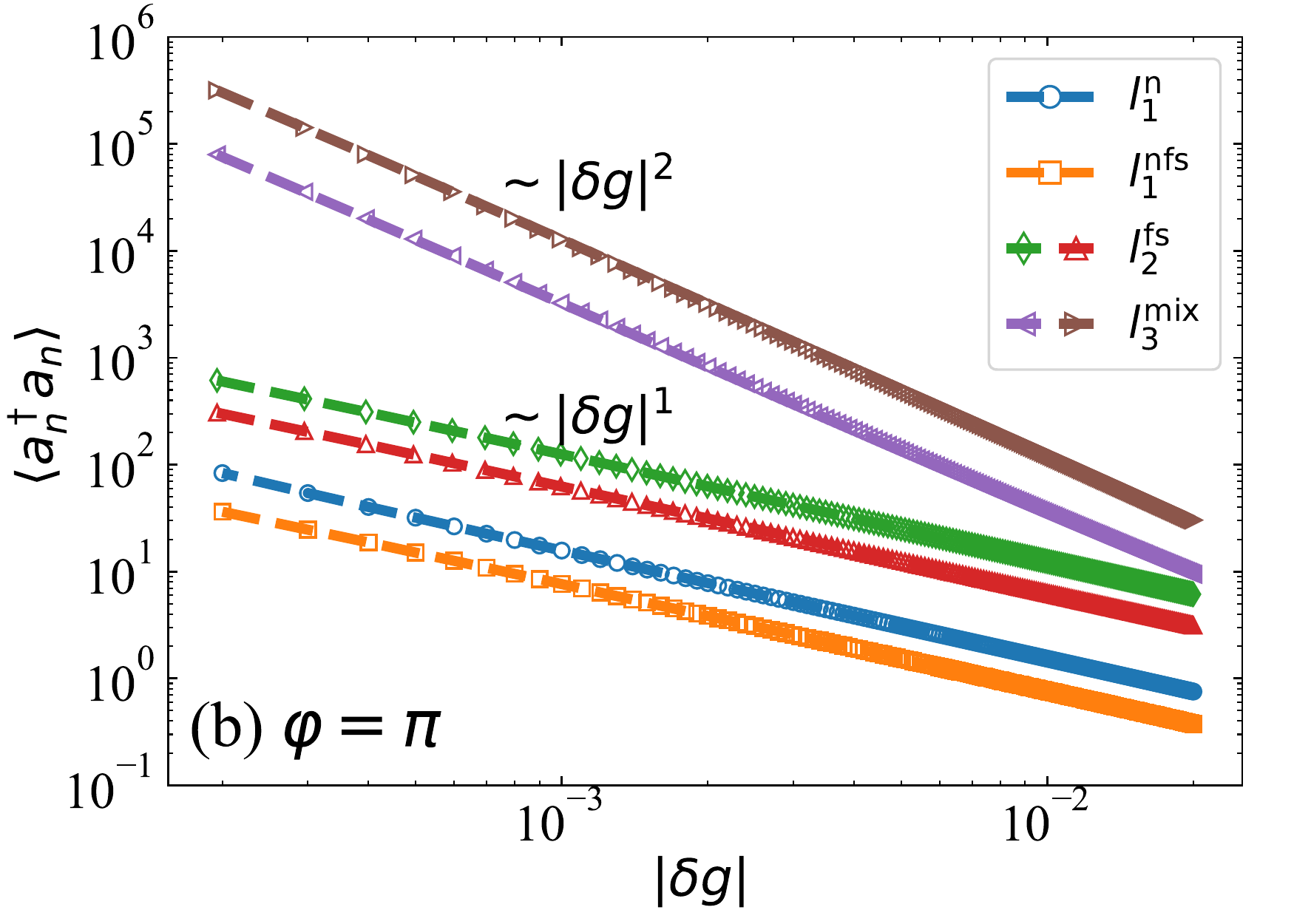}
\caption{Scaling of the cavity photon number calculated along the vertical line cuts labeled by $l_i^\text{X}$ with $i\in\{1,2,3\}$ and $\text{X}\in\{\text{n},\,\text{nfs},\,\text{fs},\,\text{mix}\}$ in Fig.~\ref{fig:openpd}. For $l_i^\text{n}$ or $l_i^\text{nfs}$ there is only one curve as all cavity photon numbers are the same; while for $l_i^\text{fs}$ or $l_i^\text{mix}$ there are two curves. 
}\label{fig:openscaling}
\end{figure}

\subsubsection{Scaling of Steady-state Fluctuations}

To conclude this section, we examine the scaling behavior of the quantum fluctuations specified by the expectation values of the quadratic observables ($a_na_m$, $a_n^\dagger a_m^\dagger$, $a_n^\dagger a_m$, $b_nb_m$, $b_n^\dagger b_m^\dagger$, $b_n^\dagger b_m$, $a_nb_m$, $a_n^\dagger b_m^\dagger$, $a_n^\dagger b_m$, $b_n^\dagger a_m$, see Appendix.~\ref{sec:ssfluc}) along the line cuts in Fig.~\ref{fig:openpd} labeled by $l_i^\text{X}$ with $i\in\{1,2,3\}$ and $\text{X}\in\{\text{n},\,\text{nfs},\,\text{fs},\,\text{mix}\}$. We numerically verified that the condition $\det\mathcal{M}_f\neq0$ holds for all these line cuts, implying the existence of a unique solution $\mathbf{f}_{ss}=\mathcal{M}_f^{-1}\mathbf{v}_f$ (here we reiterate that $\mathbf{f}_\text{ss}$ is a vector whose entries are the expectation values of the quadratic operators listed above, while $\mathcal{M}_f$ and $\mathbf{v}_f$ are the coefficient matrix and vector of the system observables, see Appendix.~\ref{sec:ssfluc}).In Fig.~\ref{fig:openscaling}(a) ($\varphi=0$) and Fig.~\ref{fig:openscaling}(b) ($\varphi=\pi$), we plot the expected photon number $\langle a_n^\dagger a_n\rangle$ for all $n$ which give non-equivalent values. The other quadratic observables  behave similarly. In Fig.~\ref{fig:openscaling}(a) we see that along the cut $l_1^{\text{fs}}$ photon number in one cavity scales as $\langle a_n^\dagger a_n\rangle\sim\lvert\delta g\rvert^2$ while in the other two (where the cavity fields are the same) it scales as $\langle a_n^\dagger a_n\rangle\sim\lvert\delta g\rvert^1$, a typical characteristics of the two critical scalings with the critical exponents ($2$ and $1$) twice that ($1$ and $0.5$) of the closed system~\cite{PhysRevLett.128.163601}. We note that doubling of critical exponents has been found in the open Dicke model~\cite{https://doi.org/10.1002/qute.201800043}. For other line cuts, we find mean-field scaling $\langle a_n^\dagger a_n\rangle\sim\lvert\delta g\rvert^1$. In Fig.~\ref{fig:openscaling}(b) we see that along the cut $l_3^\text{mix}$ the photon number now scales as $\langle a_n^\dagger a_n\rangle\sim\lvert\delta g\rvert^2$ whereas the two critical scalings are replaced by mean-field behavior.

\subsection{Case II: In the Vicinity of $\varphi=0$ and $\pi$}

\begin{figure}
\includegraphics[clip, width = 0.98\columnwidth]{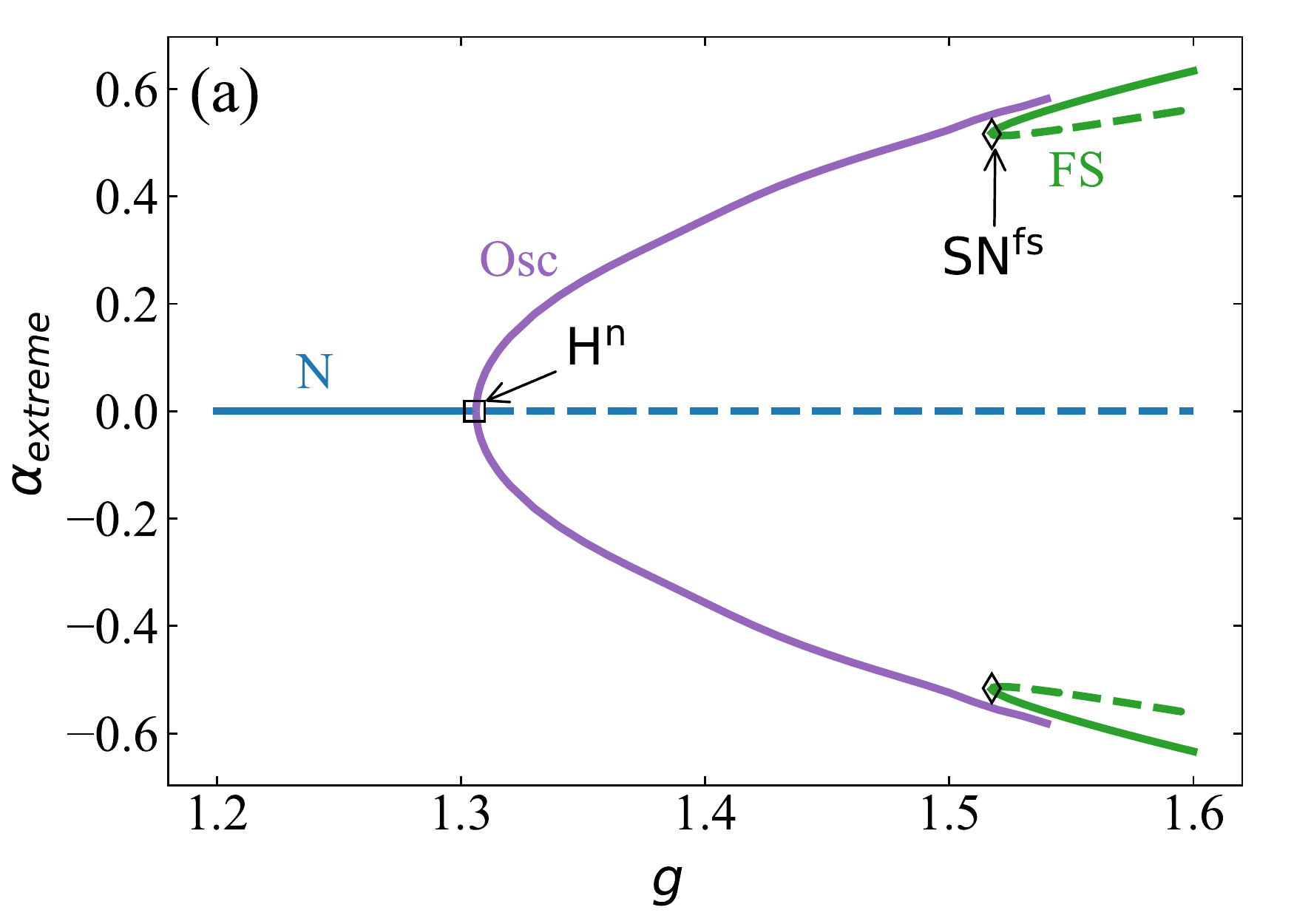}
\includegraphics[clip, width = 0.98\columnwidth]{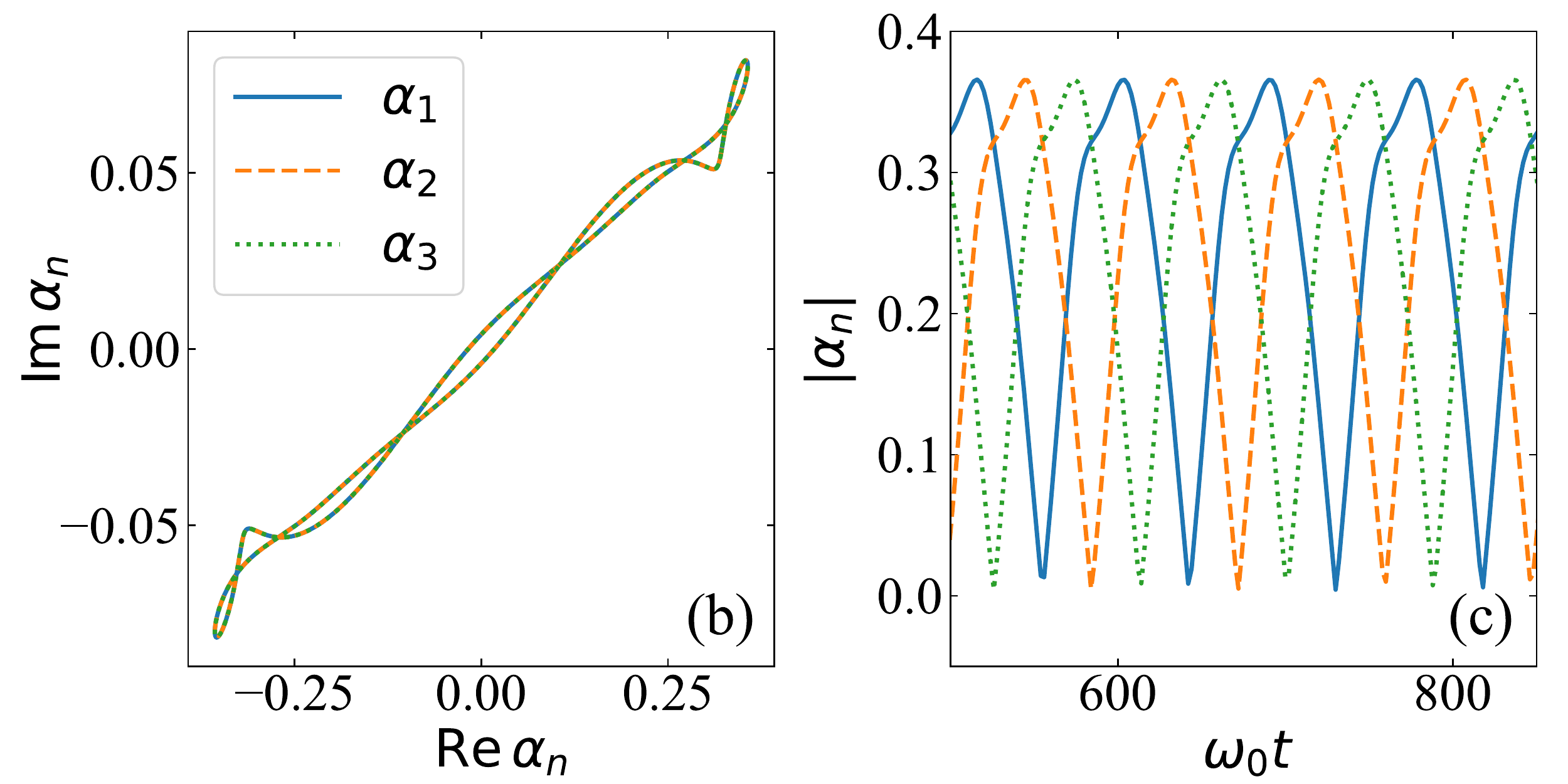}
\caption{Shown are (a) bifurcation diagram for $\varphi=0.1,\,\eta=0.3$, (b) projection and (c) temporal trace of the periodic attractor at $g=1.4$. Hopf bifurcation of the N branch and saddle-node bifurcations of the FS branches are marked by a square symbol and diamonds respectively. The label ``Osc'' stands for periodic solutions. 
}\label{fig:phi=0_vicinity}
\end{figure}

We now discuss the case $0<\varphi<\pi$, where time-reversal symmetry is lifted. In particular, we focus on the lines $\mathbf{P}_\text{s}^{\text{fs}}$ in Fig.~\ref{fig:openpd}(a) and $\mathbf{P}_\text{s}^{\text{nfs}}$ in Fig.~\ref{fig:openpd}(b) around $\varphi=0$ and $\varphi=\pi$. Near $\varphi=0$, we do not observe supercritical pitchfork bifurcations anymore. Instead, the SPTs at $\varphi=0$ are replaced by Hopf bifurcations of the N equilibria, leading to the emergence of new periodic attractors after the bifurcations, as demonstrated in Fig.~\ref{fig:phi=0_vicinity}(a) for $\varphi=0.1,\,\eta=0.3$. The projection and temporal trace of the periodic attractor are shown in Fig.~\ref{fig:phi=0_vicinity}(b) and (c) respectively. There, we observe that the periodic attractor encloses the origin in the $\alpha$ plane and oscillations in different cavities are synchronized up to a fixed time lag, similar to the behavior shown in Fig.~\ref{fig:burst_osc} and Fig.~\ref{fig:M1M2M3}. Further increasing $g$, a branch of stable FS equilibria and a branch of unstable FS equilibria appear via a saddle-node bifurcation. The periodic attractor terminates at some critical coupling, due to its collision with the basin of attraction of the FS equilibrium point. We numerically examine steady-state fluctuations of the N branch and find simple mean-field behavior (not shown), as opposed to the anomalous finite critical fluctuations in closed system. 
On the contrary, in the vicinity of $\varphi=\pi$, there still exists a supercritical pitchfork bifurcation line between the N and nFS regions with mean-field scalings on both sides.

\section{Conclusions and Outlook}\label{sec:conclusion}

We presented an analysis of a generalization of the Dicke trimer model in which both unbalanced light-matter interactions and cavity losses are considered. In the closed case, we  presented a variety of analytical results concerning the ground state structure, excitation spectra, and critical exponents associated with the underlying superradiant phase transition. In this model, we found the emergence of a zero energy mode in the Tavis-Cummings limit where the light-matter interactions possess higher $U(1)$ symmetry. We demonstrated that the appearance of this zero energy mode is related to the corresponding  phase redundancy of the ground state energy. In order to gain more intuition about the 
presence finite critical fluctuations reported in Ref.~\cite{zhao2022anomalous}, we further developed a semiclassical theory for the light degrees of freedom. This effective model suggests that such finite critical fluctuations are a direct consequence of the presence of finite quadratic contributions to the renormalized semiclassical Hamiltonian while approaching the phase transition. 

These results, together with those presented in Refs.~\cite{PhysRevLett.128.163601,zhao2022anomalous} confirm that the closed Dicke trimer model is a promising platform for exploring novel critical phenomena. Future works may proceed to  further analyze entanglement properties of the ground state~\cite{PhysRevLett.92.073602}, the effects of staggered Zeeman magnetic field~\cite{PhysRevLett.122.193201}, two-photon light-matter interactions~\cite{Garbe2020}, and other generalizations of the Dicke model~\cite{PhysRevA.104.043708,PhysRevA.104.063705}.

To analyze the open, unbalanced Dicke trimer model, we have focused on a semiclassical limit whose dynamics is described by a set of nonlinear differential equations. We revealed the existence of a stationary solution featuring superradiant and normal states in different cavities and dynamical phases characterized by periodic, quasiperiodic, or chaotic attractors and  displaying superradiant oscillations. The identification of the transitions related to the arising of these dynamical phases is one of the main contributions of the present work. 
In particular, we introduced anomalous Hopf bifurcations (characterized by bursts oscillations) and  exterior crises (featuring transient chaotic dynamics). The further observation of the robustness of the two-critical-scalings against cavity losses could be supporting its future observation in state-of-art experiments~\cite{Baumann2010,PhysRevLett.107.140402,doi:10.1073/pnas.1417132112,PhysRevLett.121.163601,doi:10.1126/science.abd4385,Zhiqiang:17,PhysRevLett.121.040503, doi:10.1126/science.abi5226,PhysRevLett.91.203001,Zhiqiang2018DickemodelSV}. 

As an outlook on possible future extensions, it would be interesting to further analyze a more general parameter regime characterized by off-resonant light-matter interaction and $\lvert\eta\rvert>1$ where potentially interesting dynamics may arise akin to the open, unbalanced Dicke model \cite{PhysRevResearch.2.033131}. It would also be interesting to generalize the presented semiclassical analysis into a pure quantum perspective, i.e., to study the Lindblad master equation in Eq.~(\ref{eq:ME}) for large, but finite, $N_a$.

\section{Acknowledgements}

C.Z. acknowledges support from the Startup Fund of Yanshan University. M.C. acknowledges support from NSFC (Grants No.~12050410264 and No.~11935012) and NSAF (Grant No.~U1930403).

{\it Note added.---} The results in Sec.~\ref{sec:lossless} for the case of zero dissipation, constituted an early version of this work and were derived independently of Zhao et.~al.~\cite{zhao2022anomalous} where the closed, balanced Dicke lattice model is considered. 

\newpage

\appendix

\section{Derivation of the Ground State Energy and Quadratic Hamiltonian}\label{sec:clq}

In this section, we present details about the derivation of the ground state energy $\bar E_\text{GS}$ and quantum fluctuation Hamiltonian $H_\text{q}$. To this end, as presented in the main text, we first displace the cavity fields by $\sqrt{N_a}\alpha_n$ using the displacement operator $D(\{\alpha_n\})=\prod_n D_n(\alpha_n)$ with $D_n(\alpha_n)=\exp(\sqrt{N_a}\alpha_na_n^\dagger-\sqrt{N_a}\alpha_n^*a_n)$ and express the Hamiltonian as
\begin{eqnarray}
&&D^\dagger(\{\alpha_n\})H_N D(\{\alpha_n\}) \nonumber\\
&=& \sum_{n=1}^N \omega_0(a_n^\dagger+\sqrt{N_a}\alpha_n^*)(a_n+\sqrt{N_a}\alpha_n) + \omega_a J_n^z \nonumber\\
&& + J\Big[e^{i\varphi}(a_n^\dagger+\sqrt{N_a}\alpha_n^*)(a_{n+1}+\sqrt{N_a}\alpha_{n+1})+\text{H. c.}\Big] \nonumber\\
&& + \frac{2\lambda}{\sqrt{N_a}}\Big[\eta_+(a_n^\dagger+a_n+\sqrt{N_a}\alpha_n^*+\sqrt{N_a}\alpha_n)J_n^x \nonumber\\
&& + i\eta_-(a_n-a_n^\dagger+\sqrt{N_a}\alpha_n-\sqrt{N_a}\alpha_n^*)J_n^y \Big].
\end{eqnarray}
Next we apply the spin rotation 
\begin{equation}
U_n^y(\theta_n)=e^{-i\theta_n J_n^y},\;U_n^z(\phi_n)=e^{-i\phi_nJ_n^z},
\end{equation}
with the angles specified by 
\begin{equation}
\cos\theta_n =  \displaystyle\frac{\omega_a}{\Omega_n},~ 
\cos\phi_n = \displaystyle\frac{\eta_+\Re\alpha_n}{A_n},~ 
\sin\phi_n = \displaystyle-\frac{\eta_-\Im\alpha_n}{A_n}, 
\end{equation}
where we defined the new atomic frequency in the rotating frame $\Omega_n = \sqrt{\omega_a^2+16\lambda^2A_n^2}$ and $A_n=\sqrt{\eta_+^2\Re^2\alpha_n+\eta_-^2\Im^2\alpha_n}$. Note that in the NP the solution $\theta_n=\phi_n=0$ should be used.  
After the rotation $U(\{\theta_n,\varphi_n\})=\prod_n U_n^y(\theta_n)U_n^z(\varphi_n)$, the Hamiltonian in the rotated frame becomes
\begin{widetext}
\begin{eqnarray}
&&\Big(U^\dagger(\{\theta_n,\phi_n\})D^\dagger(\{\alpha_n\})\Big) H_N \Big(D(\{\alpha_n\})U(\{\theta_n,\phi_n\})\Big) 
\nonumber\\
&=& \sum_{n=1}^N N_a\Big[\omega_0 |\alpha_n|^2 + J\big(e^{i\varphi}\alpha_n^*\alpha_{n+1}+\text{H. c.}\big)\Big]+\sqrt{N_a}\Big[\omega_0(\alpha_n^*a_n+\text{H. c.})+J\big[e^{i\varphi}(\alpha_n^*a_{n+1}+\alpha_{n+1}a_n^\dagger)+\text{H. c.}\big]\Big] \nonumber\\
&& + \Omega_n J_n^z + \omega_0a_n^\dagger a_n + 
\frac{2\lambda\eta_+}{\sqrt{N_a}}(a_n+a_n^\dagger)(\cos\theta_n\cos\phi_n J_n^x - \sin\phi_n J_n^y + \sin\theta_n\cos\phi_n J_n^z) \nonumber\\
&&+ i\frac{2\lambda\eta_-}{\sqrt{N_a}}(a_n-a_n^\dagger)(\cos\theta_n\sin\phi_n J_n^x + \cos\phi_n J_n^y + \sin\theta_n\sin\phi_n J_n^z) + J(e^{i\varphi}a_n^\dagger a_{n+1}+e^{-i\varphi}a_{n+1}^\dagger a_n). 
\end{eqnarray}
\end{widetext}
Finally we represent the collective spins in terms of bosons using the Holstein-Pirmakoff transformation
\begin{equation}
J_n^z = b_n^\dagger b_n-N_a/2,\; J_n^+ = b_n^\dagger\sqrt{N_a-b_n^\dagger b_n}, 
\end{equation}
which in the thermodynamic limit $N_a\to\infty$ allows to expand the Hamiltonian as a series of $1/N_a$. The ground state energy $E_\text{GS}$ of $\mathcal O(N_a)$ is 
\begin{equation}
E_\text{GS}/N_a = \sum_{n=1}^N \omega_0|\alpha_n|^2
+J(e^{i\varphi}\alpha_n^*\alpha_{n+1}+\text{H. c.}) - \frac{\Omega_n}{2}, 
\end{equation}
while the quantum fluctuations of order $\mathcal O(1)$ is
\begin{eqnarray}
H_\text{q} &=& \sum_{n=1}^N \omega_0 a_n^\dagger a_n + \Omega_n b_n^\dagger b_n \nonumber\\
&& + J(e^{i\varphi}a_n^\dagger a_{n+1}+e^{-i\varphi}a_{n+1}^\dagger a_n) \nonumber\\
&&+\lambda\eta_+\cos\theta_n\cos\phi_n(a_n+a_n^\dagger)(b_n+b_n^\dagger) \nonumber\\
&&-i\lambda\eta_+\sin\phi_n(a_n+a_n^\dagger)(b_n-b_n^\dagger) \nonumber\\
&&+i\lambda\eta_-\cos\theta_n\sin\phi_n(a_n-a_n^\dagger)(b_n+b_n^\dagger) \nonumber\\
&&-\lambda\eta_-\cos\phi_n(a_n-a_n^\dagger)(b_n-b_n^\dagger). 
\end{eqnarray}
After rescaling all bare parameters, we obtain the expressions Eq.~(\ref{eq:EGSsimplifed}) and Eq.~(\ref{eq:Hq}) given in the main text.

\section{Submatrices in the Jacobian Matrix}\label{sec:app_Jacmat}

This section gives the submatrices of the Jacobian matrix in Eq.~(\ref{eq:Jacobian}),
\begin{widetext}
\begin{eqnarray}
\mathcal{A}_n &=& 
\begin{pmatrix}
-\kappa & \omega_0 & 0 & -2\lambda\eta_- \\
-\omega_0 & \kappa & -2\lambda\eta_+ & 0 \\
0 & -4\lambda\eta_- Z_\text{ss} & \frac{4\lambda\eta_-X_\text{ss}\Im\alpha_n}{Z_\text{ss}} & \frac{4\lambda\eta_-Y_\text{ss}\Im\alpha_n}{Z_\text{ss}}-\omega_a \\
-4\lambda\eta_+ Z_\text{ss} & 0 & \frac{4\lambda\eta_+X_\text{ss}\Re\alpha_n}{Z_\text{ss}}+\omega_a & \frac{4\lambda\eta_+Y_\text{ss}\Re\alpha_n}{Z_\text{ss}} 
\end{pmatrix},\nonumber\\
\mathcal{B} &=& 
\begin{pmatrix}
J\sin\varphi & J\cos\varphi & 0 & 0 \\
-J\cos\varphi & J\sin\varphi & 0 & 0 \\
0 & 0 & 0 & 0 \\
0 & 0 & 0 & 0
\end{pmatrix},\;
\mathcal{C} = 
\begin{pmatrix}
-J\sin\varphi & J\cos\varphi & 0 & 0 \\
-J\cos\varphi & -J\sin\varphi & 0 & 0 \\
0 & 0 & 0 & 0 \\
0 & 0 & 0 & 0
\end{pmatrix}.   
\end{eqnarray}
\end{widetext}

\section{Steady-state Fluctuations}\label{sec:ssfluc}

A steady state $\rho_\text{eq}$ of Eq.~(\ref{eq:ME}) satisfies the equation
\begin{equation}
-i[H_N,\,\rho_\text{eq}] + \kappa\sum_n\Big(2a_n\rho_\text{eq}a_n^\dagger
- \{a_n^\dagger a_n,\,\rho_\text{eq}\}\Big) = 0.       
\end{equation}
Now we proceed as in the closed case by switching to a displaced, rotated frame defined with respect to $\mathcal{U}=\mathcal{U}(\{\alpha_n^\text{eq},\theta_m^\text{eq},\phi^\text{eq}_n\})$ in which parameters $\alpha_n^\text{eq},\theta_m^\text{eq},\phi^\text{eq}_n$ are those associated to $\rho_\text{eq}$. In the new frame, we have
\begin{eqnarray}
&&-i[\tilde H_N,\,\tilde\rho_\text{ss}] + \kappa\sum_n\Big(2(a_n+\sqrt{N_a}\alpha_n)\tilde\rho_\text{eq}(a_n^\dagger+\sqrt{N_a}\alpha_n^*) \nonumber\\
&&\quad- \{(a_n^\dagger+\sqrt{N_a}\alpha_n^*)(a_n+\sqrt{N_a}\alpha_n),\,\tilde\rho_\text{eq}\}\Big) = 0,      
\end{eqnarray}
where $\tilde H_N=\mathcal{U}^\dagger H_N \mathcal{U}$, $\tilde\rho_\text{ss}=\mathcal{U}^\dagger \rho_\text{ss} \mathcal{U}$ are operators in the new frame. Now we expand the left-hand side up to order $\mathcal{O}(N_a)$ terms included. Here, the $\mathcal{O}(\sqrt{N_a})$ terms vanish by construction (in the choice of the parameters $\alpha_n^\text{eq}$, $\theta^\text{eq}$, $\phi_n^\text{eq}$), while the $\mathcal{O}({N_a})$ terms are $\mathbb{C}$-numbers, leading to
\begin{equation}
-i[H_\text{q},\,\tilde\rho_\text{ss}] + \kappa\sum_n\Big(2a_n\tilde\rho_\text{ss}a_n^\dagger - \{a_n^\dagger a_n,\,\tilde\rho_\text{ss}\}\Big) = 0.       
\end{equation}
The equation above is closed in the expectations $a_na_m$, $a_n^\dagger a_m^\dagger$, $a_n^\dagger a_m$, $b_nb_m$, $b_n^\dagger b_m^\dagger$, $b_n^\dagger b_m$, $a_nb_m$, $a_n^\dagger b_m^\dagger$, $a_n^\dagger b_m$, $b_n^\dagger a_m$. To simplify the notation, we defined $\Lambda_n^{++}=\lambda\eta_+\cos\theta_n\cos\phi_n$, $\Lambda_n^{+-}=-i\lambda\eta_+\sin\phi_n$, $\Lambda_n^{-+}=i\lambda\eta_-\cos\theta_n\sin\phi_n$, $\Lambda_n^{--}=-\lambda\eta_-\cos\phi_n$ for the coefficients in Eq.~(\ref{eq:Hq}). Below we explicitly list these equations for the interested reader as    

\begin{widetext}
\begin{equation}
\begin{array}{lll}
0&=&\displaystyle - 2i\omega_0\langle a_na_m\rangle - i(\Lambda_n^{++}+\Lambda_n^{+-}-\Lambda_n^{-+}-\Lambda_n^{--})\langle a_mb_n\rangle - i(\Lambda_n^{++}-\Lambda_n^{+-}-\Lambda_n^{-+}+\Lambda_n^{--})\langle a_mb_n^\dagger\rangle  \\
&&\displaystyle - i(\Lambda_m^{++}+\Lambda_m^{+-}-\Lambda_m^{-+}-\Lambda_m^{--})\langle a_nb_m\rangle - i(\Lambda_m^{++}-\Lambda_m^{+-}-\Lambda_m^{-+}+\Lambda_m^{--})\langle a_nb_m^\dagger\rangle \\
&&\displaystyle - iJe^{-i\varphi}\langle a_{n-1}a_m+a_na_{m-1}\rangle - iJe^{i\varphi}
\langle a_{n+1}a_m+a_na_{m+1}\rangle - 2\kappa \langle a_na_m\rangle\;,     \\

0&=&\displaystyle i(\Lambda_n^{++}+\Lambda_n^{+-}+\Lambda_n^{-+}+\Lambda_n^{--})\langle a_mb_n\rangle - i(\Lambda_m^{++}+\Lambda_m^{+-}-\Lambda_m^{-+}-\Lambda_m^{--})\langle a_n^\dagger b_m\rangle \\
&&\displaystyle + i(\Lambda_n^{++}-\Lambda_n^{+-}+\Lambda_n^{-+}-\Lambda_n^{--})\langle a_mb_n^\dagger\rangle - i(\Lambda_m^{++}-\Lambda_m^{+-}-\Lambda_m^{-+}+\Lambda_m^{--})\langle a_n^\dagger b_m^\dagger\rangle \\
&&\displaystyle + iJe^{i\varphi}\langle a_{n-1}^\dagger a_m - a_n^\dagger a_{m+1}\rangle + iJe^{-i\varphi}
\langle a_{n+1}^\dagger a_m - a_n^\dagger a_{m-1}\rangle - 2\kappa \langle a_n^\dagger a_m\rangle\;, \\

0&=&\displaystyle - i\Big(\frac{\omega_a}{\cos\theta_n}+\frac{\omega_a}{\cos\theta_m}\Big)\langle b_nb_m\rangle - i(\Lambda_n^{++}-\Lambda_n^{+-}+\Lambda_n^{-+}-\Lambda_n^{--})\langle a_nb_m\rangle - i(\Lambda_n^{++}-\Lambda_n^{+-}-\Lambda_n^{-+}+\Lambda_n^{--})\langle a_n^\dagger b_m\rangle \\
&&\quad\displaystyle - i(\Lambda_m^{++}-\Lambda_m^{+-}+\Lambda_m^{-+}-\Lambda_m^{--})\langle a_mb_n\rangle - i(\Lambda_m^{++}-\Lambda_m^{+-}-\Lambda_m^{-+}+\Lambda_m^{--})\langle a_m^\dagger b_n\rangle\;, \\

0&=&\displaystyle i\Big(\frac{\omega_a}{\cos\theta_n}-\frac{\omega_a}{\cos\theta_m}\Big)\langle b_n^\dagger b_m\rangle + i(\Lambda_n^{++}+\Lambda_n^{+-}+\Lambda_n^{-+}+\Lambda_n^{--})\langle a_nb_m\rangle + i(\Lambda_n^{++}+\Lambda_n^{+-}-\Lambda_n^{-+}-\Lambda_n^{--})\langle a_n^\dagger b_m\rangle \\
&&\quad\displaystyle - i(\Lambda_m^{++}-\Lambda_m^{+-}+\Lambda_m^{-+}-\Lambda_m^{--})\langle a_mb_n^\dagger\rangle - i(\Lambda_m^{++}-\Lambda_m^{+-}-\Lambda_m^{-+}+\Lambda_m^{--})\langle a_m^\dagger b_n^\dagger\rangle\;, \\

0&=&\displaystyle - i\Big(\omega_0+\frac{\omega_a}{\cos\theta_m}\Big)\langle a_nb_m\rangle - i(\Lambda_n^{++}+\Lambda_n^{+-}-\Lambda_n^{-+}-\Lambda_n^{--})\langle b_nb_m\rangle - i(\Lambda_n^{++}-\Lambda_n^{+-}-\Lambda_n^{-+}+\Lambda_n^{--})\langle b_n^\dagger b_m\rangle \\
&&\quad\displaystyle - i(\Lambda_m^{++}-\Lambda_m^{+-}+\Lambda_m^{-+}-\Lambda_m^{--})\langle a_na_m\rangle - i(\Lambda_m^{++}-\Lambda_m^{+-}-\Lambda_m^{-+}+\Lambda_m^{--})(\delta_{nm}+\langle a_m^\dagger a_n\rangle) \\
&&\quad\displaystyle - iJe^{-i\varphi}\langle a_{n-1}b_m\rangle - iJe^{i\varphi}\langle a_{n+1}b_m\rangle - \kappa\langle a_nb_m\rangle\;,  \\

0&=&\displaystyle i\Big(\omega_0-\frac{\omega_a}{\cos\theta_m}\Big)\langle a_n^\dagger b_m\rangle + i(\Lambda_n^{++}+\Lambda_n^{+-}+\Lambda_n^{-+}+\Lambda_n^{--})\langle b_nb_m\rangle + i(\Lambda_n^{++}-\Lambda_n^{+-}+\Lambda_n^{-+}-\Lambda_n^{--})\langle b_n^\dagger b_m\rangle \\
&&\quad\displaystyle - i(\Lambda_m^{++}-\Lambda_m^{+-}+\Lambda_m^{-+}-\Lambda_m^{--})\langle a_n^\dagger a_m\rangle - i(\Lambda_m^{++}-\Lambda_m^{+-}-\Lambda_m^{-+}+\Lambda_m^{--})\langle a_n^\dagger a_m^\dagger\rangle \\
&&\quad\displaystyle + iJe^{i\varphi}\langle a_{n-1}^\dagger b_m\rangle + iJe^{-i\varphi}\langle a_{n+1}^\dagger b_m\rangle - \kappa\langle a_n^\dagger b_m\rangle\;.
\end{array}
\end{equation}
\end{widetext}

As noted in the main text, these equations  can be expressed in a matrix form as $\mathcal{M}_f\mathbf{f}_\text{ss} + \mathbf{v}_f = 0$ in terms of a vector $\mathbf{f}_\text{ss}$  listing the expectation values of the quadratic operators and in terms of a coefficient matrix , $\mathcal{M}_f$ and a inhomogenous term $\mathbf{v}_f$.

\section{Analytical Results of the nFSP Solutions}\label{sec:analy_nFSP}

In the NP and nFSP, all $\bar\alpha_n$ are identical so we can write $\bar\alpha_n=\bar\alpha$ to simplify Eq.~(\ref{eq:EGSsimplifed}) as
\begin{eqnarray}
\bar E_{\text{GS}}/N = (1+2\bar J\cos\varphi)\lvert\bar\alpha\rvert^2 - \frac 12\sqrt{1+4g^2\bar A^2}, 
\end{eqnarray}
where $\bar A=\sqrt{\eta_+^2\Re^2\bar\alpha+\eta_-^2\Im^2\bar\alpha}$. 
Minimizing $\bar E_{\text{GS}}$ over $\bar\alpha$ leads to the solutions 
\begin{equation}
    \bar{\alpha} =\pm\frac{1}{2g\eta_+}\sqrt{\frac{g^4\eta_+^4}{(1+2\bar{J}\cos\varphi)^2}-1}
\end{equation}
for $\eta>0$, 
\begin{eqnarray}
\bar\alpha = \pm i \frac{1}{2g\eta_-}\sqrt{\frac{g^4\eta_-^4}{(1+2\bar J\cos\varphi)^2}-1}, 
\end{eqnarray}
for $\eta<0$, and
\begin{eqnarray}
\bar\alpha = \frac1g\sqrt{\frac{g^4}{16(1+2\bar J\cos\varphi)^2}-1}\;e^{i\zeta}, 
\end{eqnarray}
for $\eta=0$ and with arbitrary real $\zeta$.

\section{Generic Properties of the FSP Solutions}\label{sec:app_prop_FSP}

In this section we present a semi-analytical proof of  the properties satisfied by the FSP solutions discussed in the main text. We start by simplifying the expression of $\bar E_{\text{GS}}$ at $\eta=1$. 
Substituting $\eta=1$ into Eq.~(\ref{eq:EGSsimplifed}) in the main text we obtain
\begin{eqnarray}\label{eq:EGSeta1}
\bar{E}_{\text{GS}}&=&\sum_n\bigg[ \lvert\bar{\alpha}_{n}\rvert^{2}
-
\frac 12\sqrt{1+4g^{2}\Re^2\bar{\alpha}_{n}} \nonumber\\
&&~~~~~~~~~~~~~+\bar J\left(e^{i\varphi}\bar{\alpha}_{n}^{*}\bar{\alpha}_{n+1}
+\text{H. c.}\right) \bigg]. 
\end{eqnarray}
We can further eliminate $\Im\bar\alpha_n$ by calculating
\begin{eqnarray}\label{alpha}
0=\frac{\partial\bar{E}_{\text{GS}}}{\partial\Im\bar{\alpha}_{n}} &=& 2\Im\bar{\alpha}_{n}+2\bar{J}\cos\varphi\left(\Im\bar{\alpha}_{n-1}+\Im\bar{\alpha}_{n+1}\right) \nonumber\\
&&+2\bar{J}\sin\varphi\left(\Re\bar{\alpha}_{n+1}-\Re\bar{\alpha}_{n-1}\right), 
\end{eqnarray}
and summing over $n$ 
\begin{equation}
\qquad\sum_{n}\Im\bar{\alpha}_{n}+\bar{J}\cos\varphi\sum_{n}\left(\Im\bar{\alpha}_{n+1}+\Im\bar{\alpha}_{n-1}\right)=0, 
\end{equation}
which leads to $\sum_n\Im\bar\alpha_n = 0$. Since in the present model $N=3$, we find the relation 
\begin{equation}
\Im\bar{\alpha}_{n}=\frac{-\bar{J}\sin\varphi}{1-\bar{J}\cos\varphi}\left(\Re\bar{\alpha}_{n+1}-\Re\bar{\alpha}_{n-1}\right).
\end{equation}
Substituting this expression into Eq.~(\ref{eq:EGSeta1}) leads to a simplified ground state energy function 
\begin{equation}\label{eq:EGSeta1s}
\bar E_{\text{GS}} = \sum_n \Bigg[ 
\xi_0\Re^2\alpha_n 
- \frac12\sqrt{1+4g^2\Re^2\bar\alpha_n} 
+ \xi_1\Re\bar\alpha_n \Re\bar\alpha_{n+1}
\Bigg],
\end{equation}
where we defined $\xi_0 = 1-2\bar J^2\sin^2\varphi/(1-\bar J \cos\varphi)$ and 
$\xi_1 = 2\bar J\cos\varphi + 2\bar J^2\sin^2\varphi/(1-\bar J \cos\varphi)$. Differentiating over $\Re\bar\alpha_n$ and rearranging the terms we obtain
\begin{eqnarray}\label{eq:RealphaEq}
\frac{\xi_1}{2}\sum_n\Re\bar\alpha_n = f(\Re\bar\alpha_n),
\end{eqnarray}
in terms of the function \cite{PhysRevLett.128.163601}
\begin{equation}\label{fx}
f(x)=\frac{g^2x}{\sqrt{1+4g^2x^2}}-\left(\xi_{0}-\frac{1}{2}\xi_{1}\right)x. 
\end{equation}
We can now follow the same arguments presented in Ref.\cite{PhysRevLett.128.163601} to prove the property (i) presented in the main text for the FSP solutions at $\eta=1$. We also numerically checked that the same holds for a generic $\eta>0$. We performed similar arguments in the cases $\eta<0$ and $\eta=0$ to prove the properties (ii) and (iii), respectively.

\section{The Matrices $\mathcal{H}_n$ and $\mathcal{H}_J$ in the NP and Superradiant Phase}\label{sec:app_HnHJ}
In this section we provide the explicit expressions for the matrices $\mathcal{H}_n$ and $\mathcal{H}_J$ in the NP and Superradiant Phase.
In the superradiant phase we have
\begin{widetext}
\begin{equation}
\mathcal H_n = 
\begin{pmatrix}
\omega_0 & 0 & \frac{g\sqrt{\omega_0\omega_a}\eta_+^2\Re\bar\alpha_n}{\bar A_n\sqrt{1+4g^2\bar A_n^2}} & -\frac{g\sqrt{\omega_0\omega_a}\eta_+\eta_-\Im\bar\alpha_n}{\bar A_n} \\
0 & \omega_0 & \frac{g\sqrt{\omega_0\omega_a}\eta_-^2\Im\bar\alpha_n}{\bar A_n\sqrt{1+4g^2\bar A_n^2}} & \frac{g\sqrt{\omega_0\omega_a}\eta_+\eta_-\Re\bar\alpha_n}{\bar A_n} \\
\frac{g\sqrt{\omega_0\omega_a}\eta_+^2\Re\bar\alpha_n}{\bar A_n\sqrt{1+4g^2\bar A_n^2}} & \frac{g\sqrt{\omega_0\omega_a}\eta_-^2\Im\bar\alpha_n}{\bar A_n\sqrt{1+4g^2\bar A_n^2}} & \omega_a\sqrt{1+4g^2\bar A_n^2} & 0 \\
-\frac{g\sqrt{\omega_0\omega_a}\eta_+\eta_-\Im\bar\alpha_n}{\bar A_n} & \frac{g\sqrt{\omega_0\omega_a}\eta_+\eta_-\Re\bar\alpha_n}{\bar A_n} & 0 & \omega_a\sqrt{1+4g^2\bar A_n^2}
\end{pmatrix},\;
\mathcal H_J = 
\begin{pmatrix}
\bar J\cos\varphi & -\bar J\sin\varphi & 0 & 0 \\
\bar J\sin\varphi & \bar J\cos\varphi & 0 & 0 \\
0 & 0 & 0 & 0 \\
0 & 0 & 0 & 0 
\end{pmatrix}. 
\end{equation}
\end{widetext}
In the NP, $\mathcal{H}_n$ is obtained by applying the replacements $\eta_+\Re\bar\alpha_n/\bar A_n \to -1,\,\eta_-\Im\bar\alpha_n/\bar A_n \to 0$ and by setting $\bar A_n=0$ to explicitly obtain
\begin{widetext}
\begin{eqnarray}
\mathcal{H}_n = 
\begin{pmatrix}
\omega_0 & 0 & -g\sqrt{\omega_0\omega_a}\eta_+ & 0 \\
0 & \omega_0 & 0 & -g\sqrt{\omega_0\omega_a}\eta_- \\
-g\sqrt{\omega_0\omega_a}\eta_+ & 0 & \omega_a & 0 \\
0 & -g\sqrt{\omega_0\omega_a}\eta_- & 0 & \omega_a
\end{pmatrix},\;
\mathcal{H}_J = 
\begin{pmatrix}
\bar J\cos\varphi & -\bar J\sin\varphi & 0 & 0 \\
\bar J\sin\varphi & \bar J\cos\varphi & 0 & 0 \\
0 & 0 & 0 & 0 \\
0 & 0 & 0 & 0 
\end{pmatrix}.
\end{eqnarray}
\end{widetext}

\section{Approximate FSP Ground State Solutions Near the Critical Point For $\eta=1$}\label{sec:app_fspsol}

In this section we derive approximate ground state solutions of the FSP near the critical point for $\eta=1$. We substitute the ansatz $\bar\alpha_2=\bar\alpha_3$ into Eq.~(\ref{eq:RealphaEq}) and rearrange the equation to arrive at 
\begin{eqnarray}\label{eq:1}
\begin{split}
\xi_0\Re\bar\alpha_1 + \frac12\xi_1\Re\bar\alpha_2 - \frac{g^2\Re\bar\alpha_1}{\sqrt{1+4g^2\Re\bar\alpha_1^2}} = 0, \\
\frac12 \xi_1\Re\bar\alpha_1 + (\xi_0+\frac12\xi_1)\Re\bar\alpha_2 - \frac{g^2\Re\bar\alpha_2}{\sqrt{1+4g^2\Re\bar\alpha_2^2}} = 0.
\end{split}
\end{eqnarray}
Near the critical point, we have the expansions $\Re\bar\alpha_1 \approx r_0 \delta g^{\beta_1} + r_1 \delta g^{\beta_1+1} + r_2 \delta g^{\beta_1+2}$, $\Re\bar\alpha_2 \approx s_0 \delta g^{\beta_2} + s_1 \delta g^{\beta_2+1} + s_2 \delta g^{\beta_2+2}$, where $\beta_{1,2}>0$, $r_0,s_0\neq0$ and $\delta g>0$ in the FSP. Now we insert these expansions into Eq.~(\ref{eq:1}) then get the equations order by order. For the lowest order, we have 
\begin{eqnarray}
\begin{split}
\xi_0r_0\delta g^{\beta_1} + \xi_1s_0\delta g^{\beta_2} - g_c^2r_0\delta g^{\beta_1} = 0, \\
\frac12\xi_1r_0\delta g^{\beta_1} + (\xi_0+\frac12\xi_1)s_0\delta g^{\beta_2} - g_c^2s_0\delta g^{\beta_2} = 0. 
\end{split}
\end{eqnarray}
It is easy to see that $\beta_1=\beta_2$ otherwise either $r_0$ or $s_0$ vanishes. From now on we assume $\beta_1=\beta_2=\beta$. Making use of this relation and $g_c^2=\xi_0-\xi_1/2$ at $\eta=1$, one can show that $s_0=-r_0/2$. 
The equations for the next order are 
\begin{eqnarray}
\begin{split}
&(\xi_0r_1 + \xi_1s_1 - g_c^2r_1 - 2g_cr_0 )\delta g^{\beta+1} + 2g_c^4r_0^3\delta g^{3\beta} = 0, \\
&\Big[\frac12\xi_1r_1+(\xi_0+\frac12\xi_1)s_1-g_c^2s_1-2g_cs_0\Big]\delta g^{\beta+1} \\
&~~~~~~~~~~~~~~~~~~~~~~~~~~~~~~~~~~~~~~~~~~~~+ 2g_c^4s_0^3 \delta g^{3\beta} = 0. \\
\end{split}
\end{eqnarray}
One can readily see that $\beta\neq1/2$ leads to contradiction thus only $\beta=1/2$ is possible and solving the above equation gives
\begin{eqnarray}
r_0 = \pm \frac{2}{\sqrt3g_c^{3/2}},~~~~s_0 = \mp \frac{1}{\sqrt3g_c^{3/2}}.   
\end{eqnarray}
Proceeding along similar lines, we find 
\begin{eqnarray}
r_1 = \pm \frac{1}{6\sqrt3g_c^{5/2}},~s_1 = \frac{4}{3\sqrt3\xi_1g_c^{1/2}}\mp\frac{1}{12\sqrt3g_c^{5/2}}. 
\end{eqnarray}
Summarizing, in the FSP near the critical point we have the approximate solutions
\begin{eqnarray}
\begin{split}
&\Re\bar\alpha_1\approx\pm\frac{2\delta g^{1/2}}{\sqrt3g_c^{3/2}}\pm\frac{\delta g^{3/2}}{6\sqrt3g_c^{5/2}}, \\
&\Re\bar\alpha_2\approx\mp\frac{\delta g^{1/2}}{\sqrt3g_c^{3/2}}+\bigg(\frac{4}{3\sqrt3\xi_1g_c^{1/2}}\mp\frac{1}{12\sqrt3g_c^{5/2}}\bigg)\delta g^{3/2}.
\end{split}
\end{eqnarray}

\section{Classification of the Equilibria}\label{sec:app_classification}

In this section we classify the equilibria of Eq.~(\ref{eq:EOMs}). We first classify the equilibria in terms of the number of vanishing cavity fields denoted by $\#_\text{n}$ by running over all values of $\#_\text{n}$. Obviously $\#_\text{n}=3$ corresponds to the N equilibria.
For $\#_\text{n}=2$, without loss of generality, let us assume $\alpha_1=\alpha_2=0$. Then from 
the first line in Eq.~(\ref{eq:EOMs}) we have 
\begin{equation}
(\kappa+i\omega_0)\alpha_1+2i\lambda\eta_+X_1+2\lambda\eta_-Y_1\pm iJ\alpha_2\pm iJ\alpha_3=0.    
\end{equation}
Since $X_1=Y_1=0$ as a result of $\alpha_1=0$, we get $\alpha_3=0$, which contradicts with the condition $\#_\text{n}=2$. This excludes solutions satisfying $\#_\text{n}=2$.
For $\#_\text{n}=1$, we show below that for $\varphi=0,\,\pi$ these equilibria are subject to the relations $\alpha_n=0,\;\alpha_{n-1}=-\alpha_{n+1}$.

\begin{proof}
Since $\#_\text{n}=1$, without loss of generality we assume $\alpha_n=0$, then from the first line of Eq.~(\ref{eq:EOMs}) one can see that the other two cavity fields satisfy the relation $e^{i\varphi}\alpha_{n-1}+e^{-i\varphi}\alpha_{n+1}=0$ by examining the equation for $\alpha_n$. For $\varphi=0,\,\pi$, this reduces to $\alpha_{n-1}=-\alpha_{n+1}$. From the second and third equations in Eq.~(\ref{eq:EOMs}) we can derive the relation
\begin{equation}
Z_{n\pm1}^2 = \frac{\omega_a^2}{4(\omega_a^2+16\lambda^2A_{n\pm1}^2)}, 
\end{equation}
which implies that $Z_{n-1}=\pm Z_{n+1}$. If $Z_{n-1}=-Z_{n+1}$, we have $X_{n-1}=X_{n+1}$, $Y_{n-1}=Y_{n+1}$. Substituting these into the first line in Eq.~(\ref{eq:EOMs}), we have $\alpha_{n-1}=\alpha_{n+1}$ which, in turn, implies $\alpha_{n\pm1}=0$, contradicting the condition $\#_\text{n}=1$. If $Z_{n-1}=Z_{n+1}$, we have $X_{n-1}=-X_{n+1}$, $Y_{n-1}=-Y_{n+1}$. Combining this relation with the first line in Eq.~(\ref{eq:EOMs}), we find $\alpha_n=-\alpha_n$, so that $\alpha_n=0$. As a consequence, for $\varphi=0,\,\pi$, a solution characterized by $\#_n=1$ has to satisfy the relation $\alpha_n=0,\;\alpha_{n-1}=-\alpha_{n+1}$.

\end{proof}

For $\#_\text{n}=0$, we further classify these equilibria by the number of identical cavity fields denoted by $\#_\text{s}$. Stationary solutions where all cavities have the same field amplitude ($\alpha_n=\alpha$) correspond to the nFS equilibria, while solutions breaking  translational symmetry 
are the FS equilibria with $\#_\text{s}=0,\,2$. This completes the classification of the equilibria.

\nocite{apsrev41Control}
\bibliographystyle{apsrev4-2}
\bibliography{dickerefs}

\begin{thebibliography}{49}%
\makeatletter
\providecommand \@ifxundefined [1]{%
 \@ifx{#1\undefined}
}%
\providecommand \@ifnum [1]{%
 \ifnum #1\expandafter \@firstoftwo
 \else \expandafter \@secondoftwo
 \fi
}%
\providecommand \@ifx [1]{%
 \ifx #1\expandafter \@firstoftwo
 \else \expandafter \@secondoftwo
 \fi
}%
\providecommand \natexlab [1]{#1}%
\providecommand \enquote  [1]{``#1''}%
\providecommand \bibnamefont  [1]{#1}%
\providecommand \bibfnamefont [1]{#1}%
\providecommand \citenamefont [1]{#1}%
\providecommand \href@noop [0]{\@secondoftwo}%
\providecommand \href [0]{\begingroup \@sanitize@url \@href}%
\providecommand \@href[1]{\@@startlink{#1}\@@href}%
\providecommand \@@href[1]{\endgroup#1\@@endlink}%
\providecommand \@sanitize@url [0]{\catcode `\\12\catcode `\$12\catcode
  `\&12\catcode `\#12\catcode `\^12\catcode `\_12\catcode `\%12\relax}%
\providecommand \@@startlink[1]{}%
\providecommand \@@endlink[0]{}%
\providecommand \url  [0]{\begingroup\@sanitize@url \@url }%
\providecommand \@url [1]{\endgroup\@href {#1}{\urlprefix }}%
\providecommand \urlprefix  [0]{URL }%
\providecommand \Eprint [0]{\href }%
\providecommand \doibase [0]{https://doi.org/}%
\providecommand \selectlanguage [0]{\@gobble}%
\providecommand \bibinfo  [0]{\@secondoftwo}%
\providecommand \bibfield  [0]{\@secondoftwo}%
\providecommand \translation [1]{[#1]}%
\providecommand \BibitemOpen [0]{}%
\providecommand \bibitemStop [0]{}%
\providecommand \bibitemNoStop [0]{.\EOS\space}%
\providecommand \EOS [0]{\spacefactor3000\relax}%
\providecommand \BibitemShut  [1]{\csname bibitem#1\endcsname}%
\let\auto@bib@innerbib\@empty
\bibitem [{\citenamefont {Sondhi}\ \emph {et~al.}(1997)\citenamefont {Sondhi},
  \citenamefont {Girvin}, \citenamefont {Carini},\ and\ \citenamefont
  {Shahar}}]{RevModPhys.69.315}%
  \BibitemOpen
  \bibfield  {author} {\bibinfo {author} {\bibfnamefont {S.~L.}\ \bibnamefont
  {Sondhi}}, \bibinfo {author} {\bibfnamefont {S.~M.}\ \bibnamefont {Girvin}},
  \bibinfo {author} {\bibfnamefont {J.~P.}\ \bibnamefont {Carini}},\ and\
  \bibinfo {author} {\bibfnamefont {D.}~\bibnamefont {Shahar}},\ }\bibfield
  {title} {\bibinfo {title} {Continuous quantum phase transitions},\ }\href
  {https://doi.org/10.1103/RevModPhys.69.315} {\bibfield  {journal} {\bibinfo
  {journal} {Rev. Mod. Phys.}\ }\textbf {\bibinfo {volume} {69}},\ \bibinfo
  {pages} {315--333} (\bibinfo {year} {1997})}\BibitemShut {NoStop}%
\bibitem [{\citenamefont {Sachdev}(2011)}]{qptbook}%
  \BibitemOpen
  \bibfield  {author} {\bibinfo {author} {\bibfnamefont {S.}~\bibnamefont
  {Sachdev}},\ }\href {https://doi.org/10.1017/CBO9780511973765} {\emph
  {\bibinfo {title} {\it Quantum Phase Transitions}}}\ (\bibinfo  {publisher}
  {Cambridge University Press},\ \bibinfo {address} {Cambridge},\ \bibinfo
  {year} {2011})\BibitemShut {NoStop}%
\bibitem [{\citenamefont {Cardy}(1996)}]{scalingbook}%
  \BibitemOpen
  \bibfield  {author} {\bibinfo {author} {\bibfnamefont {J.}~\bibnamefont
  {Cardy}},\ }\href {https://doi.org/10.1017/CBO9781316036440} {\emph {\bibinfo
  {title} {\it Scaling and Renormalization in Statistical Physics}}}\ (\bibinfo
   {publisher} {Cambridge University Press},\ \bibinfo {address} {Cambridge},\
  \bibinfo {year} {1996})\BibitemShut {NoStop}%
\bibitem [{\citenamefont {Kirton}\ \emph {et~al.}(2019)\citenamefont {Kirton},
  \citenamefont {Roses}, \citenamefont {Keeling},\ and\ \citenamefont
  {Dalla~Torre}}]{https://doi.org/10.1002/qute.201800043}%
  \BibitemOpen
  \bibfield  {author} {\bibinfo {author} {\bibfnamefont {P.}~\bibnamefont
  {Kirton}}, \bibinfo {author} {\bibfnamefont {M.~M.}\ \bibnamefont {Roses}},
  \bibinfo {author} {\bibfnamefont {J.}~\bibnamefont {Keeling}},\ and\ \bibinfo
  {author} {\bibfnamefont {E.~G.}\ \bibnamefont {Dalla~Torre}},\ }\bibfield
  {title} {\bibinfo {title} {Introduction to the {D}icke model: From
  equilibrium to nonequilibrium, and vice versa},\ }\href
  {https://doi.org/https://doi.org/10.1002/qute.201800043} {\bibfield
  {journal} {\bibinfo  {journal} {Advanced Quantum Technologies}\ }\textbf
  {\bibinfo {volume} {2}},\ \bibinfo {pages} {1800043} (\bibinfo {year}
  {2019})}\BibitemShut {NoStop}%
\bibitem [{\citenamefont {Dicke}(1954)}]{PhysRev.93.99}%
  \BibitemOpen
  \bibfield  {author} {\bibinfo {author} {\bibfnamefont {R.~H.}\ \bibnamefont
  {Dicke}},\ }\bibfield  {title} {\bibinfo {title} {Coherence in spontaneous
  radiation processes},\ }\href {https://doi.org/10.1103/PhysRev.93.99}
  {\bibfield  {journal} {\bibinfo  {journal} {Phys. Rev.}\ }\textbf {\bibinfo
  {volume} {93}},\ \bibinfo {pages} {99--110} (\bibinfo {year}
  {1954})}\BibitemShut {NoStop}%
\bibitem [{\citenamefont {Wang}\ and\ \citenamefont
  {Hioe}(1973)}]{PhysRevA.7.831}%
  \BibitemOpen
  \bibfield  {author} {\bibinfo {author} {\bibfnamefont {Y.~K.}\ \bibnamefont
  {Wang}}\ and\ \bibinfo {author} {\bibfnamefont {F.~T.}\ \bibnamefont
  {Hioe}},\ }\bibfield  {title} {\bibinfo {title} {Phase transition in the
  {D}icke model of superradiance},\ }\href
  {https://doi.org/10.1103/PhysRevA.7.831} {\bibfield  {journal} {\bibinfo
  {journal} {Phys. Rev. A}\ }\textbf {\bibinfo {volume} {7}},\ \bibinfo {pages}
  {831--836} (\bibinfo {year} {1973})}\BibitemShut {NoStop}%
\bibitem [{\citenamefont {Hioe}(1973)}]{PhysRevA.8.1440}%
  \BibitemOpen
  \bibfield  {author} {\bibinfo {author} {\bibfnamefont {F.~T.}\ \bibnamefont
  {Hioe}},\ }\bibfield  {title} {\bibinfo {title} {Phase transitions in some
  generalized {D}icke models of superradiance},\ }\href
  {https://doi.org/10.1103/PhysRevA.8.1440} {\bibfield  {journal} {\bibinfo
  {journal} {Phys. Rev. A}\ }\textbf {\bibinfo {volume} {8}},\ \bibinfo {pages}
  {1440--1445} (\bibinfo {year} {1973})}\BibitemShut {NoStop}%
\bibitem [{\citenamefont {Hepp}\ and\ \citenamefont
  {Lieb}(1973)}]{PhysRevA.8.2517}%
  \BibitemOpen
  \bibfield  {author} {\bibinfo {author} {\bibfnamefont {K.}~\bibnamefont
  {Hepp}}\ and\ \bibinfo {author} {\bibfnamefont {E.~H.}\ \bibnamefont
  {Lieb}},\ }\bibfield  {title} {\bibinfo {title} {Equilibrium statistical
  mechanics of matter interacting with the quantized radiation field},\ }\href
  {https://doi.org/10.1103/PhysRevA.8.2517} {\bibfield  {journal} {\bibinfo
  {journal} {Phys. Rev. A}\ }\textbf {\bibinfo {volume} {8}},\ \bibinfo {pages}
  {2517--2525} (\bibinfo {year} {1973})}\BibitemShut {NoStop}%
\bibitem [{\citenamefont {Shammah}\ \emph {et~al.}(2018)\citenamefont
  {Shammah}, \citenamefont {Ahmed}, \citenamefont {Lambert}, \citenamefont
  {De~Liberato},\ and\ \citenamefont {Nori}}]{PhysRevA.98.063815}%
  \BibitemOpen
  \bibfield  {author} {\bibinfo {author} {\bibfnamefont {N.}~\bibnamefont
  {Shammah}}, \bibinfo {author} {\bibfnamefont {S.}~\bibnamefont {Ahmed}},
  \bibinfo {author} {\bibfnamefont {N.}~\bibnamefont {Lambert}}, \bibinfo
  {author} {\bibfnamefont {S.}~\bibnamefont {De~Liberato}},\ and\ \bibinfo
  {author} {\bibfnamefont {F.}~\bibnamefont {Nori}},\ }\bibfield  {title}
  {\bibinfo {title} {Open quantum systems with local and collective incoherent
  processes: Efficient numerical simulations using permutational invariance},\
  }\href {https://doi.org/10.1103/PhysRevA.98.063815} {\bibfield  {journal}
  {\bibinfo  {journal} {Phys. Rev. A}\ }\textbf {\bibinfo {volume} {98}},\
  \bibinfo {pages} {063815} (\bibinfo {year} {2018})}\BibitemShut {NoStop}%
\bibitem [{\citenamefont {Baumann}\ \emph {et~al.}(2010)\citenamefont
  {Baumann}, \citenamefont {Guerlin}, \citenamefont {Brennecke},\ and\
  \citenamefont {Esslinger}}]{Baumann2010}%
  \BibitemOpen
  \bibfield  {author} {\bibinfo {author} {\bibfnamefont {K.}~\bibnamefont
  {Baumann}}, \bibinfo {author} {\bibfnamefont {C.}~\bibnamefont {Guerlin}},
  \bibinfo {author} {\bibfnamefont {F.}~\bibnamefont {Brennecke}},\ and\
  \bibinfo {author} {\bibfnamefont {T.}~\bibnamefont {Esslinger}},\ }\bibfield
  {title} {\bibinfo {title} {{D}icke quantum phase transition with a superfluid
  gas in an optical cavity},\ }\href {https://doi.org/10.1038/nature09009}
  {\bibfield  {journal} {\bibinfo  {journal} {Nature}\ }\textbf {\bibinfo
  {volume} {464}},\ \bibinfo {pages} {1301--1306} (\bibinfo {year}
  {2010})}\BibitemShut {NoStop}%
\bibitem [{\citenamefont {Baumann}\ \emph {et~al.}(2011)\citenamefont
  {Baumann}, \citenamefont {Mottl}, \citenamefont {Brennecke},\ and\
  \citenamefont {Esslinger}}]{PhysRevLett.107.140402}%
  \BibitemOpen
  \bibfield  {author} {\bibinfo {author} {\bibfnamefont {K.}~\bibnamefont
  {Baumann}}, \bibinfo {author} {\bibfnamefont {R.}~\bibnamefont {Mottl}},
  \bibinfo {author} {\bibfnamefont {F.}~\bibnamefont {Brennecke}},\ and\
  \bibinfo {author} {\bibfnamefont {T.}~\bibnamefont {Esslinger}},\ }\bibfield
  {title} {\bibinfo {title} {Exploring symmetry breaking at the {D}icke quantum
  phase transition},\ }\href {https://doi.org/10.1103/PhysRevLett.107.140402}
  {\bibfield  {journal} {\bibinfo  {journal} {Phys. Rev. Lett.}\ }\textbf
  {\bibinfo {volume} {107}},\ \bibinfo {pages} {140402} (\bibinfo {year}
  {2011})}\BibitemShut {NoStop}%
\bibitem [{\citenamefont {Klinder}\ \emph {et~al.}(2015)\citenamefont
  {Klinder}, \citenamefont {Keßler}, \citenamefont {Wolke}, \citenamefont
  {Mathey},\ and\ \citenamefont {Hemmerich}}]{doi:10.1073/pnas.1417132112}%
  \BibitemOpen
  \bibfield  {author} {\bibinfo {author} {\bibfnamefont {J.}~\bibnamefont
  {Klinder}}, \bibinfo {author} {\bibfnamefont {H.}~\bibnamefont {Keßler}},
  \bibinfo {author} {\bibfnamefont {M.}~\bibnamefont {Wolke}}, \bibinfo
  {author} {\bibfnamefont {L.}~\bibnamefont {Mathey}},\ and\ \bibinfo {author}
  {\bibfnamefont {A.}~\bibnamefont {Hemmerich}},\ }\bibfield  {title} {\bibinfo
  {title} {Dynamical phase transition in the open {D}icke model},\ }\href
  {https://doi.org/10.1073/pnas.1417132112} {\bibfield  {journal} {\bibinfo
  {journal} {Proceedings of the National Academy of Sciences}\ }\textbf
  {\bibinfo {volume} {112}},\ \bibinfo {pages} {3290--3295} (\bibinfo {year}
  {2015})}\BibitemShut {NoStop}%
\bibitem [{\citenamefont {Zhiqiang}\ \emph {et~al.}(2017)\citenamefont
  {Zhiqiang}, \citenamefont {Lee}, \citenamefont {Kumar}, \citenamefont
  {Arnold}, \citenamefont {Masson}, \citenamefont {Parkins},\ and\
  \citenamefont {Barrett}}]{Zhiqiang:17}%
  \BibitemOpen
  \bibfield  {author} {\bibinfo {author} {\bibfnamefont {Z.}~\bibnamefont
  {Zhiqiang}}, \bibinfo {author} {\bibfnamefont {C.~H.}\ \bibnamefont {Lee}},
  \bibinfo {author} {\bibfnamefont {R.}~\bibnamefont {Kumar}}, \bibinfo
  {author} {\bibfnamefont {K.~J.}\ \bibnamefont {Arnold}}, \bibinfo {author}
  {\bibfnamefont {S.~J.}\ \bibnamefont {Masson}}, \bibinfo {author}
  {\bibfnamefont {A.~S.}\ \bibnamefont {Parkins}},\ and\ \bibinfo {author}
  {\bibfnamefont {M.~D.}\ \bibnamefont {Barrett}},\ }\bibfield  {title}
  {\bibinfo {title} {Nonequilibrium phase transition in a spin-1 {D}icke
  model},\ }\href {https://doi.org/10.1364/OPTICA.4.000424} {\bibfield
  {journal} {\bibinfo  {journal} {Optica}\ }\textbf {\bibinfo {volume} {4}},\
  \bibinfo {pages} {424--429} (\bibinfo {year} {2017})}\BibitemShut {NoStop}%
\bibitem [{\citenamefont {Kroeze}\ \emph {et~al.}(2018)\citenamefont {Kroeze},
  \citenamefont {Guo}, \citenamefont {Vaidya}, \citenamefont {Keeling},\ and\
  \citenamefont {Lev}}]{PhysRevLett.121.163601}%
  \BibitemOpen
  \bibfield  {author} {\bibinfo {author} {\bibfnamefont {R.~M.}\ \bibnamefont
  {Kroeze}}, \bibinfo {author} {\bibfnamefont {Y.}~\bibnamefont {Guo}},
  \bibinfo {author} {\bibfnamefont {V.~D.}\ \bibnamefont {Vaidya}}, \bibinfo
  {author} {\bibfnamefont {J.}~\bibnamefont {Keeling}},\ and\ \bibinfo {author}
  {\bibfnamefont {B.~L.}\ \bibnamefont {Lev}},\ }\bibfield  {title} {\bibinfo
  {title} {Spinor self-ordering of a quantum gas in a cavity},\ }\href
  {https://doi.org/10.1103/PhysRevLett.121.163601} {\bibfield  {journal}
  {\bibinfo  {journal} {Phys. Rev. Lett.}\ }\textbf {\bibinfo {volume} {121}},\
  \bibinfo {pages} {163601} (\bibinfo {year} {2018})}\BibitemShut {NoStop}%
\bibitem [{\citenamefont {Zhang}\ \emph {et~al.}(2021)\citenamefont {Zhang},
  \citenamefont {Chen}, \citenamefont {Wu}, \citenamefont {Wang}, \citenamefont
  {Fan}, \citenamefont {Deng},\ and\ \citenamefont
  {Wu}}]{doi:10.1126/science.abd4385}%
  \BibitemOpen
  \bibfield  {author} {\bibinfo {author} {\bibfnamefont {X.}~\bibnamefont
  {Zhang}}, \bibinfo {author} {\bibfnamefont {Y.}~\bibnamefont {Chen}},
  \bibinfo {author} {\bibfnamefont {Z.}~\bibnamefont {Wu}}, \bibinfo {author}
  {\bibfnamefont {J.}~\bibnamefont {Wang}}, \bibinfo {author} {\bibfnamefont
  {J.}~\bibnamefont {Fan}}, \bibinfo {author} {\bibfnamefont {S.}~\bibnamefont
  {Deng}},\ and\ \bibinfo {author} {\bibfnamefont {H.}~\bibnamefont {Wu}},\
  }\bibfield  {title} {\bibinfo {title} {Observation of a superradiant quantum
  phase transition in an intracavity degenerate {F}ermi gas},\ }\href
  {https://doi.org/10.1126/science.abd4385} {\bibfield  {journal} {\bibinfo
  {journal} {Science}\ }\textbf {\bibinfo {volume} {373}},\ \bibinfo {pages}
  {1359--1362} (\bibinfo {year} {2021})}\BibitemShut {NoStop}%
\bibitem [{\citenamefont {Safavi-Naini}\ \emph {et~al.}(2018)\citenamefont
  {Safavi-Naini}, \citenamefont {Lewis-Swan}, \citenamefont {Bohnet},
  \citenamefont {G\"arttner}, \citenamefont {Gilmore}, \citenamefont {Jordan},
  \citenamefont {Cohn}, \citenamefont {Freericks}, \citenamefont {Rey},\ and\
  \citenamefont {Bollinger}}]{PhysRevLett.121.040503}%
  \BibitemOpen
  \bibfield  {author} {\bibinfo {author} {\bibfnamefont {A.}~\bibnamefont
  {Safavi-Naini}}, \bibinfo {author} {\bibfnamefont {R.~J.}\ \bibnamefont
  {Lewis-Swan}}, \bibinfo {author} {\bibfnamefont {J.~G.}\ \bibnamefont
  {Bohnet}}, \bibinfo {author} {\bibfnamefont {M.}~\bibnamefont {G\"arttner}},
  \bibinfo {author} {\bibfnamefont {K.~A.}\ \bibnamefont {Gilmore}}, \bibinfo
  {author} {\bibfnamefont {J.~E.}\ \bibnamefont {Jordan}}, \bibinfo {author}
  {\bibfnamefont {J.}~\bibnamefont {Cohn}}, \bibinfo {author} {\bibfnamefont
  {J.~K.}\ \bibnamefont {Freericks}}, \bibinfo {author} {\bibfnamefont {A.~M.}\
  \bibnamefont {Rey}},\ and\ \bibinfo {author} {\bibfnamefont {J.~J.}\
  \bibnamefont {Bollinger}},\ }\bibfield  {title} {\bibinfo {title}
  {Verification of a many-ion simulator of the {D}icke model through slow
  quenches across a phase transition},\ }\href
  {https://doi.org/10.1103/PhysRevLett.121.040503} {\bibfield  {journal}
  {\bibinfo  {journal} {Phys. Rev. Lett.}\ }\textbf {\bibinfo {volume} {121}},\
  \bibinfo {pages} {040503} (\bibinfo {year} {2018})}\BibitemShut {NoStop}%
\bibitem [{\citenamefont {Gilmore}\ \emph {et~al.}(2021)\citenamefont
  {Gilmore}, \citenamefont {Affolter}, \citenamefont {Lewis-Swan},
  \citenamefont {Barberena}, \citenamefont {Jordan}, \citenamefont {Rey},\ and\
  \citenamefont {Bollinger}}]{doi:10.1126/science.abi5226}%
  \BibitemOpen
  \bibfield  {author} {\bibinfo {author} {\bibfnamefont {K.~A.}\ \bibnamefont
  {Gilmore}}, \bibinfo {author} {\bibfnamefont {M.}~\bibnamefont {Affolter}},
  \bibinfo {author} {\bibfnamefont {R.~J.}\ \bibnamefont {Lewis-Swan}},
  \bibinfo {author} {\bibfnamefont {D.}~\bibnamefont {Barberena}}, \bibinfo
  {author} {\bibfnamefont {E.}~\bibnamefont {Jordan}}, \bibinfo {author}
  {\bibfnamefont {A.~M.}\ \bibnamefont {Rey}},\ and\ \bibinfo {author}
  {\bibfnamefont {J.~J.}\ \bibnamefont {Bollinger}},\ }\bibfield  {title}
  {\bibinfo {title} {Quantum-enhanced sensing of displacements and electric
  fields with two-dimensional trapped-ion crystals},\ }\href
  {https://doi.org/10.1126/science.abi5226} {\bibfield  {journal} {\bibinfo
  {journal} {Science}\ }\textbf {\bibinfo {volume} {373}},\ \bibinfo {pages}
  {673--678} (\bibinfo {year} {2021})}\BibitemShut {NoStop}%
\bibitem [{\citenamefont {Zhang}\ \emph {et~al.}(2018)\citenamefont {Zhang},
  \citenamefont {Lee}, \citenamefont {Kumar}, \citenamefont {Arnold},
  \citenamefont {Masson}, \citenamefont {Grimsmo}, \citenamefont {Parkins},\
  and\ \citenamefont {Barrett}}]{Zhiqiang2018DickemodelSV}%
  \BibitemOpen
  \bibfield  {author} {\bibinfo {author} {\bibfnamefont {Z.}~\bibnamefont
  {Zhang}}, \bibinfo {author} {\bibfnamefont {C.~H.}\ \bibnamefont {Lee}},
  \bibinfo {author} {\bibfnamefont {R.}~\bibnamefont {Kumar}}, \bibinfo
  {author} {\bibfnamefont {K.~J.}\ \bibnamefont {Arnold}}, \bibinfo {author}
  {\bibfnamefont {S.~J.}\ \bibnamefont {Masson}}, \bibinfo {author}
  {\bibfnamefont {A.~L.}\ \bibnamefont {Grimsmo}}, \bibinfo {author}
  {\bibfnamefont {A.~S.}\ \bibnamefont {Parkins}},\ and\ \bibinfo {author}
  {\bibfnamefont {M.~D.}\ \bibnamefont {Barrett}},\ }\bibfield  {title}
  {\bibinfo {title} {{D}icke-model simulation via cavity-assisted {R}aman
  transitions},\ }\href {https://doi.org/10.1103/PhysRevA.97.043858} {\bibfield
   {journal} {\bibinfo  {journal} {Phys. Rev. A}\ }\textbf {\bibinfo {volume}
  {97}},\ \bibinfo {pages} {043858} (\bibinfo {year} {2018})}\BibitemShut
  {NoStop}%
\bibitem [{\citenamefont {Black}\ \emph {et~al.}(2003)\citenamefont {Black},
  \citenamefont {Chan},\ and\ \citenamefont {Vuleti\ifmmode~\acute{c}\else
  \'{c}\fi{}}}]{PhysRevLett.91.203001}%
  \BibitemOpen
  \bibfield  {author} {\bibinfo {author} {\bibfnamefont {A.~T.}\ \bibnamefont
  {Black}}, \bibinfo {author} {\bibfnamefont {H.~W.}\ \bibnamefont {Chan}},\
  and\ \bibinfo {author} {\bibfnamefont {V.}~\bibnamefont
  {Vuleti\ifmmode~\acute{c}\else \'{c}\fi{}}},\ }\bibfield  {title} {\bibinfo
  {title} {Observation of collective friction forces due to spatial
  self-organization of atoms: From {R}ayleigh to {B}ragg scattering},\ }\href
  {https://doi.org/10.1103/PhysRevLett.91.203001} {\bibfield  {journal}
  {\bibinfo  {journal} {Phys. Rev. Lett.}\ }\textbf {\bibinfo {volume} {91}},\
  \bibinfo {pages} {203001} (\bibinfo {year} {2003})}\BibitemShut {NoStop}%
\bibitem [{\citenamefont {Zhao}\ and\ \citenamefont
  {Hwang}(2022)}]{PhysRevLett.128.163601}%
  \BibitemOpen
  \bibfield  {author} {\bibinfo {author} {\bibfnamefont {J.}~\bibnamefont
  {Zhao}}\ and\ \bibinfo {author} {\bibfnamefont {M.-J.}\ \bibnamefont
  {Hwang}},\ }\bibfield  {title} {\bibinfo {title} {Frustrated superradiant
  phase transition},\ }\href {https://doi.org/10.1103/PhysRevLett.128.163601}
  {\bibfield  {journal} {\bibinfo  {journal} {Phys. Rev. Lett.}\ }\textbf
  {\bibinfo {volume} {128}},\ \bibinfo {pages} {163601} (\bibinfo {year}
  {2022})}\BibitemShut {NoStop}%
\bibitem [{\citenamefont {Zhao}\ and\ \citenamefont
  {Hwang}(2023)}]{zhao2022anomalous}%
  \BibitemOpen
  \bibfield  {author} {\bibinfo {author} {\bibfnamefont {J.}~\bibnamefont
  {Zhao}}\ and\ \bibinfo {author} {\bibfnamefont {M.-J.}\ \bibnamefont
  {Hwang}},\ }\bibfield  {title} {\bibinfo {title} {Anomalous criticality with
  bounded fluctuations and long-range frustration induced by broken
  time-reversal symmetry},\ }\href
  {https://doi.org/10.1103/PhysRevResearch.5.L042016} {\bibfield  {journal}
  {\bibinfo  {journal} {Phys. Rev. Res.}\ }\textbf {\bibinfo {volume} {5}},\
  \bibinfo {pages} {L042016} (\bibinfo {year} {2023})}\BibitemShut {NoStop}%
\bibitem [{\citenamefont {Fallas~Padilla}\ \emph {et~al.}(2022)\citenamefont
  {Fallas~Padilla}, \citenamefont {Pu}, \citenamefont {Cheng},\ and\
  \citenamefont {Zhang}}]{PhysRevLett.129.183602}%
  \BibitemOpen
  \bibfield  {author} {\bibinfo {author} {\bibfnamefont {D.}~\bibnamefont
  {Fallas~Padilla}}, \bibinfo {author} {\bibfnamefont {H.}~\bibnamefont {Pu}},
  \bibinfo {author} {\bibfnamefont {G.-J.}\ \bibnamefont {Cheng}},\ and\
  \bibinfo {author} {\bibfnamefont {Y.-Y.}\ \bibnamefont {Zhang}},\ }\bibfield
  {title} {\bibinfo {title} {Understanding the quantum rabi ring using
  analogies to quantum magnetism},\ }\href
  {https://doi.org/10.1103/PhysRevLett.129.183602} {\bibfield  {journal}
  {\bibinfo  {journal} {Phys. Rev. Lett.}\ }\textbf {\bibinfo {volume} {129}},\
  \bibinfo {pages} {183602} (\bibinfo {year} {2022})}\BibitemShut {NoStop}%
\bibitem [{\citenamefont {Shen}\ \emph {et~al.}(2018)\citenamefont {Shen},
  \citenamefont {Zhen},\ and\ \citenamefont {Fu}}]{PhysRevLett.120.146402}%
  \BibitemOpen
  \bibfield  {author} {\bibinfo {author} {\bibfnamefont {H.}~\bibnamefont
  {Shen}}, \bibinfo {author} {\bibfnamefont {B.}~\bibnamefont {Zhen}},\ and\
  \bibinfo {author} {\bibfnamefont {L.}~\bibnamefont {Fu}},\ }\bibfield
  {title} {\bibinfo {title} {Topological band theory for non-{H}ermitian
  {H}amiltonians},\ }\href {https://doi.org/10.1103/PhysRevLett.120.146402}
  {\bibfield  {journal} {\bibinfo  {journal} {Phys. Rev. Lett.}\ }\textbf
  {\bibinfo {volume} {120}},\ \bibinfo {pages} {146402} (\bibinfo {year}
  {2018})}\BibitemShut {NoStop}%
\bibitem [{\citenamefont {Vasiloiu}\ \emph {et~al.}(2018)\citenamefont
  {Vasiloiu}, \citenamefont {Carollo},\ and\ \citenamefont
  {Garrahan}}]{PhysRevB.98.094308}%
  \BibitemOpen
  \bibfield  {author} {\bibinfo {author} {\bibfnamefont {L.~M.}\ \bibnamefont
  {Vasiloiu}}, \bibinfo {author} {\bibfnamefont {F.}~\bibnamefont {Carollo}},\
  and\ \bibinfo {author} {\bibfnamefont {J.~P.}\ \bibnamefont {Garrahan}},\
  }\bibfield  {title} {\bibinfo {title} {Enhancing correlation times for edge
  spins through dissipation},\ }\href
  {https://doi.org/10.1103/PhysRevB.98.094308} {\bibfield  {journal} {\bibinfo
  {journal} {Phys. Rev. B}\ }\textbf {\bibinfo {volume} {98}},\ \bibinfo
  {pages} {094308} (\bibinfo {year} {2018})}\BibitemShut {NoStop}%
\bibitem [{\citenamefont {Okuma}\ \emph {et~al.}(2020)\citenamefont {Okuma},
  \citenamefont {Kawabata}, \citenamefont {Shiozaki},\ and\ \citenamefont
  {Sato}}]{PhysRevLett.124.086801}%
  \BibitemOpen
  \bibfield  {author} {\bibinfo {author} {\bibfnamefont {N.}~\bibnamefont
  {Okuma}}, \bibinfo {author} {\bibfnamefont {K.}~\bibnamefont {Kawabata}},
  \bibinfo {author} {\bibfnamefont {K.}~\bibnamefont {Shiozaki}},\ and\
  \bibinfo {author} {\bibfnamefont {M.}~\bibnamefont {Sato}},\ }\bibfield
  {title} {\bibinfo {title} {Topological origin of non-{H}ermitian skin
  effects},\ }\href {https://doi.org/10.1103/PhysRevLett.124.086801} {\bibfield
   {journal} {\bibinfo  {journal} {Phys. Rev. Lett.}\ }\textbf {\bibinfo
  {volume} {124}},\ \bibinfo {pages} {086801} (\bibinfo {year}
  {2020})}\BibitemShut {NoStop}%
\bibitem [{\citenamefont {Borgnia}\ \emph {et~al.}(2020)\citenamefont
  {Borgnia}, \citenamefont {Kruchkov},\ and\ \citenamefont
  {Slager}}]{PhysRevLett.124.056802}%
  \BibitemOpen
  \bibfield  {author} {\bibinfo {author} {\bibfnamefont {D.~S.}\ \bibnamefont
  {Borgnia}}, \bibinfo {author} {\bibfnamefont {A.~J.}\ \bibnamefont
  {Kruchkov}},\ and\ \bibinfo {author} {\bibfnamefont {R.-J.}\ \bibnamefont
  {Slager}},\ }\bibfield  {title} {\bibinfo {title} {Non-{H}ermitian boundary
  modes and topology},\ }\href {https://doi.org/10.1103/PhysRevLett.124.056802}
  {\bibfield  {journal} {\bibinfo  {journal} {Phys. Rev. Lett.}\ }\textbf
  {\bibinfo {volume} {124}},\ \bibinfo {pages} {056802} (\bibinfo {year}
  {2020})}\BibitemShut {NoStop}%
\bibitem [{\citenamefont {Xu}\ \emph {et~al.}(2023)\citenamefont {Xu},
  \citenamefont {Zhou}, \citenamefont {Guo},\ and\ \citenamefont
  {Zhou}}]{PhysRevResearch.5.043004}%
  \BibitemOpen
  \bibfield  {author} {\bibinfo {author} {\bibfnamefont {X.-S.}\ \bibnamefont
  {Xu}}, \bibinfo {author} {\bibfnamefont {X.-F.}\ \bibnamefont {Zhou}},
  \bibinfo {author} {\bibfnamefont {G.-C.}\ \bibnamefont {Guo}},\ and\ \bibinfo
  {author} {\bibfnamefont {Z.-W.}\ \bibnamefont {Zhou}},\ }\bibfield  {title}
  {\bibinfo {title} {Dissipation-induced {L}iouville-{M}ajorana modes in open
  quantum system},\ }\href {https://doi.org/10.1103/PhysRevResearch.5.043004}
  {\bibfield  {journal} {\bibinfo  {journal} {Phys. Rev. Res.}\ }\textbf
  {\bibinfo {volume} {5}},\ \bibinfo {pages} {043004} (\bibinfo {year}
  {2023})}\BibitemShut {NoStop}%
\bibitem [{\citenamefont {Sieberer}\ \emph {et~al.}(2013)\citenamefont
  {Sieberer}, \citenamefont {Huber}, \citenamefont {Altman},\ and\
  \citenamefont {Diehl}}]{PhysRevLett.110.195301}%
  \BibitemOpen
  \bibfield  {author} {\bibinfo {author} {\bibfnamefont {L.~M.}\ \bibnamefont
  {Sieberer}}, \bibinfo {author} {\bibfnamefont {S.~D.}\ \bibnamefont {Huber}},
  \bibinfo {author} {\bibfnamefont {E.}~\bibnamefont {Altman}},\ and\ \bibinfo
  {author} {\bibfnamefont {S.}~\bibnamefont {Diehl}},\ }\bibfield  {title}
  {\bibinfo {title} {Dynamical critical phenomena in driven-dissipative
  systems},\ }\href {https://doi.org/10.1103/PhysRevLett.110.195301} {\bibfield
   {journal} {\bibinfo  {journal} {Phys. Rev. Lett.}\ }\textbf {\bibinfo
  {volume} {110}},\ \bibinfo {pages} {195301} (\bibinfo {year}
  {2013})}\BibitemShut {NoStop}%
\bibitem [{\citenamefont {Tonielli}\ \emph {et~al.}(2019)\citenamefont
  {Tonielli}, \citenamefont {Fazio}, \citenamefont {Diehl},\ and\ \citenamefont
  {Marino}}]{PhysRevLett.122.040604}%
  \BibitemOpen
  \bibfield  {author} {\bibinfo {author} {\bibfnamefont {F.}~\bibnamefont
  {Tonielli}}, \bibinfo {author} {\bibfnamefont {R.}~\bibnamefont {Fazio}},
  \bibinfo {author} {\bibfnamefont {S.}~\bibnamefont {Diehl}},\ and\ \bibinfo
  {author} {\bibfnamefont {J.}~\bibnamefont {Marino}},\ }\bibfield  {title}
  {\bibinfo {title} {Orthogonality catastrophe in dissipative quantum many-body
  systems},\ }\href {https://doi.org/10.1103/PhysRevLett.122.040604} {\bibfield
   {journal} {\bibinfo  {journal} {Phys. Rev. Lett.}\ }\textbf {\bibinfo
  {volume} {122}},\ \bibinfo {pages} {040604} (\bibinfo {year}
  {2019})}\BibitemShut {NoStop}%
\bibitem [{\citenamefont {Hanai}\ and\ \citenamefont
  {Littlewood}(2020)}]{PhysRevResearch.2.033018}%
  \BibitemOpen
  \bibfield  {author} {\bibinfo {author} {\bibfnamefont {R.}~\bibnamefont
  {Hanai}}\ and\ \bibinfo {author} {\bibfnamefont {P.~B.}\ \bibnamefont
  {Littlewood}},\ }\bibfield  {title} {\bibinfo {title} {Critical fluctuations
  at a many-body exceptional point},\ }\href
  {https://doi.org/10.1103/PhysRevResearch.2.033018} {\bibfield  {journal}
  {\bibinfo  {journal} {Phys. Rev. Res.}\ }\textbf {\bibinfo {volume} {2}},\
  \bibinfo {pages} {033018} (\bibinfo {year} {2020})}\BibitemShut {NoStop}%
\bibitem [{\citenamefont {Tomadin}\ \emph {et~al.}(2011)\citenamefont
  {Tomadin}, \citenamefont {Diehl},\ and\ \citenamefont
  {Zoller}}]{PhysRevA.83.013611}%
  \BibitemOpen
  \bibfield  {author} {\bibinfo {author} {\bibfnamefont {A.}~\bibnamefont
  {Tomadin}}, \bibinfo {author} {\bibfnamefont {S.}~\bibnamefont {Diehl}},\
  and\ \bibinfo {author} {\bibfnamefont {P.}~\bibnamefont {Zoller}},\
  }\bibfield  {title} {\bibinfo {title} {Nonequilibrium phase diagram of a
  driven and dissipative many-body system},\ }\href
  {https://doi.org/10.1103/PhysRevA.83.013611} {\bibfield  {journal} {\bibinfo
  {journal} {Phys. Rev. A}\ }\textbf {\bibinfo {volume} {83}},\ \bibinfo
  {pages} {013611} (\bibinfo {year} {2011})}\BibitemShut {NoStop}%
\bibitem [{\citenamefont {Soriente}\ \emph {et~al.}(2018)\citenamefont
  {Soriente}, \citenamefont {Donner}, \citenamefont {Chitra},\ and\
  \citenamefont {Zilberberg}}]{PhysRevLett.120.183603}%
  \BibitemOpen
  \bibfield  {author} {\bibinfo {author} {\bibfnamefont {M.}~\bibnamefont
  {Soriente}}, \bibinfo {author} {\bibfnamefont {T.}~\bibnamefont {Donner}},
  \bibinfo {author} {\bibfnamefont {R.}~\bibnamefont {Chitra}},\ and\ \bibinfo
  {author} {\bibfnamefont {O.}~\bibnamefont {Zilberberg}},\ }\bibfield  {title}
  {\bibinfo {title} {Dissipation-induced anomalous multicritical phenomena},\
  }\href {https://doi.org/10.1103/PhysRevLett.120.183603} {\bibfield  {journal}
  {\bibinfo  {journal} {Phys. Rev. Lett.}\ }\textbf {\bibinfo {volume} {120}},\
  \bibinfo {pages} {183603} (\bibinfo {year} {2018})}\BibitemShut {NoStop}%
\bibitem [{\citenamefont {Iemini}\ \emph {et~al.}(2018)\citenamefont {Iemini},
  \citenamefont {Russomanno}, \citenamefont {Keeling}, \citenamefont
  {Schir\`o}, \citenamefont {Dalmonte},\ and\ \citenamefont
  {Fazio}}]{PhysRevLett.121.035301}%
  \BibitemOpen
  \bibfield  {author} {\bibinfo {author} {\bibfnamefont {F.}~\bibnamefont
  {Iemini}}, \bibinfo {author} {\bibfnamefont {A.}~\bibnamefont {Russomanno}},
  \bibinfo {author} {\bibfnamefont {J.}~\bibnamefont {Keeling}}, \bibinfo
  {author} {\bibfnamefont {M.}~\bibnamefont {Schir\`o}}, \bibinfo {author}
  {\bibfnamefont {M.}~\bibnamefont {Dalmonte}},\ and\ \bibinfo {author}
  {\bibfnamefont {R.}~\bibnamefont {Fazio}},\ }\bibfield  {title} {\bibinfo
  {title} {Boundary time crystals},\ }\href
  {https://doi.org/10.1103/PhysRevLett.121.035301} {\bibfield  {journal}
  {\bibinfo  {journal} {Phys. Rev. Lett.}\ }\textbf {\bibinfo {volume} {121}},\
  \bibinfo {pages} {035301} (\bibinfo {year} {2018})}\BibitemShut {NoStop}%
\bibitem [{\citenamefont {Prazeres}\ \emph {et~al.}(2021)\citenamefont
  {Prazeres}, \citenamefont {Souza},\ and\ \citenamefont
  {Iemini}}]{PhysRevB.103.184308}%
  \BibitemOpen
  \bibfield  {author} {\bibinfo {author} {\bibfnamefont {L.~F.~d.}\
  \bibnamefont {Prazeres}}, \bibinfo {author} {\bibfnamefont {L.~d.~S.}\
  \bibnamefont {Souza}},\ and\ \bibinfo {author} {\bibfnamefont
  {F.}~\bibnamefont {Iemini}},\ }\bibfield  {title} {\bibinfo {title} {Boundary
  time crystals in collective $d$-level systems},\ }\href
  {https://doi.org/10.1103/PhysRevB.103.184308} {\bibfield  {journal} {\bibinfo
   {journal} {Phys. Rev. B}\ }\textbf {\bibinfo {volume} {103}},\ \bibinfo
  {pages} {184308} (\bibinfo {year} {2021})}\BibitemShut {NoStop}%
\bibitem [{\citenamefont {Stitely}\ \emph {et~al.}(2020)\citenamefont
  {Stitely}, \citenamefont {Giraldo}, \citenamefont {Krauskopf},\ and\
  \citenamefont {Parkins}}]{PhysRevResearch.2.033131}%
  \BibitemOpen
  \bibfield  {author} {\bibinfo {author} {\bibfnamefont {K.~C.}\ \bibnamefont
  {Stitely}}, \bibinfo {author} {\bibfnamefont {A.}~\bibnamefont {Giraldo}},
  \bibinfo {author} {\bibfnamefont {B.}~\bibnamefont {Krauskopf}},\ and\
  \bibinfo {author} {\bibfnamefont {S.}~\bibnamefont {Parkins}},\ }\bibfield
  {title} {\bibinfo {title} {Nonlinear semiclassical dynamics of the
  unbalanced, open {D}icke model},\ }\href
  {https://doi.org/10.1103/PhysRevResearch.2.033131} {\bibfield  {journal}
  {\bibinfo  {journal} {Phys. Rev. Res.}\ }\textbf {\bibinfo {volume} {2}},\
  \bibinfo {pages} {033131} (\bibinfo {year} {2020})}\BibitemShut {NoStop}%
\bibitem [{\citenamefont {Liu}\ \emph {et~al.}(2017)\citenamefont {Liu},
  \citenamefont {Chesi}, \citenamefont {Ying}, \citenamefont {Chen},
  \citenamefont {Luo},\ and\ \citenamefont {Lin}}]{PhysRevLett.119.220601}%
  \BibitemOpen
  \bibfield  {author} {\bibinfo {author} {\bibfnamefont {M.}~\bibnamefont
  {Liu}}, \bibinfo {author} {\bibfnamefont {S.}~\bibnamefont {Chesi}}, \bibinfo
  {author} {\bibfnamefont {Z.-J.}\ \bibnamefont {Ying}}, \bibinfo {author}
  {\bibfnamefont {X.}~\bibnamefont {Chen}}, \bibinfo {author} {\bibfnamefont
  {H.-G.}\ \bibnamefont {Luo}},\ and\ \bibinfo {author} {\bibfnamefont {H.-Q.}\
  \bibnamefont {Lin}},\ }\bibfield  {title} {\bibinfo {title} {Universal
  scaling and critical exponents of the anisotropic quantum {R}abi model},\
  }\href {https://doi.org/10.1103/PhysRevLett.119.220601} {\bibfield  {journal}
  {\bibinfo  {journal} {Phys. Rev. Lett.}\ }\textbf {\bibinfo {volume} {119}},\
  \bibinfo {pages} {220601} (\bibinfo {year} {2017})}\BibitemShut {NoStop}%
\bibitem [{\citenamefont {Emary}\ and\ \citenamefont
  {Brandes}(2003)}]{PhysRevE.67.066203}%
  \BibitemOpen
  \bibfield  {author} {\bibinfo {author} {\bibfnamefont {C.}~\bibnamefont
  {Emary}}\ and\ \bibinfo {author} {\bibfnamefont {T.}~\bibnamefont
  {Brandes}},\ }\bibfield  {title} {\bibinfo {title} {Chaos and the quantum
  phase transition in the {D}icke model},\ }\href
  {https://doi.org/10.1103/PhysRevE.67.066203} {\bibfield  {journal} {\bibinfo
  {journal} {Phys. Rev. E}\ }\textbf {\bibinfo {volume} {67}},\ \bibinfo
  {pages} {066203} (\bibinfo {year} {2003})}\BibitemShut {NoStop}%
\bibitem [{\citenamefont {Xu}\ and\ \citenamefont
  {Pu}(2019)}]{PhysRevLett.122.193201}%
  \BibitemOpen
  \bibfield  {author} {\bibinfo {author} {\bibfnamefont {Y.}~\bibnamefont
  {Xu}}\ and\ \bibinfo {author} {\bibfnamefont {H.}~\bibnamefont {Pu}},\
  }\bibfield  {title} {\bibinfo {title} {Emergent universality in a quantum
  tricritical {D}icke model},\ }\href
  {https://doi.org/10.1103/PhysRevLett.122.193201} {\bibfield  {journal}
  {\bibinfo  {journal} {Phys. Rev. Lett.}\ }\textbf {\bibinfo {volume} {122}},\
  \bibinfo {pages} {193201} (\bibinfo {year} {2019})}\BibitemShut {NoStop}%
\bibitem [{\citenamefont {Nayfeh}\ and\ \citenamefont
  {Balachandran}(1995)}]{Nayfehbook}%
  \BibitemOpen
  \bibfield  {author} {\bibinfo {author} {\bibfnamefont {A.~H.}\ \bibnamefont
  {Nayfeh}}\ and\ \bibinfo {author} {\bibfnamefont {B.}~\bibnamefont
  {Balachandran}},\ }\href {https://doi.org/10.1002/9783527617548} {\emph
  {\bibinfo {title} {\it Applied Nonlinear Dynamics}}}\ (\bibinfo  {publisher}
  {John Wiley \& Sons, Ltd},\ \bibinfo {address} {Cambridge},\ \bibinfo {year}
  {1995})\BibitemShut {NoStop}%
\bibitem [{\citenamefont {Serafini}(2017)}]{Serafini}%
  \BibitemOpen
  \bibfield  {author} {\bibinfo {author} {\bibfnamefont {A.}~\bibnamefont
  {Serafini}},\ }\href {https://doi.org/10.1201/9781315118727} {\emph {\bibinfo
  {title} {Quantum Continuous Variables: A Primer of Theoretical Methods}}}\
  (\bibinfo  {publisher} {Taylor \& Francis Group},\ \bibinfo {year}
  {2017})\BibitemShut {NoStop}%
\bibitem [{\citenamefont {M\"uller}\ \emph {et~al.}(1991)\citenamefont
  {M\"uller}, \citenamefont {Stolze}, \citenamefont {Leschke},\ and\
  \citenamefont {Nagel}}]{PhysRevA.44.1022}%
  \BibitemOpen
  \bibfield  {author} {\bibinfo {author} {\bibfnamefont {L.}~\bibnamefont
  {M\"uller}}, \bibinfo {author} {\bibfnamefont {J.}~\bibnamefont {Stolze}},
  \bibinfo {author} {\bibfnamefont {H.}~\bibnamefont {Leschke}},\ and\ \bibinfo
  {author} {\bibfnamefont {P.}~\bibnamefont {Nagel}},\ }\bibfield  {title}
  {\bibinfo {title} {Classical and quantum phase-space behavior of a spin-boson
  system},\ }\href {https://doi.org/10.1103/PhysRevA.44.1022} {\bibfield
  {journal} {\bibinfo  {journal} {Phys. Rev. A}\ }\textbf {\bibinfo {volume}
  {44}},\ \bibinfo {pages} {1022--1033} (\bibinfo {year} {1991})}\BibitemShut
  {NoStop}%
\bibitem [{\citenamefont {Finney}\ and\ \citenamefont
  {Gea-Banacloche}(1994)}]{PhysRevA.50.2040}%
  \BibitemOpen
  \bibfield  {author} {\bibinfo {author} {\bibfnamefont {G.~A.}\ \bibnamefont
  {Finney}}\ and\ \bibinfo {author} {\bibfnamefont {J.}~\bibnamefont
  {Gea-Banacloche}},\ }\bibfield  {title} {\bibinfo {title} {Quasiclassical
  approximation for the spin-boson {H}amiltonian with counterrotating terms},\
  }\href {https://doi.org/10.1103/PhysRevA.50.2040} {\bibfield  {journal}
  {\bibinfo  {journal} {Phys. Rev. A}\ }\textbf {\bibinfo {volume} {50}},\
  \bibinfo {pages} {2040--2052} (\bibinfo {year} {1994})}\BibitemShut {NoStop}%
\bibitem [{\citenamefont {Finney}\ and\ \citenamefont
  {Gea-Banacloche}(1996)}]{PhysRevE.54.1449}%
  \BibitemOpen
  \bibfield  {author} {\bibinfo {author} {\bibfnamefont {G.~A.}\ \bibnamefont
  {Finney}}\ and\ \bibinfo {author} {\bibfnamefont {J.}~\bibnamefont
  {Gea-Banacloche}},\ }\bibfield  {title} {\bibinfo {title} {Quantum
  suppression of chaos in the spin-boson model},\ }\href
  {https://doi.org/10.1103/PhysRevE.54.1449} {\bibfield  {journal} {\bibinfo
  {journal} {Phys. Rev. E}\ }\textbf {\bibinfo {volume} {54}},\ \bibinfo
  {pages} {1449--1456} (\bibinfo {year} {1996})}\BibitemShut {NoStop}%
\bibitem [{\citenamefont {Furuya}\ \emph {et~al.}(1998)\citenamefont {Furuya},
  \citenamefont {Nemes},\ and\ \citenamefont
  {Pellegrino}}]{PhysRevLett.80.5524}%
  \BibitemOpen
  \bibfield  {author} {\bibinfo {author} {\bibfnamefont {K.}~\bibnamefont
  {Furuya}}, \bibinfo {author} {\bibfnamefont {M.~C.}\ \bibnamefont {Nemes}},\
  and\ \bibinfo {author} {\bibfnamefont {G.~Q.}\ \bibnamefont {Pellegrino}},\
  }\bibfield  {title} {\bibinfo {title} {Quantum dynamical manifestation of
  chaotic behavior in the process of entanglement},\ }\href
  {https://doi.org/10.1103/PhysRevLett.80.5524} {\bibfield  {journal} {\bibinfo
   {journal} {Phys. Rev. Lett.}\ }\textbf {\bibinfo {volume} {80}},\ \bibinfo
  {pages} {5524--5527} (\bibinfo {year} {1998})}\BibitemShut {NoStop}%
\bibitem [{\citenamefont {Angelo}\ \emph {et~al.}(2001)\citenamefont {Angelo},
  \citenamefont {Furuya}, \citenamefont {Nemes},\ and\ \citenamefont
  {Pellegrino}}]{PhysRevA.64.043801}%
  \BibitemOpen
  \bibfield  {author} {\bibinfo {author} {\bibfnamefont {R.~M.}\ \bibnamefont
  {Angelo}}, \bibinfo {author} {\bibfnamefont {K.}~\bibnamefont {Furuya}},
  \bibinfo {author} {\bibfnamefont {M.~C.}\ \bibnamefont {Nemes}},\ and\
  \bibinfo {author} {\bibfnamefont {G.~Q.}\ \bibnamefont {Pellegrino}},\
  }\bibfield  {title} {\bibinfo {title} {Recoherence in the entanglement
  dynamics and classical orbits in the ${N}$-atom {J}aynes-{C}ummings model},\
  }\href {https://doi.org/10.1103/PhysRevA.64.043801} {\bibfield  {journal}
  {\bibinfo  {journal} {Phys. Rev. A}\ }\textbf {\bibinfo {volume} {64}},\
  \bibinfo {pages} {043801} (\bibinfo {year} {2001})}\BibitemShut {NoStop}%
\bibitem [{\citenamefont {Lambert}\ \emph {et~al.}(2004)\citenamefont
  {Lambert}, \citenamefont {Emary},\ and\ \citenamefont
  {Brandes}}]{PhysRevLett.92.073602}%
  \BibitemOpen
  \bibfield  {author} {\bibinfo {author} {\bibfnamefont {N.}~\bibnamefont
  {Lambert}}, \bibinfo {author} {\bibfnamefont {C.}~\bibnamefont {Emary}},\
  and\ \bibinfo {author} {\bibfnamefont {T.}~\bibnamefont {Brandes}},\
  }\bibfield  {title} {\bibinfo {title} {Entanglement and the phase transition
  in single-mode superradiance},\ }\href
  {https://doi.org/10.1103/PhysRevLett.92.073602} {\bibfield  {journal}
  {\bibinfo  {journal} {Phys. Rev. Lett.}\ }\textbf {\bibinfo {volume} {92}},\
  \bibinfo {pages} {073602} (\bibinfo {year} {2004})}\BibitemShut {NoStop}%
\bibitem [{\citenamefont {Garbe}\ \emph {et~al.}(2020)\citenamefont {Garbe},
  \citenamefont {Wade}, \citenamefont {Minganti}, \citenamefont {Shammah},
  \citenamefont {Felicetti},\ and\ \citenamefont {Nori}}]{Garbe2020}%
  \BibitemOpen
  \bibfield  {author} {\bibinfo {author} {\bibfnamefont {L.}~\bibnamefont
  {Garbe}}, \bibinfo {author} {\bibfnamefont {P.}~\bibnamefont {Wade}},
  \bibinfo {author} {\bibfnamefont {F.}~\bibnamefont {Minganti}}, \bibinfo
  {author} {\bibfnamefont {N.}~\bibnamefont {Shammah}}, \bibinfo {author}
  {\bibfnamefont {S.}~\bibnamefont {Felicetti}},\ and\ \bibinfo {author}
  {\bibfnamefont {F.}~\bibnamefont {Nori}},\ }\bibfield  {title} {\bibinfo
  {title} {Dissipation-induced bistability in the two-photon {D}icke model},\
  }\href {https://doi.org/10.1038/s41598-020-69704-6} {\bibfield  {journal}
  {\bibinfo  {journal} {Scientific Reports}\ }\textbf {\bibinfo {volume}
  {10}},\ \bibinfo {pages} {13408} (\bibinfo {year} {2020})}\BibitemShut
  {NoStop}%
\bibitem [{\citenamefont {Xu}\ \emph {et~al.}(2021)\citenamefont {Xu},
  \citenamefont {Fallas~Padilla},\ and\ \citenamefont
  {Pu}}]{PhysRevA.104.043708}%
  \BibitemOpen
  \bibfield  {author} {\bibinfo {author} {\bibfnamefont {Y.}~\bibnamefont
  {Xu}}, \bibinfo {author} {\bibfnamefont {D.}~\bibnamefont {Fallas~Padilla}},\
  and\ \bibinfo {author} {\bibfnamefont {H.}~\bibnamefont {Pu}},\ }\bibfield
  {title} {\bibinfo {title} {Multicriticality and quantum fluctuation in a
  generalized {D}icke model},\ }\href
  {https://doi.org/10.1103/PhysRevA.104.043708} {\bibfield  {journal} {\bibinfo
   {journal} {Phys. Rev. A}\ }\textbf {\bibinfo {volume} {104}},\ \bibinfo
  {pages} {043708} (\bibinfo {year} {2021})}\BibitemShut {NoStop}%
\bibitem [{\citenamefont {Skulte}\ \emph {et~al.}(2021)\citenamefont {Skulte},
  \citenamefont {Kongkhambut}, \citenamefont {Ke\ss{}ler}, \citenamefont
  {Hemmerich}, \citenamefont {Mathey},\ and\ \citenamefont
  {Cosme}}]{PhysRevA.104.063705}%
  \BibitemOpen
  \bibfield  {author} {\bibinfo {author} {\bibfnamefont {J.}~\bibnamefont
  {Skulte}}, \bibinfo {author} {\bibfnamefont {P.}~\bibnamefont {Kongkhambut}},
  \bibinfo {author} {\bibfnamefont {H.}~\bibnamefont {Ke\ss{}ler}}, \bibinfo
  {author} {\bibfnamefont {A.}~\bibnamefont {Hemmerich}}, \bibinfo {author}
  {\bibfnamefont {L.}~\bibnamefont {Mathey}},\ and\ \bibinfo {author}
  {\bibfnamefont {J.~G.}\ \bibnamefont {Cosme}},\ }\bibfield  {title} {\bibinfo
  {title} {Parametrically driven dissipative three-level {D}icke model},\
  }\href {https://doi.org/10.1103/PhysRevA.104.063705} {\bibfield  {journal}
  {\bibinfo  {journal} {Phys. Rev. A}\ }\textbf {\bibinfo {volume} {104}},\
  \bibinfo {pages} {063705} (\bibinfo {year} {2021})}\BibitemShut {NoStop}%
\end{thebibliography}%

\end{document}